# Perspectives on Physics of E×B Discharges Relevant to Plasma Propulsion and Similar Technologies

Igor D. Kaganovich[1*], Andrei Smolyakov[2], Yevgeny Raitses[1], Eduardo Ahedo[3], Ioannis G. Mikellides[4], Benjamin Jorns[5], Francesco Taccogna[6], Renaud Gueroult[7], Sedina Tsikata[8], Anne Bourdon[9], Jean-Pierre Boeuf[7], Michael Keidar[10], Andrew Tasman Powis[1], Mario Merino[3], Mark Cappelli[11], Kentaro Hara[11], Johan A. Carlsson[1], Nathaniel J. Fisch[1], Pascal Chabert[9], Irina Schweigert[10,12], Trevor Lafleur[13], Konstantin Matyash[14], Alexander V. Khrabrov[1], Rod W. Boswell[15], Amnon Fruchtman[16]

*1Princeton Plasma Physics Laboratory, Princeton NJ 08543, USA*
*2University of Saskatchewan, 116 Science Place, Saskatoon, SK S7N 5E2 Canada*
*3Equipo de Propulsión Espacial y Plasmas, Universidad Carlos III de Madrid, Leganés 28911, Spain*
*4Jet Propulsion Laboratory, California Institute of Technology, Pasadena, CA 91109, USA*
*5 Department of Aerospace Engineering, University of Michigan, Ann Arbor, MI 48109,  USA*
*6CNR-Institute for Plasma Science and Technology, via Amendola 122/D 70126-Bari, Italy*
*7 LAPLACE, Université de Toulouse, CNRS, INPT, UPS, 118 Route de Narbonne, 31062 Toulouse, France*
*8ICARE, Electric Propulsion Team, Centre National de la Recherche Scientifique, 45071 Orléans, France*
*9Laboratoire de Physique des Plasmas, CNRS, Ecole Polytechnique, Sorbonne Université, Université Paris Sud, 91120 Palaiseau, France*
*10George Washington University, Washington D.C. 20052, USA*
*11Stanford University, California 94305-3032 USA*
*12Khristianovich Institute of Theoretical and Applied Mechanics RAS, Novosibirsk, 630090, Russia*
*13PlasmaPotential, Entry 29, 5/1 Moore Street, Canberra ACT 2601, Australia*
*14University of Greifswald, Greifswald, D-17487, Germany*
*15Space Plasma, Power and Propulsion Laboratory, Research School of Physics and Engineering, The Australian National University, Canberra, ACT, 2601, Australia.*
*16Holon Institute of Technology, Holon 58102, Israel*

**Abstract**

This paper provides perspectives on recent progress in the understanding of the physics of devices where the external magnetic field is applied perpendicularly to the discharge current. This configuration generates a strong electric field, which acts to accelerates ions. The many applications of this set up include generation of thrust for spacecraft propulsion and the separation of species in plasma mass separation devices. These "E×B" plasmas are subject to plasma-wall interaction effects as well as various micro and macro instabilities, and in many devices, we observe the emergence of anomalous transport.  This perspective presents the current understanding of the physics of these phenomena, state-of-the-art computational results, identifies critical questions, and suggests directions for future research.

## 1.  Introduction

Igor D. Kaganovich[1], Yevgeny Raitses[1], and Andrei Smolyakov[2]

*1Princeton Plasma Physics Laboratory, Princeton NJ 08543 USA*
*2University of Saskatchewan, 116 Science Place, Saskatoon, SK S7N 5E2 Canada*

This perspective describes joint efforts by the low-temperature plasma community and the plasma propulsion community to develop a rigorous understanding of the rich phenomena observed in E×B devices, as they were summarized in presentations given at the E×B workshop 2019, Princeton [1].

In E×B configurations the external magnetic field is applied perpendicular to the discharge current in order to generate a strong electric field. Such devices are widely used for electric propulsion; most common is the Hall or Hall-effect thruster, and for plasma processing in DC magnetrons.  Some of the discussed effects are also relevant to radiofrequency (RF) plasma sources, e.g., helicon sources and thrusters which make use of a magnetic nozzle to generate thrust. It is widely recognized that many of the complex physics effects observed in these devices are not well understood. Here, we share perspectives from 25 leading experts within the field, and report on the state-of-the-art in 8 major subtopics as well as propose recommendations on directions for future research.

---

* Corresponding author; email ikaganov@pppl.gov

Like many spacecraft components, it would be highly desirable to design plasma thrusters via robust engineering models, which through step-by-step procedures, can deliver predictable performance and lifetime. The status quo is far from this ideal situation. The design and development of plasma thrusters are not fully based on first principle physics models but rather rely on a semi-empirical approach, combined with long and expensive lifetime tests. This is because plasma propulsion is one of few remaining applications where engineering developments are seriously constrained due to the lack of a complete description of plasma properties. For example, in Hall thrusters, one of the most critical challenges is to self-consistently simulate the electron cross-field current, which affects the efficiency of propellant ionization and ion acceleration. Electron scattering in turbulent processes also affects the plasma-wall interactions and erosion, because it changes the sheath potential at the walls. Sixty years of experimental and theoretical research on Hall thrusters have provided much insight into the instability-driven mechanisms which contribute to the enhanced electron cross-field transport in these plasma devices. Nevertheless, we still do not have a sufficiently clear understanding of these mechanisms, and thus cannot accurately describe their contributions to the cross-field transport. Consequently, reliable and robust methods to predictively design these thrusters do not yet exist. Such methods are especially important for new applications e.g., very low power thrusters for CubeSats or extremely high-power thrusters for human exploration. For these applications, purely engineering solutions, such as size-scaling or clustering of multiple thrusters, are not necessarily optimal for achieving the required thruster characteristics. *Any variation of established thruster design, for example, achieving thruster compactness or higher thrust density, developing operational envelopes through throttling and variable thrust, or altering the propellant from Xenon, requires significant experimental work which could be greatly aided by physics-based predictive modeling.*

To facilitate a discussion on these complex processes between experimentalists, numerical simulation experts, and theorists, a series of specialized workshops started in 2012 [2]. A focused workshop on "E × B plasmas for space and industrial applications" was organized in Toulouse by J.P. Boeuf and A. Smolyakov in June 2017 [3]. A follow up to the workshop included a collection of special topic papers titled "Modern issues and applications of E × B plasmas" published in Physics of Plasmas in 2018 [3]. The next workshop was organized in Princeton by Y. Raitses, E.Choueri, I. Kaganovich, and A. Smolyakov in October 2018 [1]. The workshop participants discussed the following topics: "Validation & Verification for discharges and Sheaths", "Mechanisms of ECDI saturation and turbulence", "Kinetic vs Fluid, Hybrid models", "Low-frequency phenomena in E×B discharges, modeling and experiments", "Experiments in turbulence", "Towards full 3D modeling and GPUs", "E×B Mass-filtering", and "Unusual effects in magnetic nozzles". Workshop participants – about 40 leading experts in the field of E×B plasma physics and devices - agreed that there is a need to prepare perspectives on each of these research directions.

Correspondingly, this perspective article discusses 9 topics, which represent major directions in electric propulsion community:

1. Plasma-Wall Interactions in E×B Discharges Relevant to Propulsion Plasma Devices
2. Low-Frequency Oscillations in E×B Discharges
3. Experiments in Turbulence in Low Temperature, E×B Devices
4. Electron Drift Instabilities in E×B plasmas: Mechanisms, Nonlinear Saturation and Turbulent Transport
5. Fluid and Hybrid (Fluid-Kinetic) Modeling of E×B Discharges
6. Towards full Three-dimensional Modeling of Hall thruster E×B Discharges
7. ExB Configurations for Plasma Mass Separation Applications
8. Validation and Verification Procedures for Discharge Modeling
9 Magnetic Nozzles for Electric Propulsion

Section 2 (topic 1) describes plasma-wall interactions. It is a well-established experimental fact for Hall thrusters that the wall material can affect the discharge current, the electric field, and the plasma plume. Thruster performance can, therefore, be affected by the wall material. This wall material effect is commonly attributed to the electron-induced secondary electron emission (SEE), which is different between materials. The secondary electron emission causes additional transport across the magnetic field, the so-called near-wall conductivity, that can increase the electron current across the magnetic field and lead to a reduction of the thrust. Sheaths near the wall determine ion energy of ions impinging the walls and therefore, the sheath structure and sheath potential affect the wall erosion and, correspondingly the thruster lifetime. Channel of the Hall thruster is narrow, and plasma is collisionless with the electron mean free path with respect to collisions with ions and neutrals is much larger than the channel width. Correspondingly, the sheath structure and the voltage drop on the sheath is determined by many kinetic processes that control electron fluxes to the walls. For example, the electron emission from the wall becomes especially strong if the secondary electron emission yield, which is the ratio of the flux of emitted electrons to the flux of primary electrons, reaches or becomes larger than unity. Under such conditions, intense electron flux can cause a non-monotonic potential profile in the sheath, either so-called space charge limited or inverse sheath structures. The sheath can also become unstable in the presence of intense electron emission, and instabilities lead to oscillations of the wall potential, the so-called relaxation sheath oscillations. The value of the sheath potential is strongly affected by the non-Maxwellian character of the electron energy distribution function (EEDF) due to the depletion of the EEDF tail. The EEDF tail can form by many processes, including scattering in electron-neutral collisions as well as in turbulent processes, such as the oblique two-stream instability. Correspondingly, the wall fluxes and the sheath potential can be affected by anomalous transport. Examples of the complex interplay between all these processes are given in Section 2.

Section 3 (topic 2) describes low-frequency (typically <100 kHz) oscillations, namely the breathing and spoke-type oscillations observed in thrusters. These constitute the most intense oscillations in E×B devices, and may reach up to 100% modulation of plasma parameters. The breathing oscillations manifest as oscillations of the plasma and neutral densities; which are coupled through ionization. These oscillations develop mostly along the direction of the external electric field. These oscillations resemble predator-prey type oscillations, but recent studies showed that an adequate description of these modes cannot be reduced to a simple zero-dimensional description of predator-prey type oscillations. Theoretical models and simulations of the breathing oscillations suggest a large sensitivity to values of the effective electron mobility along the applied electric field (and hence sensitivity to a poorly known anomalous transport), which is still not well understood as well to details of ionization zone near the anode. In contrast, the spoke develops mostly in the azimuthal or E×B direction, perpendicular to the crossed electric (E) and magnetic (B) fields. The spoke may exist without ionization and could be caused by the Simon-Hoh (SH)-type instability, driven by the combination of the applied electric field and gradient in the plasma density. Ionization can also strongly affect the spoke, especially for devices with higher pressures such as DC magnetrons. With a few exceptions, modeling of the breathing and azimuthal spoke oscillations have been performed separately and without considering possible coupling effects. Some experimental data suggest that there is a coupling between them which makes analysis complicated and will require 3D modeling for a deeper analysis.

Section 4 (topic 3) describes experimental observations of high-frequency (> 1 MHz) and short-wavelength (< 1 mm) waves. These oscillations are believed to be the main contributing mechanism responsible for the enhancement of the electron cross-field current in these devices. The electron anomalous cross-field current is directly responsible for power losses in E×B devices that heat electrons and do not contribute to propulsion thrust. The wave characteristics are measured with laser-based diagnostics, coherent Thomson scattering (CTS), fast Langmuir, and emissive probes. It is very difficult to measure inside the narrow thruster channel and therefore most measurements are performed outside, where access is possible. Measurements indicate that high-frequency and short-wavelength waves are driven by the electron drift instability (EDI), which under some conditions exhibit cyclotron resonances, (for which the frequency of modes is proportional to the electron cyclotron frequency). Under these circumstances, this instability is considered to be an Electron Cyclotron Drift Instability (ECDI). In the far plume area where the magnetic field is small, a wide spectrum of excited turbulent oscillations develops and has properties typical for ion-acoustic turbulence (IAT). The relation between ECDI and IAT is still actively debated because large scale 3D simulations are required to understand the spectrum of resulting turbulence in steady-state.

Section 5 (topic 4) continues the description of high-frequency and short-wavelength waves, but unlike Section 4 it provides a focus on the theoretical review of the current understanding of high-frequency and short-wavelength waves. In the case of a purely 2D system in the axial-azimuthal plane perpendicular to the magnetic field and with finite electron temperature, there exists a set of multiple narrow bands of unstable ECDI modes near the cyclotron resonances. The linear theory of ECDI instability is well developed, however, nonlinear saturation is still the subject of active research. Some earlier numerical work suggests that at a certain amplitude, the cyclotron resonances are washed out by nonlinear effects, and the instability proceeds similarly to the ion sound wave instability in unmagnetized plasma. Other studies indicate that the instability retains the cyclotron resonance features even in the nonlinear stage. When the direction along the magnetic field is considered, an additional instability appears due to the finite value of the wavevector along the magnetic field, the so-called modified two-stream instability, or the modified Buneman two-stream instability. This instability leads to the heating and scattering of electrons along the magnetic field. Given the complexity of mode saturation in realistic 3D geometry, only large-scale 3D simulations can provide a full understanding of the mechanism of anomalous transport and heating. This is made more challenging by the fact that several simulations indicate that secondary nonlinear processes take place resulting in the appearance of long-wavelength waves. The large-scale waves are typically expected to provide large contributions to the anomalous transport and need to be resolved for realistic parameters.

Section 6 (topic 5) summarizes the current state-of-the-art in fluid (fluid electrons – fluid ions) and hybrid (fluid electrons – kinetic ions) simulations of E×B plasma discharges. The fluid simulations are being used as a relatively non-expensive alternative for full kinetic simulations with realistic device parameters, which are outside the current capabilities of existing kinetic codes. The first principle 2D nonlinear fluid models are based on rigorous closures for the electron dynamics perpendicular to the magnetic field (in the azimuthal-axial plane), which are based on the low-frequency approximation, i.e. the mode frequency is assumed to be low compared to the electron cyclotron frequency. Such models show that fluid-type instabilities, such as the gradient-drift modes of the Simon-Hoh type and their generalizations including the lower-hybrid instabilities, result in large anomalous transport generally consistent with experimental values. A notable effect demonstrated in such simulations is a nonlinear energy transfer from the most unstable small-scale modes (of the lower-hybrid type) to the large-scale structures of the simulation-box size. These models, however, are highly simplified; typically, they are in the azimuthal-axial plane and neglect the electron motion along the magnetic field; they also neglect sheath and plasma-wall interactions, and most importantly, consider ions in the fluid approximation. Such models are not suitable for the design of practical systems but could prove useful for understanding complex nonlinear processes in anomalous transport. Motivated by engineering design needs, for many years the alternative hybrid (fluid-kinetic) approach has been employed. This method describes the ions and neutrals kinetically, while electrons are treated with the fluid model. Such 2D, radial-axial models do not include the azimuthal direction, therefore, they require ad-hoc or experimentally based models for the anomalous transport, typically parameterized by the anomalous collision frequency. The formulation and verification of the anomalous electron transport model remains a great challenge here. It is also unclear whether the concept of the anomalous collision frequency remains a good parameterization for the anomalous transport in the presence of large-scale dynamical structures such as spokes and breathing oscillations.

Section 7 (topic 6) explains why a predictive model of E×B discharges requires a kinetic three-dimensional treatment. Typical descriptions of instabilities and transport in two dimensions are performed in the axial-azimuthal or axial-radial plane. In the first case, the important effects of plasma-wall interactions, most notably wall losses, are not captured, as well as convective transport of energy from unstable regions to stable regions. In the second case, the EDI is not captured and only the breathing oscillations and other axial e.g., gradient modes are resolved. Recent simulations of the spoke in the anode region of the thruster show their intrinsic 3D structure. In general, reduced two-dimensional models always show a stronger instability characterized by large amplitude oscillations. This leads to a significantly overestimated cross-field mobility as compared with that simulated in three dimensions and observed in experimental measurements. Another finding from the recent comparison between 2D and 3D simulations is that the spectrum of excited waves in 3D does not exhibit strong peaks at a few dominant frequencies as observed in 2D. This observation is in accord with both experimental and 3D linear analytic kinetic theory predictions that the instability is excited at a wide range of wavenumbers. For 3D simulations, future development of 3D codes should embrace modern computer algorithms, legacy codes generally do not scale well on modern computing architectures and therefore cannot perform full-size 3D simulations. Future codes should implement efficient data structures for memory access, as well as hybrid parallelism via vectorization, OpenMP, MPI in order to improve scalability up to millions of processors (exascale computing). Notwithstanding the difficulty of full 3D simulations, the complete understanding of electron transport will lead to a new era in the technological development of E×B plasma devices: design based on an empirical approach will give way to code-based refined optimization. As it has done in many other engineering disciplines, predictive design and optimization via computer-based techniques will assist and eventually replace more expensive empirical methods.

Section 8 (topic 7) describes recent progress in developing E×B configurations for plasma mass separation applications. Notwithstanding a long history of crossed-field (or E×B) configurations to separate charged particles based on mass, the limitations of small throughput of current devices drive innovation in this area and have led to the development of new plasma isotope separators, where crossed-field configurations were used to produce plasma rotation in plasma centrifuges. Applications of plasma-based elemental separation based on mass include nuclear waste clean-up, spent nuclear fuel reprocessing, and rare earth elements recycling. The separation in plasmas is conditioned upon the ability to externally apply a high electric field in the direction perpendicular to the magnetic field. This is limited by anomalous conductivity, similar to electric propulsion devices. Demonstrating the practicality of crossed-field mass filter concepts, therefore, hinges on a comprehensive understanding of anomalous perpendicular conductivity, which calls for combined modeling and experimental research efforts. Another outstanding issue in the presence of neutrals is the possible upper limit set on the rotation speed by the critical ionization velocity phenomenon. However, the promise plasma separation holds for many outstanding societal challenges is a compelling motivation to tackle these questions.

Section 9 (topic 8) is devoted to verification and validation procedures. The ultimate objective for developing computer simulations of complex physical systems is to use these simulations as a predictive tool for science and engineering design. Critical to these goals is the need to verify and validate codes used for the predictive modeling of low-temperature plasmas. Though in other fields, rigorous verification and validation (V&V) procedures have been relied upon for decades, it is only recently that these procedures were applied to simulations of E×B devices. The section describes V&V efforts in low-temperature plasmas and specifically for 2D axial-radial simulations of Hall thrusters. For validation, a comprehensive set of measurements is needed. This is challenging for compact and energetic Hall thrusters where probes can strongly perturb the plasma and it becomes difficult to measure plasma parameters in the ionization and acceleration zones. A possible approach is to validate codes on specially designed plasma systems which allow for improved access for diagnostics, similar to the GEC cell for RF discharges. An example of such a study for E×B discharges is the Penning discharge. For Hall thrusters, perhaps a scaled-up device with improved diagnostic access is desired for accurate validation of simulation results. Another approach is to use a wall-less Hall thruster where the acceleration zone is outside the thruster channel. Comprehensive measurements by several laser diagnostics and fast probes are needed for the complete characterization of anomalous transport and oscillation spectra. Currently, there is a need to develop verification practices for low-temperature magnetized plasmas. It is important to emphasize that the definition of the test-cases is a significant part of the problem. Indeed, it is important to develop a comprehensive set of test cases to provide benchmarking of codes for the regimes of interest that sufficiently characterize the relevant physics important in E×B plasmas. For low-temperature magnetized plasmas, these are: anomalous transport, low - frequency oscillation such as breathing oscillations and plasma spokes. A detailed description of validation and benchmark test-cases will be the subject of future dedicated E×B workshops and the "Frontiers in Low-temperature Plasma Simulations" workshop. These workshops will facilitate further discussions and ideas.

Section 10 (topic 9) discusses magnetic nozzles. The magnetic nozzles are used to control the radial expansion of the plasma jet, in a similar way as a de Laval nozzle operates on hot ideal gas. The section discusses which part of the electron velocity distribution function (EVDF) is responsible for transforming energy from electrons into ion kinetic energy. This process can be affected by collisions in the magnetic nozzle. This phenomenon is more acute inside a vacuum facility due to the additional effects of the background pressure and of the electrical connection between the plasma beam and the metallic chamber walls, which can affect the electric potential profile, the EVDF of confined electrons, and the amount of electron cooling. Kinetic models are necessary to rigorously describe the magnetic nozzles. However, efforts were directed to develop fluid models with the closures that can be sufficient to predict thrust. Complex sets of magnetic coils can create three-dimensional magnetic structures with adaptable shapes, capable of steering the plasma jet and allow for a non-mechanical way of thrust vector control. Effectiveness of these approaches for weakly-magnetized ions remains unexplored. A variety of plasma instabilities can develop and potentially affect magnetic nozzle operation and resulting thrust. This research field has been little explored so far, both theoretically and experimentally.

Last but not least we assembled a very comprehensive list of references with the goal to provide readers with a detailed list of references that should be credited for the original work. We ask readers to cite the original papers where appropriate. In addition, individual chapters of the perspective could be mentioned in a similar way to individual chapters in a book.

## 2. Plasma Wall interaction in E×B discharges relevant to plasma propulsion devices

Eduardo Ahedo[1], Michael Keidar[2], Irina Schweigert[2], Pascal Chabert[3], Yevgeny Raitses, and Igor D. Kaganovich[4]

[1] 'Equipo de Propulsión Espacial y Plasmas, Universidad Carlos III de Madrid, Leganés 28911, Spain
[2] George Washington University, Washington D.C. 20052, USA
[3]Laboratoire de Physique des Plasmas, CNRS, Ecole Polytechnique, Sorbonne Universités, 91120 Palaiseau, France
[4]Princeton Plasma Physics Laboratory, Princeton NJ 08543 USA

*State of the art and recent progress*

Plasma-wall interactions and magnetic field effects have a strong influence on the plasma discharge operation for many plasma applications, from electric propulsion to magnetic confinement fusion, material processing, and plasma diagnostics. There is a large body of evidence that wall materials affect Hall-effect thruster (HET) operation. The early observations [4] showed that "when a ceramic stick is introduced into the channel radially, the discharge current increases significantly. The discharge current, propulsion, and efficiency coefficient of the thruster are very sensitive to the insulator material and contamination of the surface". Following A. I. Morozov fundamental theoretical predictions [4], systematic experimental studies of these effects began in 1978 in A. I. Bugrova's laboratory at MIREA [5]. Direct demonstrations of wall material on thruster performances [6], [7], [8] have shown that it can affect strongly the discharge current, both in terms of its magnitude, Fig. 1, and its fluctuation level (known as the breathing mode), Fig. 2 (a). To a lesser extent it affects thrust and thus thrust efficiency (the ratio of the thrust squared to the input electric power times twice the mass flow rate), Fig. 2 (b).

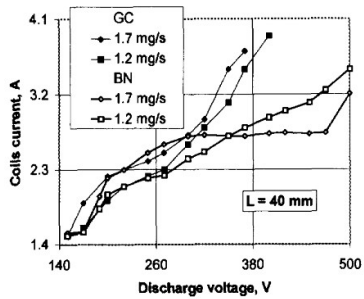

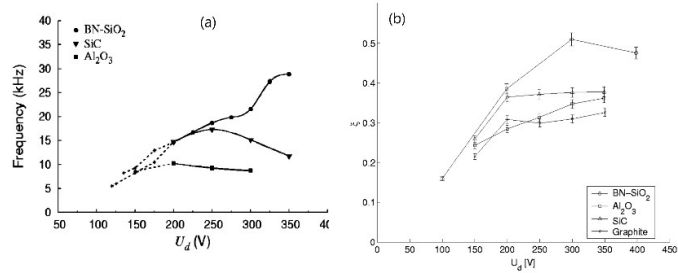

Fig. 1 Discharge current versus the discharge voltage for two channels made from machinable glass (GC) and boron nitride (BN) ceramics, respectively, with the same channel length, L = 40 mm, and for two mass flow rates, 1.2 mg/s and 1.7 mg/s, from Ref.[6]. Reproduced with permission from Proceedings of International Conference on Electric Propulsion Cleveland, OH, 1997 Electric Rocket Propulsion Society, Cleveland, OH, 1997, IEPC 97-056. Copyright International Conference on Electric Propulsion 1997.

Fig. 2 Experimental study of SPT100-ML discharge as a function of the applied voltage. (a): Frequency of the breathing mode extracted from the experimental frequency spectra as a function of discharge voltage, from Ref. [8] Reproduced from S. Barral, K. Makowski, and Z. Peradzyń´ N. Gascon and M. Dudeck, Phys. Plasmas **10**, 4137 (2003), with permission of AIP Publishing; (b): Discharge efficiency as a function of voltage for the xenon flow rate 5 mg/s, and the coil current 4.5 A, from Ref.[7]. Reproduced from N. Gascon, M. Dudeck, and S. Barral, Phys. Plasmas 10, 4123 (2003) with permission of AIP Publishing.

A detailed comparison of plasma properties and discharge operation for SPT-type thruster with boron nitride and carbon segmented walls was performed at Princeton Plasma Physics laboratory (PPPL), see Ref. [9] and references within. Fig. 3 shows the V –I characteristics, the maximum electron temperature, and the maximum electric field measured for the boron nitride ceramic channel and the channel with the nonemitting carbon-velvet walls; a big effect of wall material on these properties is evident for voltages above 350-400V. The effect was attributed to the enhanced secondary electron emission (SEE) for the boron nitride surface as compared to the carbon walls and corresponding increase in the near-wall conductivity leading to the increased electron conductivity in the channel for boron nitride ceramic walls. Detailed analysis of the discharge (and shown electron temperature) is complicated by the fact that the electric field is affected by the emission and strongly changes with the wall material as described in Ref. [9] shown in Fig. 3c. However, the relation between the emission properties and the discharge current is not always straightforward. For example, in Ref.[10] a floating graphite electrode and passive electrode configurations were implemented with two ceramic spacers made from a quartz and MACOR (machinable glass-ceramic). The measurements for the discharge currents shows the following trends: for the MACOR spacer 1.63A, for the graphite electrode (1.66A), for the boron-nitride channel 1.7A, and for a quartz spacer 1.94A. Note that the secondary electron emission from quartz is lower than that for MACOR and boron nitride [10]. The PPPL study [9] showed that metal or ceramic spacers can significantly affect the plasma potential distribution and shape of equipotential lines in Hall thrusters even though these spaces have sizes comparable, but smaller than the acceleration region. A theoretical study of these effects was performed in Refs. [12],[13]. Mechanisms behind these effects possibly involve changes in the electron temperature and transport, which are affected by the secondary electron emission and conductivity of wall materials. The effects of spacers on operating conditions of the thruster depends on the precise placement of the electrodes relative to the magnetic field distribution [10]. Another potentially important effect is the contamination of the surface; for example, quartz is not a porous material whereas boron nitride is, therefore that latter is more prone to the contamination. Reference [11] showed that contamination can greatly affect SEE.

Similar effects were observed in magnetrons, another ExB discharges. Reference [14] discusses the influence of the secondary electron yield on the property of the spokes in HIPIMS plasma, by adding nitrogen and effectively changing the target surface from metallic to compound target, affecting the secondary electron yield of the target surface.

All these experimental observations stimulated comprehensive theoretical and modeling studies aimed at the description of complex plasma processes in HETs as well as accurate measurements of secondary electron emission from dielectrics [15],[16],[17],[18]. It became clear that the theories have to be kinetic, as fluid theories cannot explain such high electron temperatures observed in HET due to fast losses to the channel walls, see Fig. 3 (b). Such high electron temperatures were also recently confirmed by measurements with incoherent Thomson scattering diagnostic [19] in addition to the previous measurements by probes [9].

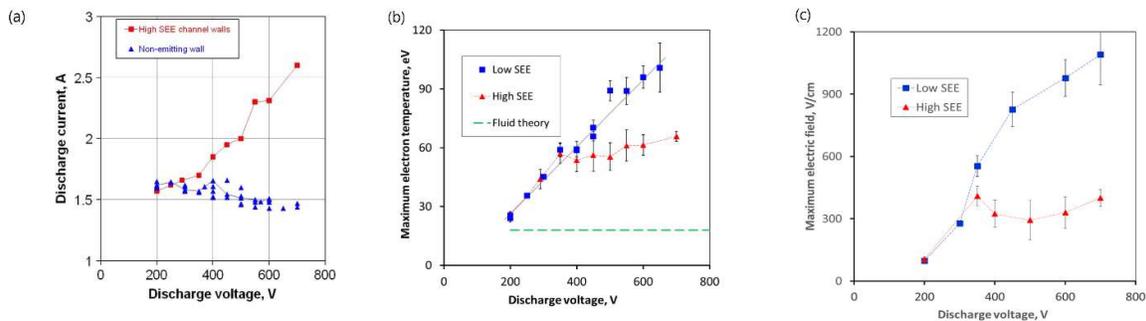

Fig. 3. Effect of wall material on discharge properties as a function of the discharge voltage for the boron nitride ceramic walls and the nonemitting carbon-velvet walls: (a) The I-V characteristics for two walls; (b) the measured maximum electron temperature deduced from floating emissive and non-emissive probe measurements; dashed green curve shows maximum temperature in the channel estimated according to the fluid theory; (c) the measured maximum electric field along the thruster channel median; the electric field was deduced by differentiating over the distribution of the plasma potential measured using floating emissive probe and the procedure - All results described in Ref. [9]; Reproduced with permission from IEEE Trans on Plasma Scie. **39**, 995 (2011). Copyright 2011 IEEE.
.

In typical electric propulsion devices and conventional annular Hall-effect thrusters, the magnetic lines intersect the walls, and therefore plasma confinement is not magnetic but mainly electrostatic; walls are typically biased negatively with respect to plasma bulk in order to collect the correct electric current (for instance zero, in the case of a dielectric wall). The absence of magnetic confinement in a conventional HET leads to high particle and energy losses at the walls, and surface erosion. Furthermore, the large secondary electron emission (SEE) induced by energetic electron impact on ceramic walls reduces the wall potential, therefore worsening the electrostatic confinement of energetic electrons, increasing the energy loss from the plasma.

The mean free path for electron collisions with neutrals or Coulomb collisions is large compared to the HET channel width. Consequently, losses of energetic electrons are expected to be higher than losses in inelastic collisions with neutrals for the conditions of conventional HET. [8],[9], [20], [23]. Moreover, the high-energy electrons that can escape to the walls with energy above the potential energy corresponding to the wall potential are not being replenished sufficiently fast. As a result, the electron velocity distribution function (EVDF) is non-Maxwellian as shown in Fig. 4. The electron flux to the walls is proportional to the EVDF for these energetic electrons escaping to the wall; whereas the EVDF in this region is small compared to the Maxwellian EVDF. The uncertainty in the EVDF makes it difficult to model HET using fluid or hybrid approaches which are based on the assumption of a Maxwellian EVDF [21].

Indeed, Particle-In-Cell (PIC) simulation studies of the EVDF formation in HET which include all necessary effects in self-consistent treatment: collisional replenishment of the EVDF for energetic electrons, the influence of the SEE, and the Debye sheath formation, are showing a rich plasma material interaction phenomena, still non-amenable to simple scaling laws of wide use. Although, scaling laws were derived for sheath potential and near wall conductivity for a case of a magnetic field perpendicular to the wall [23].

It was also shown using PIC simulations that the average energy (or effective temperature) of the electrons in the sheath due to the non-Maxwellian distribution functions mentioned above. Therefore, the usual isothermal sheath theories used to evaluate the sheath potential drop and the secondary emission yield are not correct [24],[25]. A polytropic sheath model has been proposed to overcome the limitation of isothermal models. The model works well but is not self-consistent and uses the PIC data to evaluate the polytropic index.

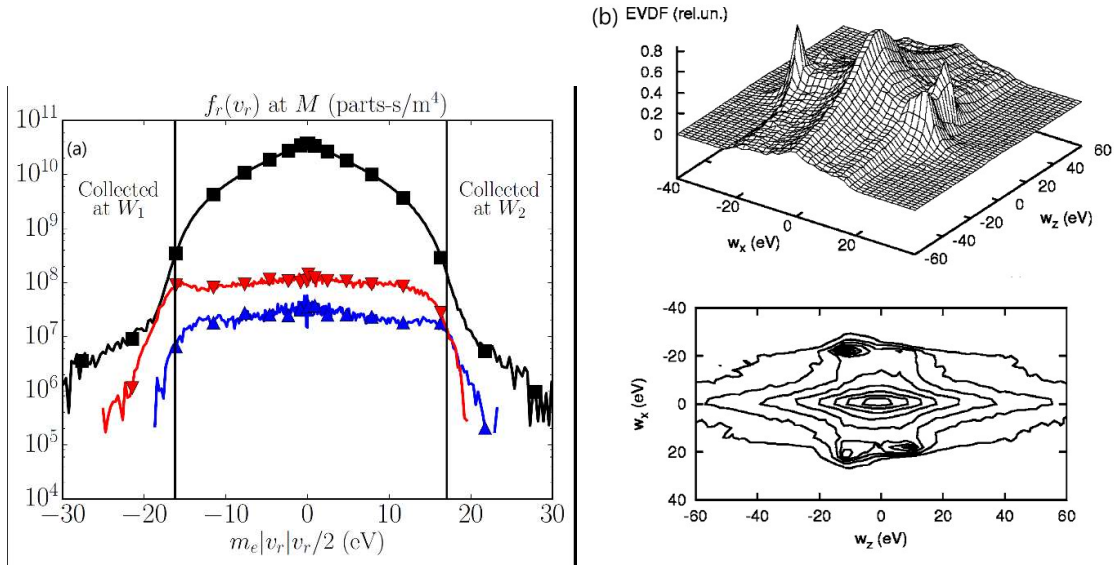

Fig. 4 (a). Electron velocity distribution functions at the thruster cylindrical channel mid-radius M, of (black) primary electrons, (blue) secondary electrons from the channel inner wall W1, and (red) secondary electrons from the channel outer wall W2, showing depleted tails, vertical lines mark the wall potential value, from Ref. [26]; Reproduced with permission from Plasma Sources Sci. Technol. **27**, 064006 (2018). Copyright 2018 IOP Publishing. Fig. 4 (b) EVDF over $v_x$ and $v_z$ shown as 3D plot and as 2D plot with contour lines; the plasma potential relative to the wall is 20 V, x-axis is normal to the wall and z-axis is parallel to the walls, from Ref.[22]; Reproduced from D. Sydorenko, A. Smolyakov, I. Kaganovich, and Y. Raitses, Phys. Plasmas **13**, 014501 (2006) with permission of AIP Publishing.

Two-stream instabilities of the SEE beams injected into the plasma can reduce the energy of SEE due to exchange with a colder bulk electrons and increase the number of energetic SEE electrons trapped by a potential well and therefore make the total EVDF closer to a Maxwellian [27]. The evolution of SEE within the plasma needs to take into account possible collection by the walls [26],[28]. Asymmetries in the EVDF caused by cylindrical effects can be significant too [29], [64]. The sheath structure and plasma-wall interaction processes are quite sensitive to the detailed description of SEE features (e.g., to the amount of true-secondary versus elastically or inelastically backscattered electrons) [27], [30].

At SEE yields of about 100%, the sheath becomes space-charge limited, and in this regime, the sheath may become unstable. [31],[32],[33],[34]. Furthermore, in the center of the acceleration region of a HET, the axial electric field is maximum and the resulting E×B drifts can be on the order of the electron thermal velocity, thus enhancing the impact energy of electrons and the transition to a space-charge limited unstable sheath [30].

The sheath instability (in the radial direction) may work as a trigger for the azimuthal fluctuations. An example of the temporal evolution of the total current collected on the outer wall and the corresponding floating potential are shown in Fig. 5. It should be pointed out that in this case the instability is detected only on the outer wall, where the secondary electron emission coefficient reaches a value larger than 1.

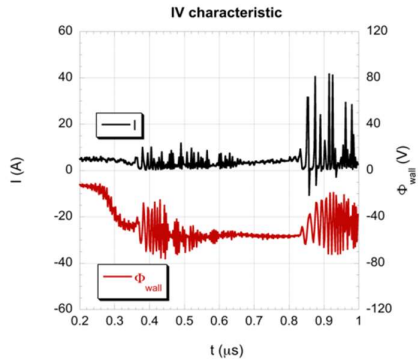

Fig. 5. Temporal evolution of current-voltage I-V characteristic at the outer wall of a HET. From Ref [35]; Reproduced from F Taccogna, S Longo, M Capitelli and R Schneider, Applied Physics Letters **94**, 251502 (2009), with the permission of AIP Publishing.

In the case of high SEE yield, the transition was observed from a space-charge limited sheath to an inverse sheath [36]. The important difference between the two regimes is that ions are not accelerated towards the wall in the inverse sheath case and thus improve ion confinement (but deteriorate energy losses). This phenomenon has been studied in the context of large thermionic emission, mainly related to emissive probes and arcs [38],[39].

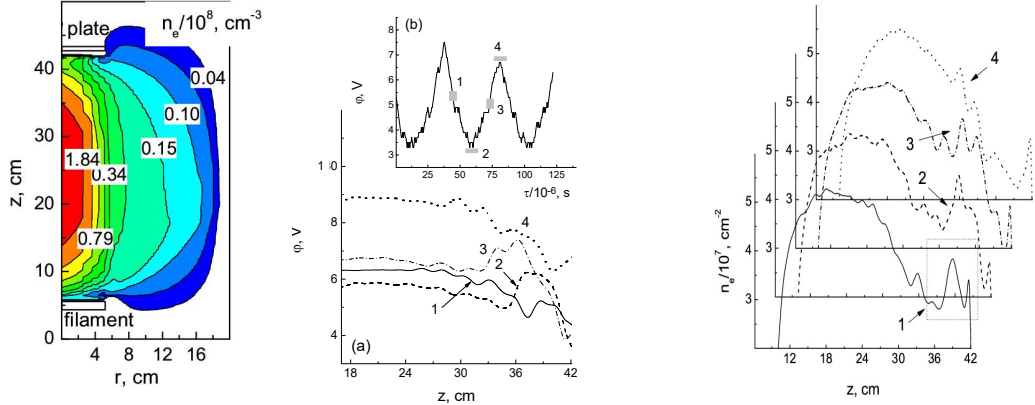

Fig. 6(a) Calculation domain with electron density distribution for the electron current $j = 30$mA from the cathode, which is a filament, $U = -70$V. The emissive $Al_2O_3$ plate is at z=42 cm from Ref. [42]; Reproduced with permission from Plasma Sources Sci. Technol. **24**, 025012 (2015). Copyright 2015 IOP Publishing.

Fig. 6(b) Electrical potential profiles at different time moments (a) and fragment of floating plate potential oscillations with time (b) for $j = 10$mA, $U = -120$V from Ref. [42]; Reproduced with permission from Plasma Sources Sci. Technol. **24**, 025012 (2015). Copyright 2015 IOP Publishing.

Fig. 6(c) Electron density profiles at different time moments of oscillating floating potential shown in figure 6(b) for $j = 10$ mA, $U = -120$V from Ref. [42]; Reproduced with permission from Plasma Sources Sci. Technol. **24**, 025012 (2015). Copyright 2015 IOP Publishing.

Sheath oscillations were experimentally observed when a beam of electrons strikes a biased plate [41]. The oscillating sheath near the floating emissive plate bombarded by a beam of energetic electrons from cathode was also observed in 2D particle-in-cell (PIC) Monte Carlo Collisions (MCC) simulations in Ref. [42] (see Fig. 6a). The virtual cathode appears, as expected, when the total electron flux from the plasma produces a larger flux of secondary electrons from the emissive surface. The potential profiles for different times of sheath oscillation cycles shown in Fig. 6 (a) are related to the periodical accumulation of secondary electrons near the emissive surface. A fragment of oscillating floating potential is shown in Fig. 6 (b). The numbers in Fig. 6 (b) point out the time of snapshots of the potential and electron density profiles shown in Fig. 6 (a) and Fig. 6 (c). The sheath oscillation frequency of about 25 kHz is set by the rate of production of secondary electrons and the ion velocity.

Excitation of ion-acoustic waves in presence of very intense SEE when an inverse sheath forms was observed in the HET channel for a high electric field [43].

One of the important aspects of HET application is the lifetime which is determined primarily by ion-induced erosion of the thruster channel walls [4],[44],[45]. It's important to note that wall erosion is quite visible (few mm in depths) after a few hundred hours of thruster operation in SPT-100 type thrusters [46],[47],[48] leading to the exposure of magnetic circuit and eventually ill-functioning. The rate of erosion is governed by the ion flux to the wall and the sputtering yield of the material, which depends on the ion energy and the incidence angle [4],[44]. An example is shown in Fig. 7 (a), where wall erosion causes significant changes to the channel shape. Due to big variation in the channel wall shape the erosion rate decreases, see Fig. 7 (b). In addition to the above volumetric erosion there exists abnormal erosion forming striations (shown in Fig. 8 b) on the material surface which could be related to plasma or sheath oscillations but is largely unknown [4]. Due to the high economical value of both increasing and qualifying lifetime, simulation tools for lifetime characterization and extension are an important research effort.

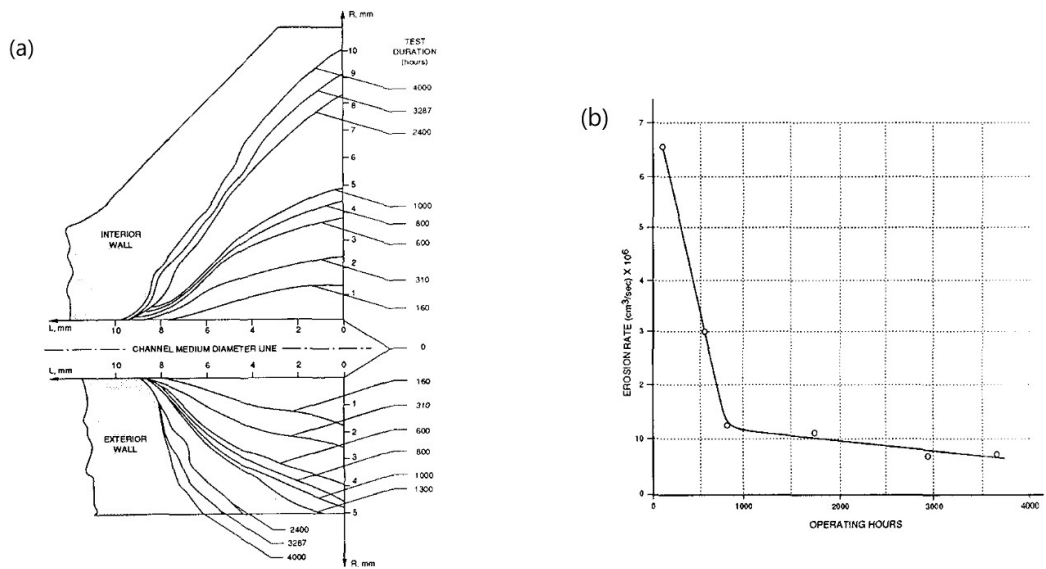

Fig. 7 (a) *SPT-100* erosion profiles were experimentally determined during *4000* hours of life testing; (b) SPT-100 erosion rate vs. time from Ref. [52]; Reproduced with permission from AIAA/SAE/ASME/ASEE 28th Joint Propulsion Conference and Exhibit July 6-8, 1992 / Nashville, TN AIAA-92-31 56. Copyright 1992 AIAA/SAE/ASME/ASEE.

Most models of plasma in HET consider a simple configuration of a magnetic field that is normal to the channel walls. If the magnetic field is oblique to the wall, the characterization of the partially-depleted tail of the EVDF becomes more complex and the E×B drift can be altered significantly[4], [40]. Consequently, wall erosion might be mitigated by applying such an oblique magnetic field. This is particularly relevant for innovative designs of HETs bearing magnetically-shielded topologies, (i.e. with magnetic lines near-parallel to the walls around the thruster exit). The magnetic-shielded topologies appeared after the discovery that the erosion of the BN walls in the BPT-4000 Hall thruster designed and developed by Aerojet had essentially stopped after 5600 h of operation during life testing, a similar observation was made in the Soviet Union [52], [53] see Fig. 7. A follow-up extensive investigation into the mechanism responsible for this effect was initiated at the NASA Jet Propulsion Laboratory (JPL). The magnetic shielding reduces, of course, the plasma ion fluxes to the walls, but, indirectly, the mean impacting ion energy as well, because the acceleration region moves downstream [49], [50]. Then, there is the THALES High-Efficiency Multi-Stage Plasma Thruster (HEMPT), a propulsion technology rather similar in operational principles to the HET where magnetic shielding is applied through magnet-based cusped magnetic fields [54]. Different conceptions of the cylindrical HET lie between HET and HEMPT designs [55]. Much remains to be known on characterizing the local E×B drifts, the EVDF, and plasma fluxes to walls in these complex topologies. As discussed above, there were many efforts on optimization of magnetic topology to better focus the plasma and minimize plasma-wall interactions resulting in the development of high-performance Hall thrusters of SPT and TAL type. However, the magnetic shielded thruster developed by JPL and Aerojet is currently a culmination of these efforts in terms of the extended lifetime [56].

Interesting effects are associated with the formation of filaments in a magnetic field under conditions similar to those in HET. The plasma-wall interactions in the external oblique magnetic field was studied in cases of different magnetic field strengths and incidence angles for the conditions similar to the Hall thruster but for a simpler plasma source. To this end PIC MCC simulations were employed [59],[60]. The results shown in Fig. 8 illustrate the electric potential distribution and alignment of electron density peaks along the magnetic field for different angles. The magnetic striation due to the instability was theoretically predicted in Refs. [61], [62] and experimentally observed in Ref. [63]. These striations can possibly explain striations in erosion pattern observed at the exit of the thruster, shown in Fig.9(c) bottom, from Ref. [61].

*Current and future challenges*

Secondary electron emission at the wall is predicted to be a key phenomenon in E×B discharges, particularly in Hall Thrusters. However, there is still no direct evidence that it is the SEE alone that is responsible for the wall materials effect. It will be important to validate modeling predictions by a comparison with measurements of electron velocity distribution function or energy distribution function in the thruster channel using, for example, Laser Thomson Scattering (LTS) and/or electrostatic Langmuir probes, respectively. Implementation of both diagnostics inside the thruster channel where the predicted effects are important is challenging.

For example, for the LTS, key challenges are 1) to get the laser beam and scattered light from the channel inside, and 2) to resolve the tail of EVDF. For probes, probe-induced perturbations of the plasma are hard to avoid unless the probe is installed on the channel wall that may change the wall effect.

For modeling, in order to appropriately model the SEE effects, the challenge is to correctly evaluate the EVDF. This is naturally done in PIC simulations, but unfortunately, these simulations are computationally expensive and therefore not yet adapted for thruster design optimization. It is, therefore, necessary to propose simplified sheath models, which account for non-Maxwellian EVDF and would allow better evaluation of the secondary emission yield. This is a very challenging task and a large research effort is needed in this area.

As it was shown above oblique magnetic field adds considerable complexity in the form of magnetic striations. How the magnetic striations form in the HET geometry is not yet explored.

The biggest complexity is due to the mutual interaction of turbulence and plasma-wall effects because turbulence leads to electron heating and scattering and can populate depleted parts of EVDF due to loss cone. These phenomena, though observed in PIC simulations, are not thoroughly studied to the degree that it can be accounted for in simplified sheath models as discussed above.

Finally, it would be important to study changes in the material properties due to exposure to the thruster plasma and how that change affects the thruster operation.

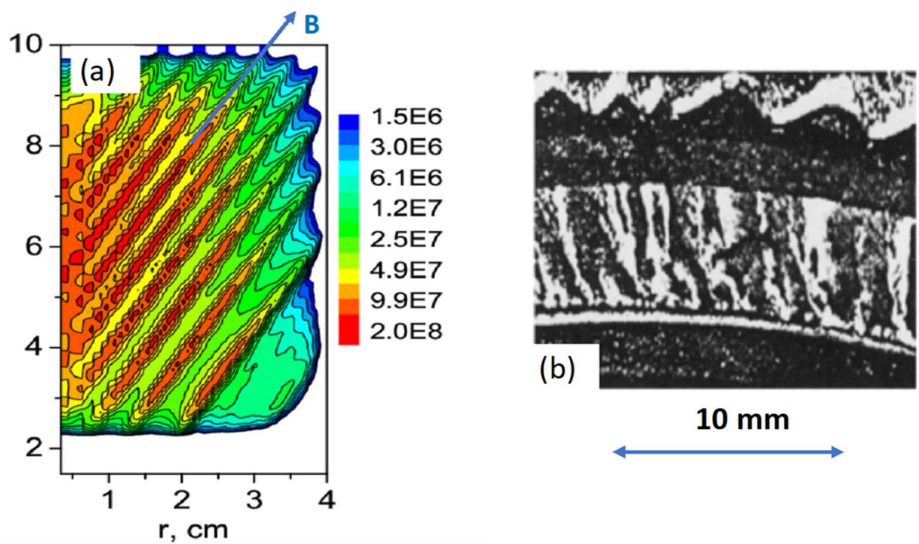

Fig. 8 Striations in the oblique magnetic field. (a) electron density profile (in cm⁻³) for $T_e$=5 eV and B=50 G, from Ref. [59]; Reproduced with permission from Plasma Source Sci. & Technol. **26**, 064001 (2017). Copyright 2017 IOP Publishing. and (b) an enlarged photo of a fragment of about 10 x 15 mm from the outer part of M-100 SPT thruster after a 5000-h lifetime test, from Ref. [66] Reproduced with permission from Plasma Physics Reports **29**, 235 (2003). Copyright 2003 Springer Nature Pleiades Publishing.

*Advances in science and technology to meet challenges*

From the technological point of view, two possible directions could be followed in parallel. The first is to continue the research efforts in material science to improve the performances of dielectric walls, both in terms of intrinsic properties (for example, by reducing secondary electron emission) while providing sufficiently long lifetime and aging properties. The progress in this direction has been already done and research in this area probably needs to be maintained. The second avenue is to propose new designs that would minimize the role of plasma-surface interactions. This is already happening with the development of magnetic shielded or wall-less thrusters. Other ideas that would allow controlling the sheath properties could be proposed.

From the scientific point of view, combined efforts are needed in theory, numerical simulations and experiments to validate predictions of theory and simulations. In theory, new sheaths models that account for non-Maxwellian EVDF are needed to be able to predict the total secondary electron emission yield at the surface.

On the simulation side, a large community effort is needed to develop massively parallelized and optimized codes able to simulate turbulence and plasma-wall interaction in 3D geometries. These codes will also have to be benchmarked and verified.

Finally, optical (non-invasive) diagnostics, able to achieve both short time and space resolution, are required to investigate the plasma-wall interaction effects in these high-density and high-energy plasmas where probes hardly survive. This is difficult and is a major challenge for the community. Moreover, it is also important to continue the development of electrostatic diagnostics suitable for measurements of EEDF in harsh plasma environments of the thrusters. Like in high-temperature fusion research which pioneered and use many sophisticated RF, optical and laser diagnostics of plasmas, probes and energy analyzers remain important diagnostics for edge physics especially at the divertors of fusion reactors.

*Conclusions*

There was a large progress in understanding plasma-wall interaction in the recent decade. A future challenge is to generalize the developed understanding for realistic devices in 3D and accounting for self-consistent interaction of turbulence and plasma –wall interactions and experimental validation of these effects using advanced diagnostics.

# 3. Low-Frequency Oscillations in E×B Discharges

Yevgeny Raitses,[1] Andrei Smolyakov,[2] Mark Cappelli,[3] Kentaro Hara,[3] Jean-Pierre Boeuf,[4] and Igor Kaganovich[1]

[1] Princeton Plasma Physics Laboratory, Princeton NJ 08543 USA
[2] University of Saskatchewan, 116 Science Place, Saskatoon, SK S7N 5E2 Canada
[3] Stanford University, Stanford, California 94305-3032 USA
[4] LAPLACE, University of Toulouse, CNRS, INPT, UPS, 118 Route de Narbonne, 31062 Toulouse France

*State of the art and recent progress*

Low-frequency (typically <100 kHz) oscillations, including spokes and breathing modes, are some of the most prominent examples of self-organization in partially-magnetized quasi-neutral plasmas of crossed-field devices at a wide range of pressures ∼ 0.1-100 mTorr, such as Hall thrusters, Penning discharges, and sputtering magnetrons [7],[65],[67],[68],[69],[70]. The spoke mode manifests itself as strong perturbations in plasma density that propagate in the E×B direction perpendicular to the crossed electric (**E**) and magnetic (**B**) fields, generating substantial components in electric field in this E×B direction. [65],[67],[68],[69] The breathing mode propagates in the direction of the external electric field, and is one of the powerful modes observed in Hall thrusters. It reveals itself in oscillations of the discharge current (see Fig. 9, and Fig. 2a), often reaching amplitudes comparable to the mean discharge current itself. [7], [65], [71] While these modes have been known and studied for some time, only recently have coordinated efforts been made towards self-consistent modeling, [72],[73],[74],[75],[76],[77] detailed experimental validation [78],[79],[80],[81],[82], and their control [83],[86],[87],[88].

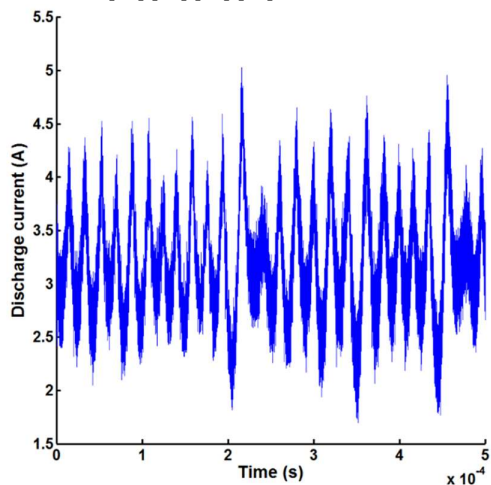

Fig. 9 Breathing oscillations measured for a 1.5 kW PPS-1350ML Hall thruster of SPT type: traces of the discharge current measured at 330 V, the gas is Xe and the flow rate is 3.5 mg/s. The characteristic breathing mode frequency is 55 kHz. "Courtesy: ICARE, CNRS".

For cylindrical devices, the spokes rotate azimuthally in the direction of the ExB drift, but with a speed that is an order of magnitude smaller than the ExB drift velocity [67],[68],[78],[80],[89],[90],[91],[92]. The mode number of these spokes is usually low, *m* =1 − 8. [67],[68],[81], [93], [94]. It has been reported that anomalous (turbulent) electron current may be enhanced in the spoke, carrying 20-90% of the total discharge current.[91],[92]. Although the mechanism for spoke formation is still debated, one candidate is the Simon-Hoh (SH)-type instability,[95],[96] driven by the combination of the applied electric field and the gradient in plasma density. A modified theory of this instability for partially magnetized collisionless plasmas was developed [74],[79],[97],[98],[107] and experimentally verified for some conditions [67],[79],[92],[93],[97].

Recent results of large-scale Particle-In-Cell (PIC) simulations of a Penning discharge [75],[108] were found to be in good agreement with experimental data,[92] showing the formation of a $m = 1$ spoke rotating with a frequency of a few kHz (Fig. 10 ), generating anomalous current due to fluctuations. The scaling of the spoke frequency deduced from both the simulations and experiments was consistent with theoretical predictions for SH instability in collisionless plasmas,[97] i.e., $f \propto \sqrt{E_r L_n / m_i}$, where $E_r$ is the radial electric field, $L_n$ is the gradient density scale, and $m_i$ is the ion mass.[75] According to these PIC simulations, the ionization of the working gas has a minor effect on the spoke formation.

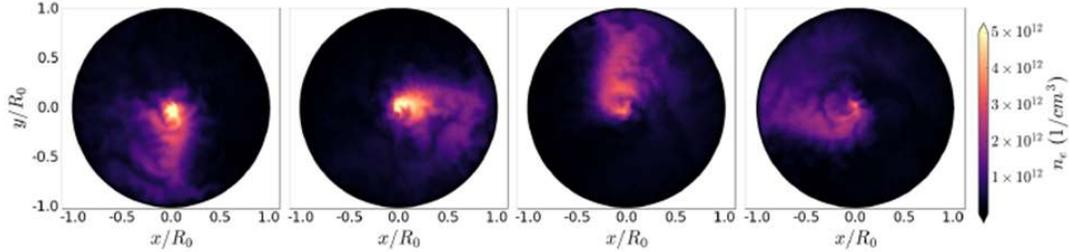

Fig. 10. PIC simulations of electron density contours in a real scale ExB Penning system with diameter 5cm, showing spoke rotation, at simulation times, from left to right; 61.4 μs, 65.8 μs, 70.7 μs, and 74.2 μs.[75] Reproduced from A. Powis. J. Carlsson, I. Kaganovich, Y. Raitses, and A. Smolyakov, Phys. Plasmas **25**, 072110 (2018), with the permission of AIP Publishing.

Unfavorable magnetic field gradients can lead to violent perturbations, but typical magnetic profiles in modern Hall thrusters, are designed to partially stabilize the most dangerous spoke modes.[65] The theory of collisionless SH instability has been modified to include the gradients of the magnetic field and electron temperature. With these modifications it is generally referred to as the gradient-drift instability.[74] Linear analysis [109],[110] as well as recent particle-fluid hybrid simulations show that gradient-drift effects are critical for the formation of azimuthally rotating spoke-like structures in the typical conditions of Hall thrusters.[81]

Although simulations have shown that spokes can form in collisionless plasmas without ionization [75], there is strong evidence from numerous experiments in high power pulsed magnetron discharges (High Power Impulse Magnetron Sputtering -HiPIMS discharges) and in lower power, DC magnetron discharges [99],[100],[103] that ionization can play an important role in the formation and dynamics of spokes. An important similarity between magnetron and Hall thruster discharges is the existence of a region where the radial component of the magnetic field decreases axially toward the anode. Recent experiments in HiPIMS and DC magnetron discharges have shown that the observed spokes are associated with an ionization instability rotating in the azimuthal (E×B ) direction. The measurements reveal the existence of a double layer structure with large electric field at the leading edge of the ionization zone and that this double layer plays a crucial role in the energization of electrons [103]. References [101],[102] discuss the experimentally determined potential structures of spokes in HIPIMS. The presence of a double layer and enhanced ionization at the spoke front is reminiscent of the Critical Ionization Velocity model of [105] and of the PIC simulations of [63], [106], although the velocity of the spoke measured in HiPIMS and magnetron discharges does not necessarily match the theoretical critical ionization velocity. Moreover experiments [79],[99],[100],[103], and recent modelling [104] have evidenced that spokes can rotate in the -E×B direction (at low power) as well as in the +E×B direction (at higher power).

The very detailed and recently published measurements of the spoke properties in HiPIMS and DC magnetron discharges should open the way for a better understanding of these structures and of their dynamics based on modeling and simulations. Although the physics of HiPIMS is more complex (e.g. ionization of the sputtered atoms) it seems very likely that spokes in Hall thruster and magnetron discharges share several common properties.

Ionization plays a key role in the axial (breathing) mode oscillations.[70],[81],[82],[111],[112],[113] Experiments demonstrated the dependence of these oscillations on the thruster operating conditions such as discharge voltage, magnetic field strength and topology, gas flow, cathode operation, and background pressure.[7],[81],[82],[86] A zero-dimensional predator-prey model for the coupled evolution of the ion and neutral density[112] predicts the oscillations, however, the equivalent one-dimensional model shows no oscillations. It was also realized that self-consistent electron dynamics is important.[69],[113] The resistive instability[114],[115] due to the phase shift in the response of the electron current (resistive) and ion current (inertial) to the perturbation of the electric field was proposed to be a triggering element of the breathing mode.[81],[116],[117], Besides, the electron temperature evolution was also shown to be important and need to be included as into fluid or hybrid models.[69],[81],[114] Recent studies that include fluctuations of the electron temperature have shown that temperature may render zero-dimensional predator-prey models unstable[81], which otherwise are stable without taking into account electron temperature. Without taking into account the variation of the electron temperature and ionization rate, the one-dimensional modeling shows no excitation of the breathing oscillations.[117],[118]

*Current and future challenges*

There is a growing realization that in conditions relevant to the practical operation of the E×B plasma devices, the low-frequency azimuthal and axial modes result from a complex interplay of various phenomena. An understanding of these complex interactions remains far from complete and they present a critical challenge for the development of predictive modeling tools needed for existing and future applications. For stronger magnetic fields and faster ion and electron rotations, it is important to investigate onset of a Simon-Hoh instability taking into account centrifugal forces [119],[120] as especially relevant to novel mass separation devices discussed in Sec. 8.

As for experiments, it is challenging to independently control the discharge parameters such as local electric field and density gradients – the properties important to the instability. This coupling makes it difficult to identify the cause for the ubiquitous presence of large-scale azimuthal disturbances while the smaller scale azimuthal modes typically have larger growth rates: for the low-$m$ modes the mode growth rate increases almost linearly with the wavenumber.[67],[97] Fluid simulations have shown that the low-$m$ modes can be formed as a result of nonlinear inverse energy cascade from small scale modes.[74], [107]. Furthermore, magnetic field, temperature and plasma gradients, which are present in many engineering devices (e.g., Hall thrusters and magnetrons),[65][77][121] and the presence of physical boundaries and associated sheaths[92] all need to be taken into account.[122] Combined effects of gradients, non-local electron kinetics, and the coupling between large and small-scale plasma structures (energy cascade) should continue to be explored theoretically, numerically, experimentally. It is also important to further explore the effects of ionization and neutral depletion in such complex, highly non-uniform plasma systems.[123],[124],[125]

Theoretical models and simulations of the breathing mode suggest a large sensitivity to the values of the effective electron mobility along the applied electric field or even wall material as was shown in Fig. 2a. Empirical values of the anomalous mobility are typically used for modeling of the breathing oscillations with values adjusted to achieve reasonable agreement with experiments.[121] In addition to anomalous mobility, the models should include other anomalous transport coefficients, including anomalous heating and energy losses. For the most part, physical mechanisms governing these anomalous effects are not well-understood. For example, it is unknown whether the anomalous heating can be described by the same effective collision frequency as the anomalous electron mobility. Understanding these anomalous effects is required for further progress in this research field.

With a few exceptions, modeling of the breathing and azimuthal spoke modes has been performed separately and without considering possible coupling effects. Some experimental data suggest that there is a coupling between them.[126] A coupling between the two modes is expected because the azimuthal mode is driven by axial gradients in plasma parameters such as the density and electric field, which experience large spatial and temporal variations during breathing mode oscillations. Control of the axial oscillations modifies the driving forces for the azimuthal modes.[127] Recent control experiments have used varying cathode electron emission[86] as well as external voltage modulation, [127] demonstrating a suppression of the oscillations associated with both of these modes.

*Advances in science and technology to meet these challenges*

Modern computational tools to study E×B plasma devices include fluid, hybrid (PIC for ions and atoms, fluid for electrons) and full PIC (PIC for all plasma species) simulation codes. Benchmarking should be performed for simulations with identical conditions, specifically relevant to low-frequency phenomena. Codes should be benchmarked against each other to better understand the limitations of numerical algorithms and physics approximations. For instance, a fluid model with electron pressure closure [128] has been proposed to eliminate the numerical uncertainties in conventional quasineutral models. Alternative approaches such as continuum grid-based kinetic models, or direct kinetic simulations,[129],[130] can be used to understand the issue of numerical noise in PIC codes. PIC codes can be used to verify the numerical diffusion issue in direct kinetic simulations.

Experiments capable of measuring plasma properties with the spatial and temporal resolution, including energy distribution functions (EDFs) of the electron, ion, and neutral species in directions along with the electric and magnetic fields and in the E×B direction, and time-resolved (<10$^{-4}$ s) electric field measurements are crucial for validating simulation results. Experiments involving active control of low-frequency oscillations [86],[87],[128] are also important as they may reveal underlying physical mechanisms of instabilities needed to be captured by truly predictive models.

The use of fast-sweeping electrostatic probes and energy analyzers for measurements of electron and ion EDFs, and plasma potential is appropriate when these invasive diagnostics induce minor plasma perturbations (e.g. Penning systems). However, for Hall thrusters and magnetron discharges, with non-uniform magnetic fields and strong potential gradients, non-intrusive diagnostics, such as Laser-Induced Fluorescence (LIF) of electronically excited ions and atoms[131],[132],[133],[134],[135],[136] and Laser Thomson Scattering (LTS),[137] may be more quantitative in measuring species EDFs in these devices. A critical challenge for time-resolving LIF of ion EDF is accounting for electronically excited states produced during spoke and breathing oscillations by direct ionization of neutral atoms as well as ions in other electronically excited states.[138],[139] For measurements of the electron EDF using LTS, a key challenge is a relatively low electron density detection limit of 10$^{10}$ cm$^{-3}$ making it difficult to characterize the EDF high energy tail that may develop and be affected in low frequency spoke and breathing mode cycles.

*Conclusions*

Low-frequency oscillations occurring in E×B discharges are the most powerful oscillations and therefore, may significantly affect the performance of these devices in several different ways including but not limited to power losses on electron transport and heating, defocusing of ions, and mismatch between the device and the power supply. Therefore, it is important to understand these oscillations and their control. The progress made in understanding the low-frequency phenomena, including spoke and breathing oscillations, has been driven by combined efforts of modern experimentation, theory, and simulation studies. However, there is an emerging need for dedicated efforts to compare and benchmark various numerical codes. Experiments capable of measuring spatially and temporally-resolved plasma properties during low-frequency oscillations will be crucial to advance our understanding of these phenomena. Continued advances in computational capabilities will eventually allow simulations to be carried out at scale and over the times needed to resolve these low-frequency structures. The ultimate challenge and the long-term goal would be the development of experimentally validated predictive computational tools that self-consistently model anomalous electron mobility and heat conduction, because these transport phenomena play a critical role in low-frequency oscillations as observed during nominal operation of E×B plasma devices. Then, the next step would require the implementation of modeling including electric circuits with passive and active control of these oscillations and other means of their control such as segmented electrodes, [83] cathode gas flow, [84], [85] and electron emission [90].

# 4. Experiments in Turbulence in Low Temperature, E×B Devices

Benjamin Jorns[1] and Sedina Tsikata[2]

[1] Department of Aerospace Engineering, University of Michigan, Ann Arbor, MI, USA
[2] ICARE, Electric Propulsion Team, Centre National de la Recherche Scientifique, Orléans, France

*State of the art and recent progress*

While there are many types of plasma oscillations in low-temperature, E × B devices, including the longer-wavelength modes linked to plasma inhomogeneities and ionization (see Sections 3 and 6) [70], [141], there is a growing interest in experimentally characterizing the role of short wavelength (< 1 mm) plasma turbulence in these devices. This interest largely has been motivated by numerical studies (starting with the work of Ref. [143] and more recently, in work such as Ref. [142] and references therein) that have shown these oscillations may be a dominant driver of anomalous electron transport in low-temperature E × B systems. To this end, experiments have investigated which waves are present in these devices, how much energy is contained in these waves, and how these measured wave properties can be related to anomalous transport. These experimental studies have been facilitated by the development of new diagnostic techniques in the past fifteen years — such as coherent Thomson scattering (CTS), developed in response to numerical work of Ref. [143] — that have allowed for unprecedented levels of non-invasive access and resolution.

Most experimental efforts on turbulent measurements to date have focused on fluctuations formed in the E×B direction (although other short wavelength fluctuations—the axially-directed two-stream instability— have been detected and characterized as well [140]). For example, Fig. 11 shows a set of representative experimental results for a Hall thruster discharge. In these experiments, which are described in Refs. [145] and [149], CTS was employed to interrogate the short-wavelength oscillations in the electron density in the direction of the Hall drift. Fig. 11(a) illustrates the geometry of the system and the measurement location, downstream of the thruster exit, with investigations performed at an observation wave vector $k$ with variable orientation in the (E × B, E) plane. Fig. 11(b) shows the resulting measured wave dispersion from CTS in the Hall (azimuthal) direction. It was proposed in Refs. [145] and [149] that the features of the measured dispersion– specifically, of MHz, mm-scale density fluctuations in the azimuthal direction – were indicative of the presence of the electron cyclotron drift instability (ECDI). The characteristic physics of the ECDI (also referred to more generally as the electron drift instability, or EDI) are discussed in more detail in Sec. 5.

To further support the link between experiment and ECDI, Refs. [145] and [149] drew parallels between their measurements and previous 2D numerical and linear kinetic theory studies [143], [156] [187]), that had predicted that the ECDI would propagate in the Hall thruster plasma. Subsequent measurements on other Hall thruster configurations using different probing techniques [146] as well as studies on a similar E × B device, the planar magnetron operating in the pulsed, high-current, high-density regime known as HiPIMS [157] also have shown dispersions which the authors have attributed to the presence of ECDI. The growing evidence that the ECDI exists in these devices is physically impactful as this mode was the first one linked (in simulations [143]) to turbulence-driven anomalous cross-field transport.

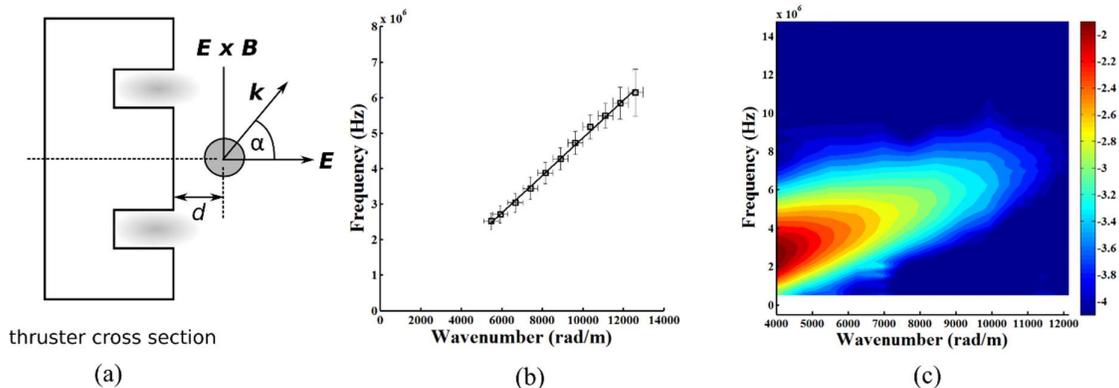

thruster cross section

(a)                    (b)                    (c)

Fig. 11. a) Schematic for Hall thruster showing measurement location for the Coherent Thomson Scattering system relative to the Hall thruster; the measurement location of the data of figures (b) and (c) was 7.5 mm. The diagnostic allows variation of the magnitude of the observation wave vector k and also its orientation, for example, via rotation through an angle α in the (E × B, E) plane, as illustrated in the figure. a) Dispersion relation for the electron cyclotron drift instability in the E × B drift direction measured using CTS. The corresponding group velocity is 3.3 km/s. Figure adapted from Ref. [145] Reproduced from S. Tsikata, N. Lemoine, V. Pisarev, and D. M. Grésillon, Phys. of Plasmas **16**, 033506 (2009), with the permission of AIP Publishing. c) Energy scaling with wavenumber for the ECDI, determined using CTS. Figure shows the log of the calibrated density fluctuation amplitude (known as the dynamic form factor $S(k,\omega)$) as a function of frequency and wavenumber.

With that said, as discussed in more detailed in Sec. 5, there is some debate in the community about the appropriateness of labeling the measured waves as so-called ECDI/EDI. This stems from the fact that the linear dispersion in Fig. 11(b) exhibits features that could also be described simply as ion sound (group velocities in the Hall direction commensurate with the local ion sound speed). This is an intriguing result as the classical ion sound is unmagnetized and therefore would not be expected to appear in the Hall direction. To reconcile this apparent incongruity, some theories (Sec. 5) posit that when wave amplitudes become sufficiently large for the ECDI, the waves become effectively unmagnetized, leading to ion sound-like behavior. Alternatively, experimental measurements have motivated a different explanation. In particular, measurements have shown the existence of a finite wave component along the magnetic field [159]. It subsequently has been demonstrated (Refs. [158],[160] and in Sec. 5) that the presence of 3D effects like this parallel propagation also may account for the sound-like dispersion.

Another question that experiments have tried to address is how much energy is contained in the excited waves. This is a critical consideration as the anomalous transport depends both on the wavelengths of the excited modes and the energy in each mode (Sec. 5). Early simulation work based on 1D and 2D kinetic theory indicated that energy could be concentrated at the electron cyclotron resonance corresponding to maximum growth (e.g., see Fig. 14 in Sec. 5) with large amplitudes (exceeding > 10% of the thermal background). This type of energy concentration is manifested in numerical results as a well-defined, characteristic length scale for the instability (e.g. Fig. 16 in Sec. 5)). The first experimental measurements, in contrast, indicated that the shape and amplitude of the power spectrum measured in actual thrusters is significantly different.

Fig. 11(c) illustrates this contrast with a plot of the distribution of energy over wavelengths from the CTS measurements performed on the same device where the dispersion was measured in Fig. 11(b). This representation shows the energy spectrum scaling (using the calibrated density fluctuation amplitude, known as the dynamic form factor $S(k,\omega)$) over the range of wavenumbers $k$ and frequencies $\omega$ found in experiments in Ref. [145]. Such information has been used to establish scaling laws (see Ref. [145]) relating the integrated fluctuation intensity $S(k)$ to $k$. An exponential decay with wavenumber has been observed, with an $e$-decrement on the order of the electron Larmor radius (< 1 mm), and amplitudes < 1% of the thermal background. The shape of this spectrum is consistent with a classically turbulent distribution in which the growth of the energy of the waves is governed by nonlinear processes, and significantly, wave energy is not limited to a single wavenumber. This departure from the predictions of early simulation results in part motivated more recent numerical studies. These new simulations have led to an updated hypothesis for the energy content of the ECDI where it has been proposed that the energy spectrum may be subject to an inverse energy cascade [167].

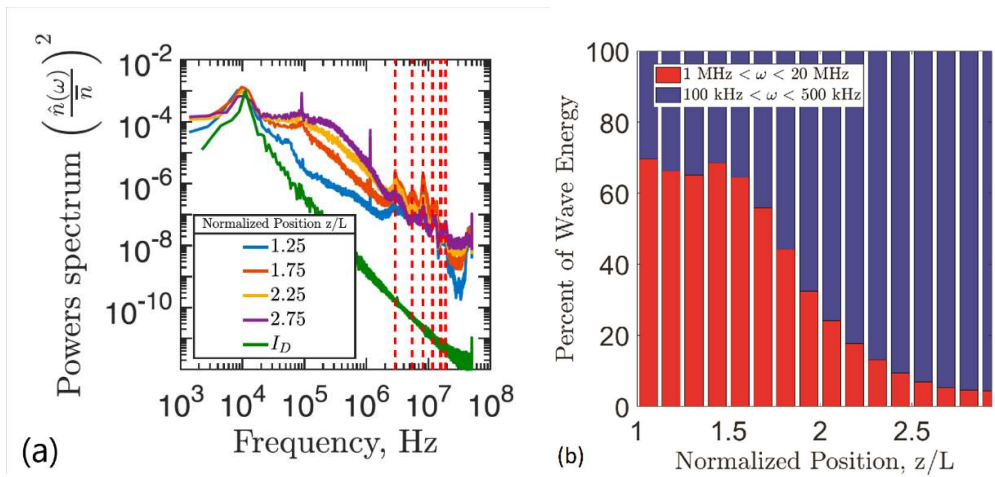

Fig. 12 (a) Power spectra of the relative fluctuations in plasma density as a function of frequency in a 9-kW class Hall thruster, determined using ion saturation probes. The power spectra are shown at multiple normalized distances from the thruster anode. L denotes the thruster channel length. The red dashed lines are drawn to mark locations of peaks in the power spectra that are assumed to be ECDI resonances. The green trace is the power spectra of fluctuations in discharge current. (b) Percentage of total energy in the density fluctuations as measured in a 9-kW Hall thruster as a function of spatial location in the 100-500 kHz regime and 1 MHz – 20 MHz. This shows a transition in the energy from high frequency (short wavelengths) to low frequency (longer wavelengths), which may be indicative of an inverse energy cascade. Both figures are taken from Ref. [146] Reproduced from Z. Brown and B. Jorns, Phys. Plasmas **26**, 113504 (2019), with the permission of AIP Publishing.

In an effort to better elucidate the processes that could lead to a nonlinear state, Ref. [146] explored the spatial evolution of the energy spectrum of oscillations in the Hall direction. Since the Hall thruster plasma convects at high speed and the ECDI moves with the plasma, it was argued in this work that measurements of the energy spectra at different distances from the thruster might show the transition of the ECDI from its initial growth upstream in the plasma to a nonlinear state. To this end, a set of translating ion saturation probes were used to measure the energy spectra of ion density fluctuations in a 9-kW class Hall thruster as a function of normalized axial distance (z/L) measured from the thruster anode.

The results are shown in Fig. 12(a), where for reference, the energy spectrum of the discharge current oscillation, $I_D$, is also plotted. At low frequencies (~10 kHz), there is a peak in the energy spectra that is consistent with the so-called breathing mode (Sec. 3). At higher values of frequency, there are number of marked features that are qualitatively consistent with the nonlinear growth processes outlined Ref. [167]. Closer to the thruster, there are a series of discrete peaks (denoted with dotted vertical dashed lines). These were shown in Ref. [146] to correspond to the resonant frequencies of maximum growth predicted from linear theory (Fig. 14 in Sec. 5). With increasing distance from the thruster exit plane, the amplitude of these peaks (1 MHz – 20 MHz) decreases. Concurrently, there is an increase in energy from 100 kHz – 500-kHz where the spectrum is broadband, showing an inverse decay in frequency. The exchange in energy from discrete structures at high frequency to these broadband lower frequencies is captured by Fig. 12(b), which shows the percentage of total wave energy in each frequency range as a function of position.

The transition of energy from the high frequency peaks near the thruster to a more broadband, lower frequency spectrum is consistent with the interpretation that ECDI waves are born in the upstream region and grow as they are convected downstream. The growth of the spectrum transitions from linear to non-linear, giving rise to an inverse energy cascade to lower frequency (longer length scales). With that said, we note that although the data shown in Fig. 12 is consistent with the idea of an inverse energy cascade, experimental measurements of this transition to date have been confined to frequency and not wavelength (a limitation of the measurement technique). To fully explore the mechanisms leading to the non-linear energy spectrum, future measurements will need to measure the full range of wavelengths.

In addition to studying the dispersion and energy content of the drift-driven waves, experimental efforts to date also have focused on trying to link the measured wave properties to anomalous transport known to exist in these devices. To this end, most experimental methods have relied on a quasilinear approximation (c.f. Ref. [152]), which can be leveraged to relate transport coefficients such as an anomalous collision frequency to wave amplitude and growth rate. Ideally, this method requires simultaneous measurements of the electric field and density fluctuations associated with the waves. In practice, this type of measurement has not been feasible in these devices, and instead, linear approximations are employed to relate what can be measured (typically density) to the potential fluctuations in the waves. For example, in low temperature electrostatic modes, it is common to use the Boltzmann relation [150], [151], [160].

Subject to these simplifying assumptions about transport and the wave properties, experimental measurements have been used to calculate effective collision frequencies for electrons in the plumes of $E \times B$ devices. It was shown in Ref. [151], for example, that depending on the assumptions about the properties of the ECDI, the electron transport from waves can account for 25-100% of the measured electron transport in a Hall thruster. Similarly, an attempt was made in Ref.[160] to determine the electric field directly from CTS by using calculations on the electron density fluctuation to determine a corresponding fluctuating potential. This gave an ECDI field amplitude that was 25% of the background electric field in the far-field thruster plume. 2D numerical simulations performed in Ref. [146] [143] demonstrated that fields this large could account for all of the anomalous electron transport.

With that said, there are a number of reservations about the validity of this preliminary, quasilinear analysis. For example, the evident turbulent spectrum of the fluctuations already suggests that nonlinear effects cannot be ignored. Similarly, the simplifying assumption that the electron density fluctuations can be related to the potential fluctuations through a Boltzmann relation is not consistently supported by findings from available PIC simulations. The use of the Boltzmann relation also relies on the assumption of thermal electrons, which does not apply to much of the region where the instability is generated. Recent PIC work (Sec. 5) also shows that the applicability of quasilinear analysis requires further study. Linking experimental measurements of turbulence to anomalous transport, in light of the complex nature of the particle properties and behavior in the plasma region of interest, therefore poses a considerable challenge and remains an active area of research.

While the exact relationship between turbulence measurements and transport remains an open question, experimental measurements of plasma turbulence are informing a new, correlational understanding of how turbulence-induced transport may impact self-organized behavior in thrusters and magnetrons (Sec. 3). For example, experimental measurements have indicated that the ECDI can coexist with lower-frequency (kHz) breathing-mode oscillations [147],[148], and experimental studies on magnetrons have revealed the presence of the ECDI and its coexistence with kHz-frequency rotating oscillations or spokes discussed in the previous section.

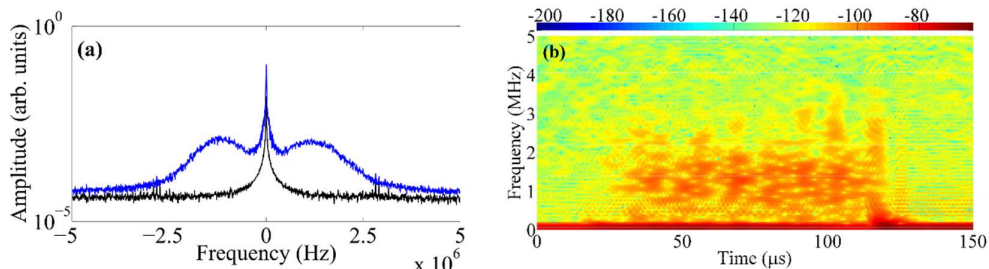

Fig. 13. Measurements of ECDI behavior in a pulsed planar magnetron: in (a), the averaged symmetric MHz frequency spectrum characteristic of the azimuthally-propagating ECDI; (b) a spectrogram of the ECDI over a single 150 μs operating pulse, showing MHz and kHz-scale behavior. Images from Ref. [149]; Reproduced with permission from Phys. Rev. Lett. **144**, 185001 (2015). Copyright 2015 American Physical Society.

An example of such a result obtained in recent years from CTS experiments on pulsed planar magnetrons is shown in Fig. 13, from Ref. [149]. In Fig. 13 (a), this image shows the averaged symmetric MHz frequency spectrum characteristic of the azimuthally-propagating ECDI measured at Larmor-radius observation scales. A time-resolved analysis of the ECDI over a single 150 µs operating pulse, shown in Fig. 13 (b), illustrates how these MHz fluctuations exhibit an additional kHz range modulation (measurable due to the fluctuations in density as spokes traverse the measurement region). Recent experimental work on hollow cathodes operating in conjunction with a Hall thruster also showed that not only does plasma turbulence dominate the electron transport in the cathode, but it also may be linked to lower-frequency oscillations [151], [152]. Taken together, these experimental results from cathodes, Hall thrusters, and magnetrons now point to the conclusion that turbulence-driven transport has a critical role in governing both the steady-state plasma properties as well as the transient, large scale oscillations discussed in Sec. 3.

*Current and future challenges*

The most pressing need for future experimental efforts is to clarify the relationship between different instabilities and anomalous transport. Although the ECDI is now considered a key factor in this process, supported by simulations showing its appearance leading directly to an axial electron drift (c.f. Ref. [156]), it is unlikely to be the only instability concerned. The balance of contributions from different unstable modes may depend on the nature of the operating regime. However, evaluating the contribution of each mode to transport poses a number of experimental and theoretical challenges.

One of the most significant challenges is estimating the electron transport directly from plasma wave properties measured experimentally. As discussed in the preceding, the nonlinear state of the measured instabilities calls into question [148] the current practice of using simplified expressions for transport coefficients derived from quasilinear theory. The least ambiguous approach to overcome these challenges is to measure the amplitude and relative phase of electric field and density fluctuations simultaneously across the energy spectrum. Ideally, future diagnostics would offer this capability.

This need also could be addressed by finding methods or validated theories to map more precisely the fluctuating electric field amplitude to density. While the very nature of the phenomena under study (weak fluctuations, high-frequency modes) renders this a significant challenge from a theoretical perspective, detailed simulations may offer a potential solution. For example, one possible future technical path could involve developing higherfidelity simulations, validating the predictions of these simulations against measurements that can be made with current techniques (e.g. CTS or probe-based measurements of density fluctuations), and then using the simulation results to determine the appropriate map from density to electric field for the experiments. To be effective, this type of approach would need to address the role of significant non-linear effects (such as the coupling of wall emission to the ECDI [148], which affects the overall electron current) and 3D effects present in experimental devices that are often absent from the linear kinetic theory and certain model descriptions.

Another key future challenge for experimental measurements will be to document the growth and energy flow of the excited instabilities. As discussed in the preceding and Sec. 5, the earliest notions regarding ECDI-driven transport as an electron Larmor-scale phenomenon are now being revised in light of observations in simulations and experiments of the development of large (cm-scale) modes, which have been linked to an energy transfer from smaller scales. Future efforts will need to assess this type of energy flow experimentally. This will require improvements in both the spatial and wavelength resolution of current diagnostics.

Building on the work in Ref. [146], characterizing the spatial evolution of the turbulence as it convects out of the Hall thruster can provide critical clues about the growth and saturation of the turbulence. However, this type of spatial interrogation poses a problem for current, non-invasive optical techniques that cannot reach the internal parts of the discharge chamber in thrusters. This latter issue regarding optical diagnostic access may potentially be alleviated through the study of modified architectures, such as so-called magnetically shielded thrusters, which shift much of the acceleration and ionization region beyond the confines of the thruster channel. However, such architectures induce modifications to the plasma which further complicate physical interpretation.

Future efforts to study energy flow also will require an expanded capability to interrogate a wider range of length scales (1 mm $< \lambda < 1$ cm) than currently can be accessed with state-of-the-art methods. These measurements similarly should have the ability to characterize all three dimensions of propagation of the waves in order to assess the spatial direction of energy flow. Given that there is evidence of coupling between the turbulence and low-frequency, self-organized structures, these future diagnostics also should have the capability to measure turbulence properties on the time scale of the lower frequency modes discussed in Sec. 3 (i.e. ~ 100 kHz).

As the sophistication of diagnostics continues to grow, the insight that emerges from experiments must continue to be leveraged to inform improved, predictive kinetic modeling. The comparison between experiments (measuring 3D phenomena) and numerical simulation results can only truly be considered valid if the simulations can eventually reach a sufficient level of sophistication, accounting for (i) the time scales needed for the study of fast and slow waves, (ii) the full range of spatial scales (from electron Larmor radius-scales to centimetric scales of large-scale fluid turbulence), and (iii) the 3D plasma environment itself. An example of the importance of this convergence is illustrated in the following way. Numerical work from a large number of groups (involving 1D, 2D PIC codes) uniformly show two persistent features for the ECDI-like fluctuations developing in the region of the E×B drift: (i) the establishment of a single dominant length scale for the mode (< 1 mm), and (ii) the alignment of these fluctuations primarily along the azimuthal direction (but with a positive axial wave vector component). In contrast, more recent 3D simulations (see Sec. 7) show clearly (i) the weakening of the coherent nature of the ECDI fluctuations (which become more compatible with a broadband structure, already established in CTS experiments) and (ii) the disappearance of a preferred propagation direction outwards. The differences already observable between 1D, 2D and 3D simulation results are strong evidence of the need for 3D simulation codes which can provide a more solid basis for future numerical/experimental comparisons.

Beyond kinetic simulations, experimental measurements also will be critical for guiding fluid-based attempts to approximate the effects of the turbulence on background plasma properties (Sec. 6). Indeed, while fluid simulations cannot capture the kinetic effects of the waves directly, these processes can be represented with approximate closure models. The fidelity of these closure models in turn depends directly on an understanding on the evolution and growth of macroscopic properties of the turbulence such as the total wave energy. Recent work on the study of anomalous electron transport in the hollow cathode, the electron source for Hall thrusters, could serve as a potential roadmap for future efforts.

In these previous works, experimental measurements of the evolution of a turbulence and anomalous transport in both space and time were leveraged to guide the development of fluid codes that included equations to approximate the effects of plasma turbulence on electron resistivity [84], [151], [152],[155],[168],[169]. These models in turn have been successfully used to simulate the plasma in both the interior and exterior of the cathode. A similar approach could be applied for modeling the effects of turbulence-induced anomalous transport in $E \times B$ devices, though, unlike in the cathode studies where the turbulence lent itself to a simple 1D description, closure models for the effects of $E \times B$ modes likely will be more nuanced. Before it is possible to develop approximations for these effects in a fluid framework, many unresolved questions related to the turbulence discussed above (3D propagation, inverse energy cascades, etc.) will need to be resolved.

*Advances in science and technology to meet challenges.*

Many of these technical challenges outlined in the preceding section may be addressed by building on existing diagnostics techniques. For example, the current implementation of CTS was designed for the purpose of accessing shorter wavelengths (< 2 mm) as this is where PIC simulations suggested the ECDI should exist. In principle, the CTS technique can be modified to access a wider range of wavelengths and therefore capture the relevant content at longer wavelengths. This increase in wavelength range comes at the expense of spatial resolution (as there is an approximate inverse relationship between measurable wavelength and spatial resolution). Future efforts should focus on expanding the capabilities of CTS while maintaining high spatial resolution. With that said, assuming CTS still will have a practical lower bound in wavelength, probes can be used to fill in information about the energy spectra of oscillations at longer length scales.

The recent development of a sensitive incoherent Thomson scattering (ITS) platform [137] allowing investigations of cathode [161], thruster [162], and magnetron plasmas[164] gives access to information that has long been lacking regarding background electron properties (temperature and drift) in these sources. This provides crucial information that can impact the predicted growth and density fluctuations in the measured waves. It is hoped that the coupling of improved CTS and ITS can be an important tool for validating basic physical understanding regarding the conditions in which certain instabilities arise. An understanding of the velocity distribution of electrons also may help lead to refined expressions for relating measured wave properties to transport [181]. Similarly, there have been recent advances in techniques based on laser induced fluorescence that have allowed, for the first time, the non-invasive measurement of anomalous electron transport inside Hall thrusters [170]. This data will be crucial for providing "ground truth" for the local transport that can be used as a point of comparison to determine if the contributions from instabilities, if any, are sufficient to explain the electron dynamics.

As discussed in the preceding, in order to directly estimate transport from the measured instabilities, it is necessary to make measurements of the electric field fluctuations and electron density fluctuations simultaneously. Both CTS and probe-based measurements that have been employed to date, however, have focused on characterizing oscillations in the plasma density. In principle, probes could be used to measure both potential and density (with limited spatial resolution), though probes are known to perturb the local plasma state. Alternatively, there are diagnostic techniques from the study of higher energy plasmas such as Stark broadening that may offer a potential technical path for characterizing the electric field fluctuations non-invasively and at small wavelengths. There are several technical hurdles stemming from accessibility and signal to noise ratio that must first be overcome, however, before these methods can be applied to low temperature systems.

Numerical simulations (and, as described above, 3D simulations in particular) may offer a key capability for bypassing the need to make direct measurements of electric field. Before these can be used, these simulations must first be validated against measurements of both the background and wave properties. With this in mind, establishing the error bars of invasive and non-invasive techniques will be important for future progress in the use of models to inform experiment. It will be necessary to explain and quantify differences observed between probes and optical techniques for the measurement of electron properties, and to establish on what data future simulation efforts should be validated, given the limitations and challenges of different diagnostics. As an example, measurements of the electron energy distribution function using ITS rely on obtaining a sufficiently high signal-to-noise ratio, which is difficult for plasma densities in the range of $10^{16}$ m$^{-3}$, even now with recent diagnostics improvements. This requirement would ultimately affect the detection of features pertaining to the spectral wings such as high-energy electron populations.

Combining information from complementary diagnostics – on electron density fluctuations, EEDFs, absolute electron density, electron temperature, and ion properties – is required to refine models and theory, and gradual progress is being made on this front. The combination of multiple diagnostics running simultaneously on the same test device would provide an ensemble of useful data for simulations. As noted above, a concerted attempt to bridge the gap between measurements at large scales using probes, and small-scale measurements using CTS, could be an important contribution towards establishing the presence (or absence) of dominant wavenumbers in anomalous transport. As these experimental capabilities become available, it will be necessary to analyze experimental findings coupled with sufficiently-advanced numerical codes, still under development, which are capable of capturing 3D plasma features.

*Conclusion*

There has been substantial progress in the past decade in experimental methods and diagnostics that have yielded new insights into the role of plasma turbulence in E × B devices. The detection and in-depth study of different types of instabilities present in such devices, such as the ECDI, ion acoustic turbulence, spokes and high-frequency, long-wavelength modes [175],[176] have contributed to our understanding of basic physics. Several challenges remain, including reconciling experimental data with simulation predictions, understanding the limitations of invasive and non-invasive diagnostics, and studying the growth and nonlinear evolution of the observed turbulence and anomalous transport in E × B devices. Efforts in diagnostic development and analysis are underway to help address these challenges. Looking to the future, the development of standardized test devices, particularly ones that can be shared internationally, could enable coordinated and more impactful efforts that leverage the diagnostic capabilities of multiple institutions. Similarly, closer collaborations between experimentalists and modelers will be critical for establishing the validity of codes currently under development, understanding the role played by different instabilities in thruster operation, and guiding the design of future E × B devices and modeling efforts to study plasma turbulence. It has become evident that understanding features of plasma turbulence measurable experimentally, and the contribution of different instabilities to transport, will also ultimately require 3D numerical code results for comparison. Parallel progress in both diagnostic development and numerical simulation capability is a clear objective for the future.

# 5. Electron Drift Instabilities in E×B plasmas: Mechanisms, Nonlinear Saturation and Turbulent Transport

Andrei Smolyakov[1], Jean-Pierre Boeuf[2], and Trevor Lafleur[3]

[1] University of Saskatchewan, 116 Science Place, Saskatoon, SK S7N 5E2 Canada
[2] LAPLACE, Université de Toulouse, CNRS, INPT, UPS, 118 Route de Narbonne, 31062 Toulouse, France
[3] PlasmaPotential, Entry 29, 5/1 Moore Street, Canberra ACT 2601, Australia

*State of the art and recent progress*

Electron transport across the magnetic field in devices employing magnetic filter configurations is typically anomalous and exceeds the transport due to classical collisions of electrons with other particle species in the plasma. Typically, the inclusion of wall-collisions (near-wall conductivity) is not sufficient to explain experimental values inside the channel. Moreover, the electron current significantly exceeds the classical value outside the channel, where the near-wall conductivity is absent. Thus, it is widely believed that the anomalous transport enhancement is due to convective transport and scattering of electrons due to turbulent plasma fluctuations. The nature of such fluctuations is neither well understood, nor does there exist any validated theoretical model that can be used to predict the turbulent electron transport for specific thruster conditions.

The electron drift instability (EDI) due to the electron drift has attracted a lot of attention recently as a possible instability and anomalous transport mechanism in Hall thrusters [142],[143],[177],[178],[179],[180],[181]. Such interest was also stimulated by experimental observations of small scale fluctuations where the wave-frequency is found to be linear with the wave-vector, which is consistent with the ion sound wave dispersion relation, see Fig. 11.

Studies of this instability were started much earlier in relation to the problem of anomalous resistivity to explain the width of collisionless shock waves in space and turbulent heating experiments [182],[183],[184],[185]. The instability occurs as a result of the differential drift of electrons with respect to unmagnetized ions, when the Doppler frequency shift results in an overlap of the kinetic resonances of electron cyclotron (Bernstein) type modes with the ion sound branch. As a kinetic instability, it does not require plasma or magnetic field gradients, and is purely based on the electron E×B drift in crossed magnetic and electric fields. When applied to Hall thrusters, the EDI is of great interest in the region of large electric field, where one can expect that the kinetic instability due to this large electric field will dominate over other instability mechanisms, such as plasma gradients and collisions [74].

The genesis of EDI mechanisms can be tracked down to the magnetized Buneman instability in cold plasmas [186]. This is the reactive instability which occurs as a result of the interaction of two stable modes: the upper hybrid resonance, and the low-frequency ion oscillations, $\omega^2 = \omega_{pi}^2$ (which is the short wavelength limit of the ion sound mode). The upper hybrid mode is Doppler shifted by the electron $\mathbf{v}_E = \mathbf{E} \times \mathbf{B} / B^2$ drift so it moves into the ion plasma $(\omega_{pi})$ frequency range. In the plasma with a finite electron temperature, the overlaping modes are the Bernstein and ion sound modes, $\omega_{ce} - k_y v_E = \omega$, so that the approximate resonance codition is $k_y v_E = \omega_{ce}$, where $k_y$ is the wavevector in the direction of the $\mathbf{E} \times \mathbf{B}$ drift, which corresponds to the azimuthal direction in Hall thruster and magnetron geometries, and $\omega_{ce}$ is the electron cyclotron frequency. When one allows electron motion along the magnetic field, an additional instability appears due to the finite value of the wavevector along the magnetic field, $k_z \neq 0$; the so-called Modified Two-Stream Instability. This instability was also considered in the original paper [186], so below it will be called the modified Buneman two-stream instability (MBTSI), which has the largest growth rate for relatively small values of $k_z \ll k_y = \omega_{ce} / v_E$.

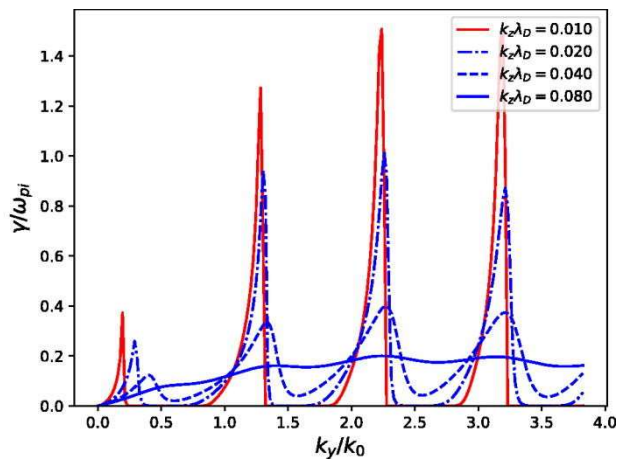

Fig. 14. The linear growth rate of the ECDI, for various values of the wave-vector along the magnetic field $k_z \lambda_D$ of 0.01, 0.02, 0.04 and 0.08, respectively. The first root from the left is the MTSI (m=0), and subsequent roots are the kinetic resonance modes with $k_y / k_0 = m = 1, 2, \ldots$, where $k_0 = \omega_{ce} / v_E$. The transition to unmagnetized ion-sound instability corresponds to $k_z \lambda_D \leq 0.08$ for the parameters in [195]. Reprinted with permission from [195]; Reproduced from S. Janhunen, A. Smolyakov, D. Sydorenko, M. Jimenez, I. Kaganovich, and Y. Raitses, Phys. Plasmas **25**, 082308 (2018), with the permission of AIP Publishing.

For purely perpendicular propagation and finite electron temperature, there a exists a set of multiple narrow bands of reactively unstable modes near the resonances $\omega - k_y v_E - m\omega_{ce} = 0$, $m = 1, 2, \ldots$. When electron parallel motion is included, for finite $k_z$, kinetic (Landau) resonances become possible leading to dissipative instabilities at $\omega - k_y v_{E0} - k_z v_{\parallel} - m\omega_{ce} = 0$. For $k_z v_{Te} \geq \omega_{ce}$, the kinetic resonances broaden and eventually overlap, resulting in an instability equivalent to the ion sound instability in an unmagnetized plasma, but driven instead by the electron E×B beam flow [158], [179], [187], see Fig. 14. The various limits of this linear instability (usually referred to as the electron cyclotron drift instability, or ECDI) are well studied, however nonlinear behavior and saturation are much less understood, especially for small or zero $k_z$ when linear Landau interactions due to particle motion along the magnetic field are not possible. Some earlier studies [184] have concluded that at a certain amplitude, the cyclotron resonances are washed out by nonlinear effects and the instability proceeds as an ordinary ion sound wave instability as in unmagnetized plasmas until it is saturated by linear or nonlinear Landau damping by ions. Other studies have argued [188],[189] that electron trapping in the magnetic field remains important, making the nonlinear stage different from that of an unmagnetized plasma ion sound instability. Alternative mechanisms (different from cyclotron and kinetic resonances) have also been proposed [190] to explain the nonlinear instability, so below we refer to the general nonlinear regime of this instability as the electron drift instability (EDI).

The ECDI/EDI is a very robust instability and can easily be seen even in the simplest versions of 1D PIC simulations of magnetized plasmas, e.g. see the cold plasma example in Ref. [191]. In the context of the anomalous transport in Hall thrusters, it has been studied in 1D simulations [142],[167],[179], 2D axial-azimuthal simulations [143],[178],[181],[192] and 2D radial-azimuthal simulations [148],[193], [195]. Typically, in such simulations, significant anomalous transport in the axial (along the direction, as well as very effective electron heating, are observed. In 1D versions, the instability occurs for perturbations propagating strictly in the direction of the electron drift, and typically shows up in the simulations as a fairly coherent non-quasineutral mode with strongly peaked ion density perturbations which are noticeably larger in amplitude than those of the electron density which remains relatively smooth [142], see Fig. 15.

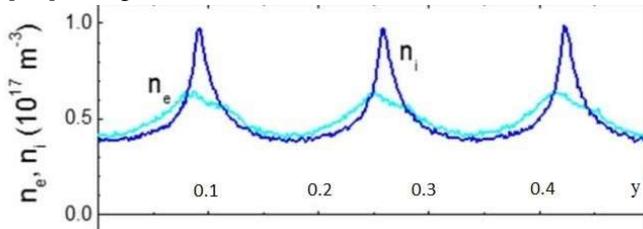

Fig. 15. Ion and electron density perturbation as a function of the azimuthal position (cm) in 1D simulations of EDI. Reprinted from [142]; Reproduced from J-P. Boeuf, Journal of Applied Physics **121**, 011101 (2017), with the permission of AIP Publishing.

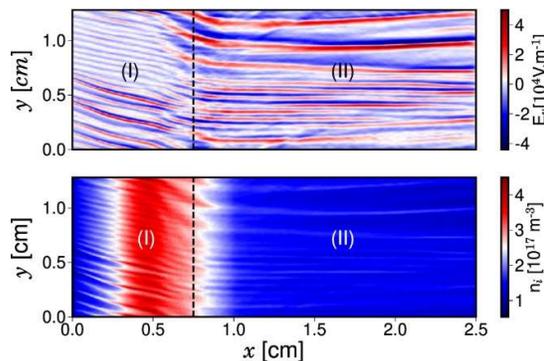

Fig. 16. 2D axial-azimuthal maps of the azimuthal electric field (top) and ion density (bottom), reprinted from Ref. [209]; Reproduced with permission from Plasma Source Sci. & Technol. **28**, 105010, (2019). Copyright 2019 IOP Publishing.

Quasilinear models of electron transport due to the EDI have been proposed [180],[198],[199],[200] based on the assumption that the underlying turbulence is the ion-acoustic as is in the case without the magnetic field. In the quasilinear approximation, the magnitude of the turbulent flux was estimated as $\Gamma = \left\langle \tilde{n} \tilde{E}_y \right\rangle / B_0$, using relations between density and electric field fluctuations from the linear theory of ion-sound instability in unmagnetized plasma. The wave amplitude at saturation in these models [180], [198],[199],[200], was estimated from the linear wave kinetic equation for the evolution of the wave energy $W_k = \left| E \right|_k^2$ convected axially by the ion flow (with the wave group velocity in the axial direction), and a combination of the linear Landau damping from warm ions and nonlinear ion trapping. However, since the wave kinetic equation is linear in the wave energy $W_k$, the result is sensitive to the choice of the initial "reference" value at some given location. A further refinement was done in Ref. [198], based on the comparison with PIC simulations which show substantial non-Maxwellian distribution functions for the electrons. Using non-Maxwellian distribution functions [198] reduces the ion-sound growth rate bringing the estimate of the fluctuation level down and in closer agreement with magnitudes expected experimentally.

Experimentally, it was found that fluctuations near the acceleration region exhibit discrete cyclotron resonances characteristic of the ECDI while downstream the turbulence evolves into the linear dispersion relation typical of the long-wavelength ion sound modes [146],[201].

*Current and future challenges*

There are significant challenges in numerical simulations of the electron drift instability, as well as in the theoretical understanding and interpretations of the results of such simulations. Because of computer resource limitations, many simulations have to be performed for parameters far realistic: reduced dimensionality, low resolution, and limited spatial coverage. Several model approximations for ionization sources and energy particle losses are also often used. The EDI has proved to be a very effective mechanism of electron heating leading to a fast rise in the electron temperature. In the 1D periodic domain (and in some 2D) simulations a "virtual" axial acceleration length and/or inelastic collision losses have to be introduced to achieve a stationary temperature state. The level of anomalous transport may become dependent on the mechanism of how the electrons are "cooled down" in these simulations. Concerns have been raised [173] that the strong electron heating may be in part due to numerical instabilities. Many simulations are performed over a limited azimuthal region (e.g. in a domain around 1cm wide versus 30 cm in reality). Such narrow widths may prevent the development of long-wavelength azimuthal modes that typically dominate the anomalous transport. Investigations of the effects of larger azimuthal simulation boxes, and extending them to realistic values is difficult because of computer resource limitations, but will be required to become relevant to experiments. The inclusion of elastic and inelastic collisions, ionization, as well as full 3D effects make such simulations very resource demanding.

Among basic questions are the effects of the magnetic field on the electrons, in particular, whether the description in terms of fully unmagnetized ion sound is appropriate for modeling the EDI in the nonlinear regime, and on the saturation mechanisms. The differences between the ion sound type turbulence in unmagnetized plasmas and in a plasma with a magnetic field have been much debated previously [188], [203], [204] without definitive conclusions that can be directly applied to Hall thruster parameters.

In the linear case, the transition to the ion sound occurs for finite values of the wave vector along the magnetic field with $k_z v_{Te} \simeq \omega_{ce}$. Several PIC simulations were performed however, in 1D (azimuthal) and 2D (azimuthal-axial) geometry, where the direction along the magnetic field is ignored, $k_z = 0$, and thus the linear mechanism of the transition to the ion sound is absent. The cyclotron resonances can be smoothed out by collisions for $(\nu / \Omega_{ce}) k^2 \rho_e^2 > \pi / 2$, forcing the instabilty into the ion-sound regime. However for most typical regimes of interest, simulations were collisionless or almost collisionless. For the most part, numerical noise, which can play the role of collisions, was estimated to be in the range of $\nu_n = 5 \times \left(10^{-5} - 10^{-6}\right) \omega_{pe}$ [202],[203]; therefore, according to the criteria above, it is not expected that numerical noise can result in electron demagnetization. The statistical fluctuations $\sim 1 / \sqrt{N}$ however can be essential in simulations with low numbers of particles per cell, $N$, and can affect the level of anomalous transport.

Another reason for smoothing out the cyclotron resonances is nonlinear resonance broadening [205] which can effectively demagnetize electrons. For short-wavelength regimes, with $k \rho_e > 1$, a simple estimate for the electron demagnetization has the form $\Xi > \left(k \rho_e\right)^{-1}$ [167], where $\Xi \equiv \left(\omega_{pe}^2 / \Omega_{ce}^2\right) W / \left(n_0 T_e\right)$, $W = E^2 / 8\pi$. It has been reported that in some 1D and 2D simulations [167], [195] this criteria is not satisfied (though not with a large margin). Consistent with this, the nonlinear evolution of the electron cyclotron instability in these simulations show distinct electron cyclotron resonances at $k_y v_E = m \omega_{ce}$, [167], [195], [196], [197] as well as the ion sound modes features, e.g. mode propagation with a phase velocity of the order of the ion sound speed and oscillations (and nonlinear harmonics) at $\omega = \omega_{pi}$ in the short wavelength part of the spectrum.

Azimuthal-radial simulations, in which the $k_z$ (in the radial direction) is finite, and the linear mechanism to the transition to ion sound should remain operative, also show inconclusive results. Quasi-coherent nonlinear waves at the cyclotron resonances moving with roughly the ion sound velocity have been observed in highly resolved simulations [195], similar to the 1D case, with strongly nonlinear cnoidal waves peaked at short wavelengths. The important role of the modified two-stream instability, due to a finite $k_z$, leading to strongly anisotropic heating and large scale radial structures in the anomalous current, was emphasized [195]. Various values of the effective $k_z$ were reported in different simulations [195],[193], [194],[24]. Effects of Secondary Electron Emission (SEE) and different propellants were studied [194], [24] and it was shown that the EDI activity is affected by the reduced electron temperature due to sheath cooling. Besides, SEE induces large electron transport due to near-wall effects.

The 2D azimuthal-axial simulations, while remaining relatively simple, are the closest to real Hall thruster configurations (in some ways) and allow one to test the effects of the axial profile of the electric field (the main EDI driver), and the effects of axial mode propagation. In these simulations, the direction along the magnetic field is ignored and the linear mechanism of the transition to the ion sound does not apply. Nevertheless, these simulations reveal strongly coherent nonlinear waves propagating with a velocity of the order of the ion sound velocity, and a wavelength that scales with the Debye length, and a wave amplitude that seems to be consistent with estimates from the ion trapping mechanism [192]. The azimuthal-axial simulations in Ref. [198], [181] report transport levels generally consistent with the model in Refs [179], [206] and some experimental results.

It appears that in the nonlinear stage, the EDI typically reveals itself as a coherent strongly nonlinear wave, akin to the periodic cnoidal wave with characteristic features of wave breaking manifested by sharp peaks in the ion density, and with smaller and much smoother electron density perturbations. This is consistent with the regime of unmagnetized ions, which therefore exhibit the tendency of wave breaking (similar to the neutral gas sound wave modes), while the electrons remain largely magnetized and show a smoother density. The coherent, almost single mode, emerging in many numerical simulations, e.g. see Fig. 16 presents a challenge to the application of quasilinear theory, which assumes an ensemble of weakly nonlinear, wide spectrum overlapping modes. The anomalous transport estimate in the form $\Gamma = \left\langle n\widetilde{E}_y \right\rangle / B_0$ assumes that the dominant electron current is due to the E×B drift of magnetized electrons, which appears to be inconsistent with the regime of unmagnetized electrons for the ion sound instability in the absence of a magnetic field, although simulations [192], [198] seem to show that this estimate is roughly valid.

*Conclusion*

For typical parameters in the acceleration region, the electron drift velocity $v_E = E_0 / B_0$ becomes comparable to the electron thermal velocity (and much larger than the ion sound velocity, $v_E \simeq v_{Te} \gg c_s$).. In such regimes, nonlinear electron and ion dynamics become strongly nonlinear and numerical modeling is critical for theoretical advancement and validation. This regime is far from the marginal stability criteria, and normally a broad spectrum of excited modes is expected. Nevertheless, many PIC simulations demonstrate the excitation of a highly coherent mode and very effective electron heating. The mechanism of anomalous transport and heating in the electron interaction with quasicoherent waves and the role of numerical noise has to be understood. Several simulations indicate that secondary nonlinear processes take place resulting in the appearance of long-wavelength modes (similar to modulational instabilities) on top of the quasicoherent mode. The large scale modes are typically expected to provide large contributions to the anomalous transport and need to be resolved for realistic parameters. Therefore simulations with larger spatial boxes representative of realistic geometrical dimensions are important, and will require substantial computer resources.

High-performance large scale simulations need to explore how the current results (with low resolution and narrow simulation boxes) can be extrapolated to real-sized devices. Many reported simulations from different groups have been performed with varying approximations and assumptions, which makes the comparison more difficult and inconclusive. Recently, a broad collaborative effort between several groups has been initiated to investigate the accuracy and convergence of the results of different numerical codes [208], [209] under the same conditions. 2D axial-azimuthal Particle-In-Cell benchmark for low-temperature partially magnetized plasmas instabilities has been recently completed [209]. The results obtained for this benchmark show good agreement between the different codes from several groups.

One has to note also, that the ECDI is only one of a number of instabilities that could be relevant to Hall thrusters, and more generally, to E×B devices; e.g. resistive axial and azimuthal instabilities can be important as well as azimuthal Simon-Hoh type modes [74], [75], [107] that are decsribed in Sections 3 and 6.

# 6. Fluid and Hybrid (Fluid-Kinetic) Modeling of E×B Discharges

Andrei Smolyakov[1] and Ioannis G. Mikellides[2]

[1] University of Saskatchewan, 116 Science Place, Saskatoon, SK S7N 5E2 Canada
[2] Jet Propulsion Laboratory, California Institute of Technology, Pasadena, CA, 91109, USA

*State of the art and recent progress*

A major challenge in the modeling of E×B discharges such as those in the HET has been that the governing processes are inherently three-dimensional (3-D) and span multiple scales. Specifically, the spatiotemporal resolution must span the device length down to the electron Larmor radius ($\rho_e$) as well as the long timescales associated with the motion of atoms down to those for the electron motion. Moreover, it has been argued that the species velocity distribution function can be far from Maxwellian, implying that the nature of plasma instabilities in these discharges may be strongly affected by kinetic effects. If true, fully kinetic simulations would be the most comprehensive approach to model such conditions. However, in many cases, such simulations are as complex as experiments and can be difficult to interpret. Though petascale computing is now possible, kinetic simulations with realistic parameters that span the entire HET domain remain computationally very expensive. Although particle methods for all three species in z-θ [143],[206],[211] and r-θ [148],[212],[213] domains, and even in 3-D [173],[214], have indeed been used successfully and have yielded critical insight in some cases, they are constrained to restricted spatial and/or temporal domains, which limits our ability to capture the entire range of physical processes. In HETs for example, though on one hand simulations in the z-θ plane can provide detailed insight into instabilities with wave vectors in the E×B and axial directions, they provide no information about plasma-wall interactions which are known to affect the operation of the device. On the other hand, in the r-θ domain the simulations cannot take into account the axial variation of the plasma properties and therefore cannot provide insight into how the transport physics evolves throughout the different regions of the channel.

Fluid theory predicts several instabilities that may be responsible for turbulence and transport in E×B discharges. In general, fluid simulations that account for plasma turbulence and thus produce self-consistent transport are computationally faster and cheaper compared to kinetic simulations. Such simulations are also easier to interpret, provide much greater flexibility in separating various physics elements and are vital in developing intuition of the complex processes that occur in plasma discharges. Over the years, fluid models have been providing indispensable contributions to the development of nonlinear physics of hot and dense magnetized plasmas such as in space and laboratory devices for fusion applications [233],[239],[247].

Generally, in rarefied plasmas where the particle collision mean free paths are large, the fluid models suffer from the fundamental problem of closure. When the magnetic field is strong enough such that the wavelength (λ) of the modes of interest and other characteristic length scales (L) are large compared to $\rho_e$, $[\rho_e/L, \rho_e/\lambda] \ll 1$, the dynamics in the plane perpendicular to the magnetic field is well represented by the fluid model. In fact, reduced fluid models taking into account the higher-order gyroviscosity tensor can capture well the behavior of the plasma even in regimes of short wavelengths ($\rho_e > \lambda$) [252]. Many phenomena in E×B discharges occur with a characteristic frequency well below $\omega_{ce}$. Therefore, another restriction of fluid models, namely, the low-frequency approximation, $\omega \ll \omega_{ce}$, is typically well satisfied for many processes of interest.

Specifically, for E×B discharges, fluid models and insights from the theory of fluid instabilities have been used successfully to characterize a wide range of fluctuations, especially in HETs [253]. For example, using the proper magnetic field profiling the stabilization of the most violent large-scale azimuthal modes driven by plasma density and magnetic field gradients [248],[249] have been predicted [235],[243],[244],[245]. The generalization of the fluid theory to shorter length scales (of the order of $\rho_e$ and smaller) leads to the lower-hybrid modes which are destabilized in E×B plasmas by density gradients and collisions [118]. The gradient-drift instabilities (i.e. modes driven by density gradients and electron E×B drift) have been invoked to explain turbulence in magnetron configurations [251] and various regions along the acceleration channel of HETs [74], [107], [240],[246] ,[255]. A theoretical model of such waves has been developed [74] that accounts for plasma density, temperature, and magnetic field gradients as well as electron-neutral collisions and sheath boundary effects [237], [122]. This model also predicts another type of unstable modes: resistive axial instabilities that propagate in the direction of the applied electric field [117],[232],[236]. These modes saturate due to ion non-linearities (similar to the ion sound waves) and therefore can grow to large amplitudes [117], though they have generally smaller growth rates compared to the azimuthal modes. It has been suggested that the resistive axial modes play an important role as a trigger for the low-frequency ionization modes involving plasma and neutral density dynamics also known as the breathing modes [232]. Nonlinear 2-D z-θ simulations of gradient drift modes demonstrate complex self-organization of turbulence coexisting at small and large scales with a large level of anomalous transport with the effective Hall parameter , $H = \omega_{ce} / \nu_{eff} \simeq 10 - 30$ [74], thus providing a self-consistent first principle anomalous transport without any additional closures. Fluid turbulence of partially magnetized plasmas (at the scales larger than $\rho_e$) demonstrates another important property: the inverse energy cascade toward longer wavelengths when the fastest instabilities occur on small scales which subsequently merge into large scale nonlinear structures such as shear flows, vortices, and streamers. Fig. 17 shows one typical example of plasma turbulence at an intermediate state in which the most unstable modes at the length scale of the order of $\lambda \leq 10\rho_e$ gradually develop into the large-scale structures (eventually saturating at the box length scales). The large-scale structures typically provide dominant contributions to the anomalous transport and may lead to the intermittent (bursting) avalanche-like transport events. These simulations suggest that large anomalous transport is possible, yielding an effective (time-averaged) $\Omega_e \lesssim 15$. However, because they have been performed in simplified 2-D geometry and have neglected many important factors such as 3-D effects, sheath boundaries, temperature and neutral component evolution, they cannot be used yet to model real devices at the full scale.

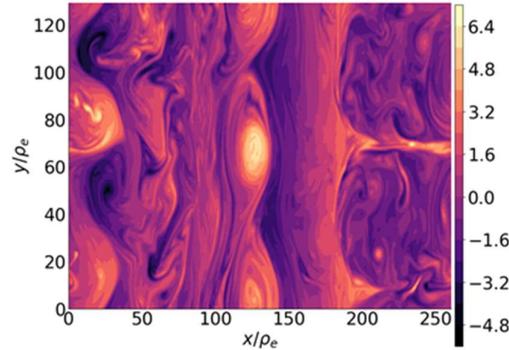

Fig. 17. Vorticity structures in axial-azimuthal (z-θ) fluid simulations [74]. Reproduced with permission from Plasma Phys. Control. Fusion 59, 014041 (2017). Copyright 2017 IOP Publishing.

In some E×B discharges the disparate length and time scales between the plasma species allow for modeling methodologies that circumvent some of the aforementioned limitations of fully kinetic and fully fluid approaches. In HETs for example, the Knudsen number for the heavy species can, in many cases, exceed unity and the ions are not magnetized. Thus, the ion and neutral dynamics may be tracked easily and relatively inexpensively with particle-based methods, within a computational framework that still treats electrons as a distinct fluid. These so-called "hybrid" (fluid-particle) methods gained substantial popularity, especially during the early attempts to model the HET discharge over three decades ago [215][216][217][218]. Typically, the electron fluid equations employ formulations that include an anomalous contribution to account for the effects of turbulence on the transport of mass, momentum and heat. One of the first hybrid codes was developed by Fife in the late 90s. Dubbed HPHall [215], the code assumed Bohm's 1/B scaling for the cross-field mobility [219]. As more plasma measurements became available however, it became clear that this scaling could not explain the behavior of the plasma in most regions' interior to the thruster channel. The argument against Bohm-driven transport was later strengthened by Mikellides [220] and by Lopez Ortega [221]. They employed a multi-variable spatial model of the anomalous collision frequency in a 2-D axisymmetric (r-z) code called Hall2De [222] and combined it with detailed plasma measurements in a HET to obtain a near-continuous, empirically-derived, piecewise spatial variation of the frequency everywhere in that thruster. The latest mathematical formulation of the model is described in detail in [229]. Typical Hall2De solutions for the electron collision frequency along the channel centerline of a magnetically shielded Hall thruster and comparisons with ion velocity measurements for two different magnetic field strengths are illustrated in Fig. 18 [257]. The comparisons underscore the dominance of the anomalous contribution to the total collision frequency over that from (classical) electron-ion (e-i) and electron-neutral (e-n) collisions. They also demonstrate the importance of the transport model in capturing spatial shifts of the discharge that have been well observed in the laboratory as operating conditions and/or the background in ground test facilities are varied (e.g. see [258] and references therein). Similar attempts with empirically-derived anomalous resistivity models were made in other codes in the past as reported for example by Hagelaar [223] and by Scharfe [224]. In the Hall2De work however the combination of simulations and extensive measurements yielded some of the closest agreement we have achieved with time-averaged plasma measurements so far, at a relatively low computational cost and without statistical noise for neither ions nor neutrals. That is because in this code ions are treated using multi-fluid conservation laws and neutrals are tracked using line-of-sight formulations, not particle methods. The concern with statistical noise has been that it may interfere with the true plasma dynamics in these devices, which, as we alluded to earlier, span a wide range of spatiotemporal scale lengths. On the other hand, hydrodynamic approaches are unable to capture the details of the ion velocity distribution function (IVDF). The multi-fluid approach in Hall2De was an attempt to provide a marginally better approximation of the effects of the IVDF. More recently, Jorns [225] increased the efficiency and speed of this approach by employing machine-learning methods to obtain a solution for the anomalous resistivity. The approach takes advantage of the many advances made recently in machine-learning algorithms to replace with computer iteration the human iteration required in the determination of the anomalous collision frequency from plasma measurements. At this time, such 2-D r-z (axisymmetric) solvers with empirical or machine-learning models for the electron transport offer the only feasible approach in providing practical support to the design and development of HETs.

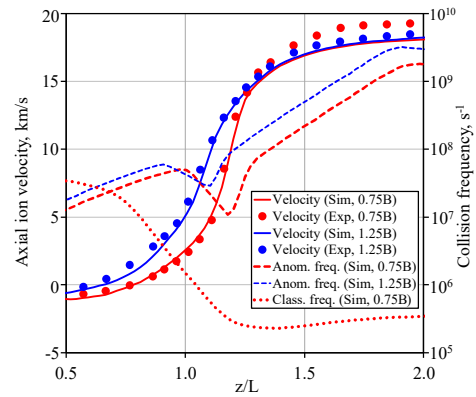

Fig. 18. Comparisons between 2-D (r-z) simulations (Sim) and measurements (Exp) of the axial ion velocity along the channel centreline of a magnetically shielded Hall thruster, for two strengths of the nominal magnetic field strength (B), ±25%[257]. Also plotted are the corresponding anomalous collision frequencies in the simulations, and the classical collision frequency (from e-i and e-n collisions) for one of the two cases. The measurements were obtained using Laser-Induced Fluorescence diagnostics [258].

Limited description of the kinetic phenomena is still possible within some fluid models by adding linear kinetic closures for the higher-order terms in moment equations. For example, it has been argued that the effects of linear Landau damping can be included in the fluid equations via the Hammett-Perking type closures for the viscosity terms [239], and effective schemes have been developed to solve them numerically [234]. Such closures were shown to be effective in fluid modeling of turbulence and anomalous transport due to temperature-gradient modes in magnetized plasmas as well as in the treatment of ion-acoustic turbulence (IAT) and transport of heat in laser plasmas [231]. It is worth mentioning that in these closure schemes the turbulence and anomalous transport is modeled self-consistently with modified fluid equations. Other kinetic processes however have proven to be more challenging. In recent years, for example, the electron cyclotron drift instability (ECDI) [182],[183],[184],[187], [143] has been proposed by Adam [143] and later Coche [211] as a potential source of fluctuations and anomalous transport in HETs due to the strong E×B drift in these devices, see also Section 4. This is a micro-instability which, in principle, requires a kinetic description though, to the best of our knowledge, no attempts have been made to apply Hammett-Perkins type closures within a fully-fluid framework. An alternative closure approach was attempted recently in 2-D (r-z) simulations by Lopez Ortega [210] who modeled the ECDI in Hall2De [222] with an additional fluid-like equation for evolution of the wave energy by the mean ion drift and a wave production term that was proportional to the growth of ion-acoustic waves and accounted for Landau damping. Closure to the hydrodynamics equations in Hall2De was achieved by relating the wave action to an anomalous collision frequency. The simulations predicted well the location of the acceleration region in unshielded and magnetically shielded versions of a HET. However finer details, such as changes in the plasma potential gradient within the acceleration region were not captured. The work was an extension of previous attempts by Mikellides [227] to incorporate the ECDI in Hall2De simulations based on the hypothesis that the instability excites IAT which, in turn, enhances the effective collision frequency. Other larger scale hybrid approaches are being pursued, such as that by Joncquieres, et al. [228] who are developing a 3-D unstructured massively parallel Particle-in-Cell/fluid solver. In this work, the fluid part for electrons and ions is based on a 10-moment model, while the kinetic simulations are used as a reference solver for the sheath boundary conditions.

*Current and future challenges*

Fully-kinetic simulations in three dimensions probably hold the greatest promise of resolving explicitly the wide-ranging spatial and temporal scales that persist in these devices. The major challenge here is computational resources which continue to limit our ability to perform global multiscale simulations. Fully-fluid models as well as hybrid models require the information and input (possibly from supplementary kinetic models in reduced dimensions) on sheath boundary conditions and related particle and heat fluxes to the walls, as well as possible kinetic distortions of the electron distribution function at high energies. However, the challenge here is how to properly incorporate such kinetic effects within the fluid formulations.

A physical basis for the underlying mechanisms of the electron transport in E×B devices such as the HET has not been established yet. Hybrid simulations that have had the greatest impact on the design of these devices are performed largely using empirically-derived models of the anomalous resistivity. In recent attempts to employ first-principles closure models for the ECDI modes, the influence of the magnetic field on the transport was not fully captured and the effects of linear and nonlinear ion Landau damping remained unclear. Moreover, the entire concept of anomalous transport as a diffusive process that can be characterized by an effective collision frequency may be challenged by some experimental and computational data that indicate anomalous transport is strongly intermittent and avalanche-like. The characterization of non-diffusive but "blobby" transport events would then require different approaches such as self-organized-criticality [256]. It is also worth noting that the electron transport in regions of low electric field has been found to remain highly anomalous. This is evident for example in Fig. 18. In such regions instabilities that are excited by a strong E×B drift are not expected and, hence, it has been argued that mechanisms other than the ECDI must be acting [255],[259].

Finally, as is the case with any transport theory, a major challenge is experimental validation. In the HET specifically, it has been shown in 2-D (r-z) simulations that large changes in the cross-field anomalous transport produce only small changes in the plasma in some regions of the acceleration channel and near-plume (Fig. 19). Because such changes are typically too small or impossible to detect by the current state of the art in plasma diagnostics, the verification of a transport model by experiment then becomes, very challenging, at best, ambiguous at worst. This underscores the need for more advanced plasma diagnostics in laboratory investigations of these devices.

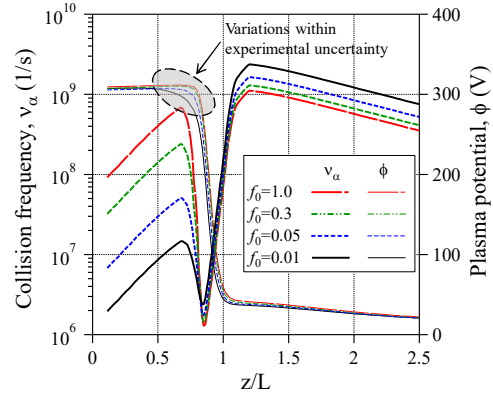

Fig. 19 Sensitivity investigations on the effects of large variations of the anomalous collision frequency in the interior of the acceleration channel in a HET [229]. Reproduced with permission from Plasma Sources Science & Technology **28**, 075001 (2019). Copyright 2019 IOP Publishing. The investigations were performed using 2-D (r-z) multi-fluid simulations with the Hall2De code [222]. The anode is at z/L=0 and the channel exit is at z/L=1. The results (plotted along the channel centreline) underscore the challenges associated with the experimental validation of anomalous transport closure models in fluid-based simulations.

# 7.  Towards full Three-dimensional Modeling of Hall thruster E×B  Discharges

Francesco Taccogna[1], Johan A. Carlsson[2], Andrew Tasman Powis[2], Sedina Tsikata[3], and Konstantin Matyash[4]

[1]CNR-Institute for Plasma Science and Technology, via Amendola 122/D 70126-Bari, Italy
[2] Princeton Plasma Physics Laboratory, Princeton NJ 08543 USA
[3]ICARE, CNRS, 1C avenue de la Recherche Scientifique, 45071 Orléans, France
[4]University of Greifswald, Greifswald, D-17487, Germany

*State of the art and recent progress*

A predictive model of E×B  discharges  requires a kinetic three-dimensional representation, see Sect.4 and Ref. [260]. The electron transport across the magnetic field lines involves all three coordinates in a complex way. In particular, Hall thrusters (HTs) [4],[142],[261],[44], characterized by a large electron current flowing along the azimuthal direction, are subject to the electron drift instability (EDI), see Sect.4 and Refs.[180][206]. Recent work has shown how this azimuthal instability is influenced by the non-Maxwellian character of the electrons [149] and connected to the behavior of the plasma along the radial (parallel to the magnetic field line) and axial (accelerating field) directions. A collective Thomson scattering experiment [149] has demonstrated that the azimuthal mode has a non-negligible wave vector component along the magnetic field, while theory predicts that the presence of non-zero components along the magnetic field direction will modify the resonant comb-like nature of the dispersion relation towards the ion-acoustic type see Sect.4 and [180],[206]. In addition to the broadening of cyclotron resonances in the mode dispersion relation, the inherently 3D nature of the instability is associated with lower growth rates in comparison to the 1D and 2D cases in linear kinetic theory. This has profound consequences: in order to correctly determine the contribution of this particular instability to electron transport (particularly in comparison to other instabilities) a 3D treatment of the instability, both in theory and in simulations, appears to be critical. Although it would seem that the Near-Wall Conductivity (NWC) results in a minor contribution to the anomalous electron transport across the magnetic field lines, secondary electron emission, when modeled in Particle-in-Cell (PIC) simulations [24], [148], [174],[262], may represent a saturation mechanism for EDI. Resolution of the axial direction represents a more natural way for the instability to be convected into the near plume region [143],[181],[192]. Finally, a fully kinetic representation is required due to the fact that ion kinetics contribute to the saturation of the wave by ion-wave trapping and ion axial acceleration. The situation is further complicated by the emergence of large scale low frequency azimuthal spoke modes [73],[75],[76],[78],[93],[108],[173] (see Fig. 20). Due to their highly non-linear, turbulent, and global nature these structures have continued to evade a clear theoretical understanding. However, their observed correlation with enhanced electron transport makes them an important feature that must be captured by any numerical model.

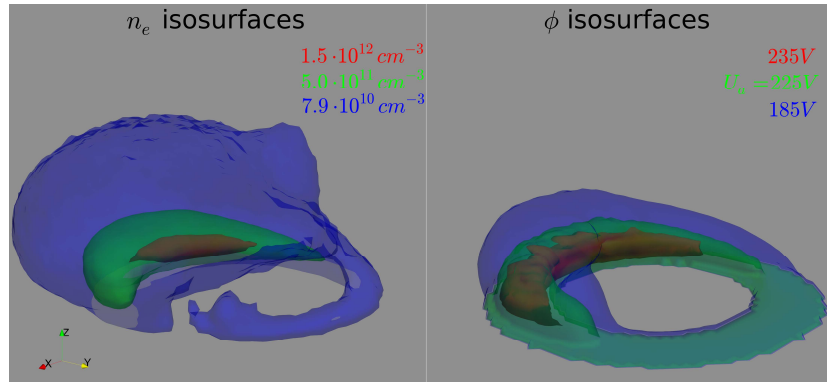

Fig. 20 Isosurfaces of the electron density $n_e$(m-3) and the plasma potential $\phi$(V) at the spoke position in the ISCT200-WL thruster [173]. Reproduced with permission from Plasma Sources Sci. Technol. **28**, 044002 (2019). Copyright 2019 IOP Publishing.

Along with the need to capture the physics in three dimensions, there is also evidence that low-dimensional (1D and 2D) models often show their limitations by leading to inconsistent and controversial results. 1D azimuthal [167], [106], [196] and 2D azimuthal-radial [24], [148],[195],[262] models must mimic the particle transport along the axial coordinate: to avoid unphysical energy growth, particles are re-injected as new colder particles when they have travelled a prescribed axial distance. This numerical artifact strongly distorts the final results and is a possible candidate to induce discordant results about the transition from electron cyclotron resonances towards the ion-acoustic type. Moreover, the accelerating axial electric field $E_z$ is externally imposed and fixed, while one expects that an increase of the local conductivity due to the anomalous transport would lead to a decrease of the local axial electric field, which in turn would reduce the local electron azimuthal drift velocity and thus significantly change the instability behavior. 2D (axial-azimuthal) models [143],[181],[192] are also subject to important shortcomings, such as neglecting components of the wave parallel to the magnetic field and effects of the non-linear coupling of the EDI with secondary electron emission from the walls. Generally, reduced dimensional models always show stronger instability characterized by large amplitude oscillations, as shown in Fig. 21 comparing the electron and ion densities from 3D and 2D PIC models. This finally leads to an artificially larger cross-field mobility (more than five times) with respect to 3D model estimations and experimental measurements. Another observation of note is that the single dominant length scale for the instability, established in Fig. 21 for the 2D case, is no longer as sharply-defined in the 3D case. This observation may be compatible with both experimental and 3D linear kinetic theory analyses showing the instability to be excited non-discretely, across a range of wavenumbers.

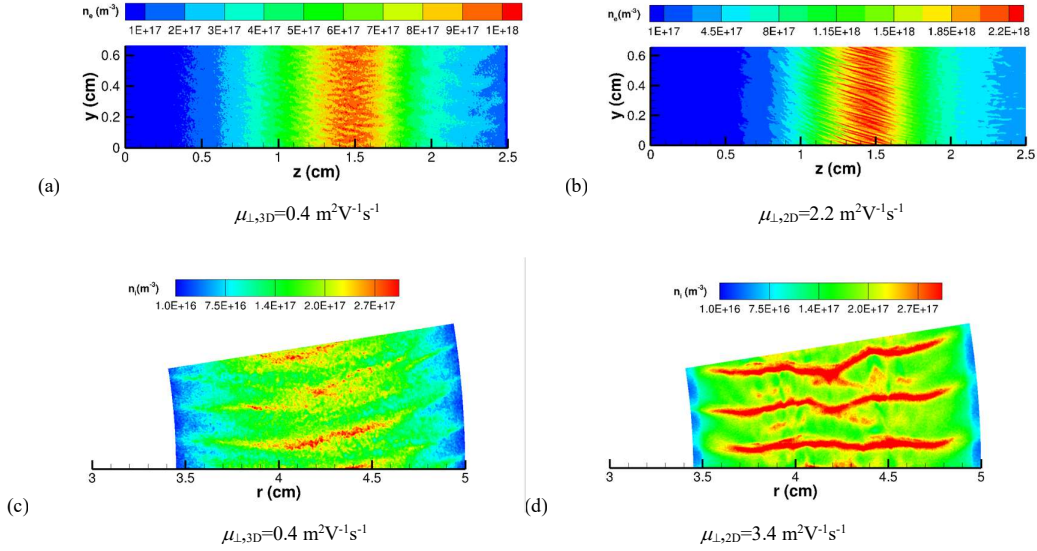

(a) $\mu_{\perp,3D}=0.4$ m$^2$V$^{-1}$s$^{-1}$

(b) $\mu_{\perp,2D}=2.2$ m$^2$V$^{-1}$s$^{-1}$

(c) $\mu_{\perp,3D}=0.4$ m$^2$V$^{-1}$s$^{-1}$

(d) $\mu_{\perp,2D}=3.4$ m$^2$V$^{-1}$s$^{-1}$

Fig. 21  Comparison of 2D and 3D simulations of plasma properties for the parameters of the SPT 100M of Fakel [142] discharge channel. A sector of 1/40 of the entire azimuthal domain has been used. Top: Electron density $n_e$(m$^{-3}$) (a) as a result of (a) 3D (in the middle of the radial plane) and (b) 2D azimuthal-axial PIC models. The cross-field electron mobility $\mu_\perp$ computed at $z$=2 cm and averaged over the azimuthal direction is also reported for the two models. Bottom:  Xenon ion density $n_i$(m$^{-3}$) as a result of (c) 3D (at an axial location $z$=2.4 cm, where the magnetic field and neutral density values correspond to that used as input parameters for 2D model)) and (d) 2D radial-azimuthal PIC models. The cross-field electron mobility $\mu_\perp$ computed as averaged value over the entire domain is also reported for the two models.

### Current and future challenges

The quantitative description of the intrinsically non-equilibrium, non-local nature of anomalous transport in HTs requires the development of an efficient kinetic numerical tool capable of capturing both electron and ion time scales under the electrostatic (ES) approximation in a three-dimensional spatial domain which includes both the discharge channel and the near-field plume regions.

In order to resolve the Debye length under typical HT physical parameters ($\lambda_D \approx 50$ μm), a three-dimensional PIC model requires ~$10^3$ mesh nodes per coordinate and time steps on the order of picoseconds to satisfy the particle-cell-transit CFL condition. The total simulation time necessary to capture the non-linear and saturation phases of the E×B  EDI is about 50 μs, even longer simulation may be required to resolve the temporal dynamics of large scale azimuthally propagating structures [73],[75],[76],[78],[93],[108],[173] (up to 100's of μs: $10^8$ time steps). In addition, an acceptable statistical noise level corresponds to a number of particles per cell $N_{ppc}$>$10^3$ [167],[192],[196]; numerical heating [263],[264],[265] is often responsible for the destruction of electron cyclotron resonances and facilitates an artificial transition to the ion acoustic-type instability. A 3D ES-PIC algorithm, using linear interpolation, requires ~200 floating-point operations (FLOPs) per particle (Lagrangian phase) and ~40 FLOPs per grid cell (Eulerian phase), which leads to a total number of required calculations on the order of ~$10^{21}$ FLOPs. Finally, simulation of a time-dependent 6-dimensional problem produces a large and complex set of output data whose analysis (spectral), visualization and animation require highly efficient and dedicated tools.

### Advances in science and technology to meet challenges

The natural progression of increasing computational power (exascale, i.e. $10^{18}$ FLOP per second) and improved algorithm efficiency will enable the development of full-resolution 3D ES-PIC models of HT within the coming decade, as is routinely performed within the laser-plasma accelerators PIC modeling community with OSIRIS [266], PICADOR [267] SMILEI [268] and PICSAR [269].

A major hurdle that has been already overcome in the past decade is the vastly improved scalability of Poisson equation solvers and their efficient implementation [270],[271],[272],[273],[274]. However, in PIC simulation, the lagrangian phase (loops over particles, i.e. projector, pushing, interpolator and Monte Carlo collision) is usually much more expensive than the eulerian phase (loops over grid nodes). In particular, the projection operator (deposition of the charge density from the particle position to the mesh nodes) can take up to 60% (depending on the order of the particle interpolation used) of the whole particle advancing time. In order to speed up the lagrangian phase two different approached can be adopted.

The first approach can already be applied to serial versions of PIC codes by implementing an efficient data structure and vectorization ideal for single instruction multiple data (SIMD) registers, in order to optimize memory access. This goal can be achieved by particle tiling and sorting methods [275],[276]. The first consists of the fact that particles are placed in tiles that fit in cache, while the second allows particles belonging to the same cell to be contiguous in memory in order to maximize cache efficiency (avoiding random memory access and facilitating data reuse) and to be easily vectorized.

The second approach is an efficient use of hybrid parallelism: the particle decomposition with Open Multiprocessing (OpenMP) within a node and the domain decomposition with Message Passing Interface (MPI) between the nodes, in order to improve the scalability up to $10^6$ processors. This will be crucial on future architectures that will have less available memory per core. Strictly related to efficient parallelism is the dynamic load balancing technique in order to balance the computational load among the different compute units. As seen above, since most of the computational load is proportional to the number of particles, the effort mainly consists of balancing the number of particles per compute unit.

Generally, most ES-PIC algorithms need to be completely rewritten and data structures need to be completely revolutionized in order to use the advantages of new architectures.

Immediately, two simpler approaches are readily available. One way to reduce the highly demanding computational requirements of a 3D PIC model of HT is to use ion-subcycling routines routines (pushing ions with a larger time step) and to decrease the azimuthal length by imposing a smaller periodicity (Fig. 22 shows results corresponding to an azimuthal sector reduced by 40). However, this can go against the evidence that E×B EDI evolves towards low wavenumbers (e.g. wavelength comparable to the entire azimuthal size) through an inverse cascade process [143],[195]. Another option is to perform simulations on miniaturized HT [277] that also meets the need of the community to scale down HT for CubeSat micro-propulsion [278].

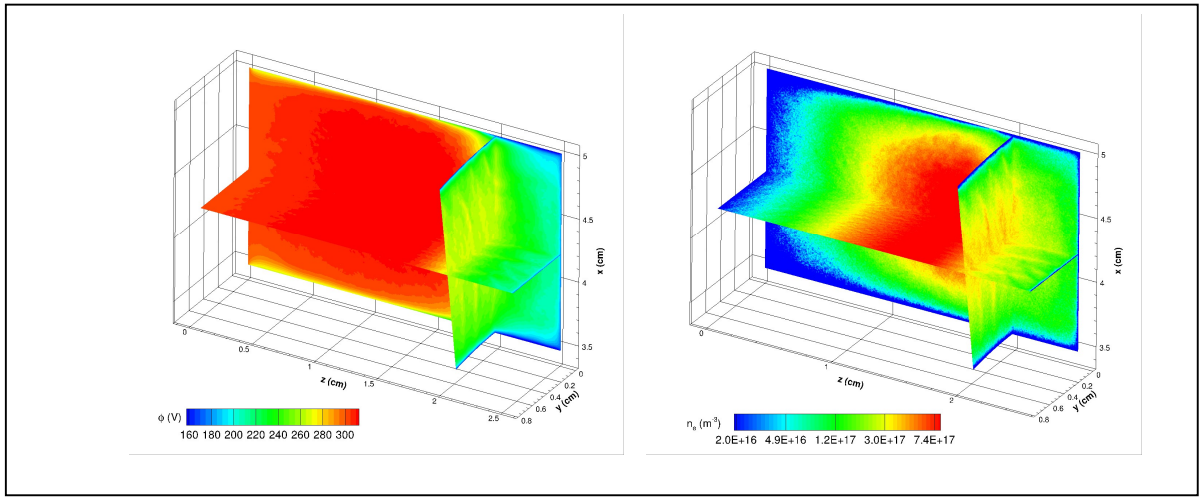

Fig. 22. Three-dimensional map of (a) electric potential $\phi(V)$ and (b) electron density $n_e(m^{-3})$ in the SPT100M of Fakel [6] discharge channel as a result of 3D PIC model. A sector of 1/40 of the entire azimuthal domain has been used.

*Conclusion*

The physics characterizing E×B plasmas, and in particular, Hall thrusters has a non-equilibrium, non-local character and involves self-organized three-dimensional structures. A predictive numerical model has to describe electron and ion time scales over a region that extends from the discharge channel up to the near field plume region.

One- and two-dimensional models are affected by their limitations and often show contradictory results. Progress in high-performance computing (HPC) and the availability of more powerful supercomputers and algorithms will enable the development of 3D fully kinetic PIC models, representing real numerical experiments devoid of any artificial hypothesis. The complete understanding of electron transport will lead to a new era in the technological development of E×B plasma devices, and designs based on an empirical approach will give way to code-based refined optimization. As it has been done in many other engineering disciplines, predictive design and optimization via computer-based techniques will assist and eventually replace empirical methods.

## 8. E×B Configurations for Plasma Mass Separation Applications

Renaud Gueroult[1] and Nathaniel J. Fisch[2]

[1] LAPLACE, Université de Toulouse, CNRS, INPT, UPS, Toulouse, France
[2] Princeton Plasma Physics Laboratory, Princeton University, Princeton, NJ 08543, USA

*State of the art and recent progress*

There is a long history of crossed-field (or E×B) configurations to separate charged particles based on mass. Although crossed-field mass spectrometers [279] first demonstrated these capabilities, it was quickly realized that the throughput achievable with non-neutralized ion beams (such as those employed in the calutron [280]) was severely limited by space charge effects and instabilities. The smaller throughput necessary for isotope separation, however, led to the development of plasma isotope separators [281], where crossed-field configurations again played a role. Specifically, crossed-fields and their associated drift were used to produce plasma rotation [282] in plasma centrifuges [283],[284], as illustrated in Fig. 23 (a).

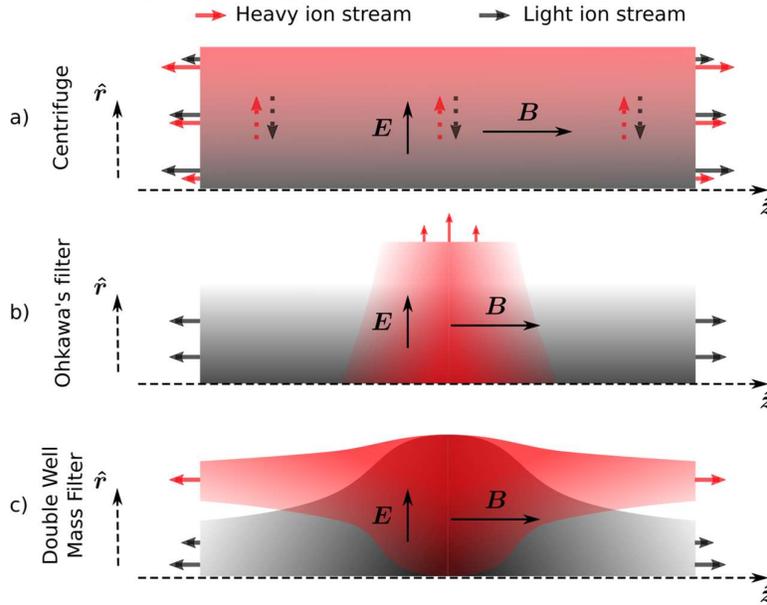

Fig. 23 Illustration of how E x B rotation can lead to mass separation in (a) a plasma centrifuge [283], (b) Ohkawa's filter [294], and c) the Double Well Mass Filter[296]. Although all three concepts utilize the same generic crossed-field configuration [E=Er; B=B0z] and operate in magnetized ion regime, the different radial potential profile used in each concept translates into very different separation flows. Thick red and gray arrows represent heavy and light ion flows, respectively. The longer the arrow, the larger the flow. Figure taken from Ref. [293]. Reproduced from R. Gueroult, S. J. Zweben, N. J. Fisch and J.-M. Rax, Phys. Plasmas 26, 043511 (2019) , with the permission of AIP Publishing.

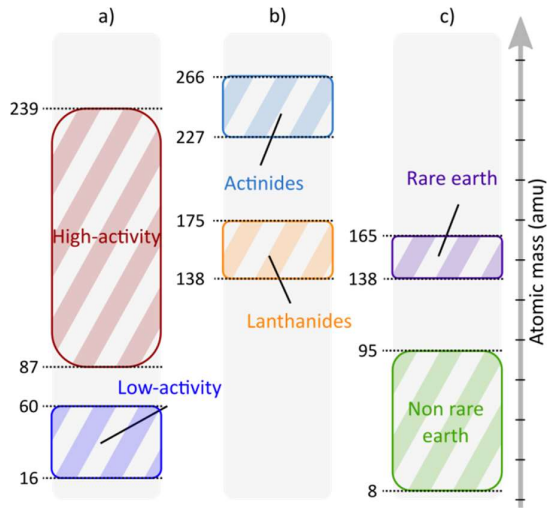

Fig. 24 Composition of the input feed as a function of atomic mass: (a) separation of high-activity waste from low activity waste in nuclear waste cleanup [285]; (b) actinides/lanthanides separation in spent nuclear fuel reprocessing [288]; and (c) rare earth separation in rare earth recycling from NdFeB magnets [291].

More recently there has been a growing recognition that elemental separation based on mass offers highly promising conceptual solutions to several societal high-impact applications, including nuclear waste cleanup [285], spent nuclear fuel (SNF) reprocessing [286]- [290] and rare earth elements (REEs) recycling [291]. For these applications, conventional chemical separation techniques are inefficient since one needs to separate feeds of varying chemical composition (REEs recycling and nuclear waste clean-up), or elements with similar chemical properties (SNF reprocessing). Further to these inefficiencies, chemical separation often leads to the production of secondary, and possibly hazardous, waste. On the other hand, as illustrated in Fig. 24, it happens that the feed conveniently splits into two groups based on atomic mass, supporting the interest of elemental mass separation. Compared to chemical techniques, another anticipated advantage of plasma separation lies in its much smaller environmental footprint. Thanks to these characteristics, plasma separation may be economically competitive for certain applications [285],[291], particularly considering that energy costs from solar may become significantly cheaper. However, the separation needs of these new applications differ significantly from those of isotope separation, thereby offering new challenges [292].

*Current and future challenges*

To be useful for these new applications, plasma separators not only have to separate elements with large mass differences ($\geq$ 10s of amu), but also, and importantly, to operate at high throughput ($\geq 10^4$ kg/yr). It is also advantageous if the separation methods do not perform well on isotopes, to lessen proliferation risks. Since these capabilities are well beyond those accessible to isotope separators, new plasma mass filter concepts are called for.

Here also, the rich physics of crossed-field configurations offers many opportunities, and various conceptual solutions have been suggested in the last decade [286], [294]. In regimes where both ions and electrons are magnetized, crossed-field configurations can be designed to drive rotation. By exploiting rotation in novel ways, conceptual solutions for mass separation can then be found beyond plasma centrifuges [294],[295], [296] as shown in Fig. 23(b) and Fig. 23(c). Alternatively, in regimes where ions are unmagnetized, but electrons are magnetized, crossed-field configurations can in principle be designed to separate the ions of a rotating annular ion beam [297],[298], or to further exploit differences in gyro-radius [299].

In all of these concepts, large mass differences are advantageous, and, in view of proliferation risks, usefully necessary. On the other hand, high-throughput operation has been shown to bring a new set of physics and engineering challenges [300], which have yet to be addressed.

*Advances in science and technology to meet challenges.*

Upstream of the elemental separation stage discussed above, one needs to consider how to produce the high-density plasma required for separation from solids or liquid input streams. Many challenges are found at this stage. First, how to feed into a vacuum chamber, and then ionize, grams of material per second while minimizing energy cost? Laser evaporation of solid targets and injection of micron size powder have been suggested so far, but this question ought to be addressed in detail. Another constraint on the plasma source is the need to maximize charge uniformity. Indeed, since separation often results from differences in the ion charge to mass ratio $Ze/m_i$ and not in the ion mass alone, effective separation requires producing a plasma with $Z$ as close as possible to homogeneous and uniform across ion species. Furthermore, low $Z$ materials are to be preferred to minimize radiation losses, which translates into electron temperature $T_e$ of at most a few eV for most atoms. Finally, related to this challenge is the open question of how molecules will impact separation. While significant molecular ion components are expected due to the low $T_e$, the detailed composition of plasmas formed from complex mixtures is far from understood, and so is the dynamics of these compounds. In addition, cross-section data for molecular processes is likely not readily available for some of the uncommon elements (*e. g.* the sesquioxides) found in these applications.

The separation stage also offers challenges. Assuming a singly ionized atomic plasma could be produced, separation in crossed-field concepts is conditioned upon externally applying a specific potential profile in the direction perpendicular to the magnetic field. Fundamentally, the ability to support this perpendicular potential profile depends on the ratio of perpendicular to parallel conductivity $\sigma_\perp/\sigma_\parallel$. However, many different driving mechanisms (*e. g.* collisions with neutral, instabilities and turbulence [302], magnetic fluctuations [303], ion viscosity [304]) are known to contribute to $\sigma_\perp$ depending on the operating plasma conditions, and new driving mechanisms continue to be uncovered [305],[306]. Demonstrating the practicality of crossed-field mass filter concepts hence hinges on a comprehensive understanding of perpendicular conductivity, which we believe will be best obtained by a combination of modeling and experiments [293],[307]. Another outstanding issue in the presence of neutrals is the possible upper limit set on the rotation speed by the critical ionization velocity phenomenon [308].

*Conclusion*

The challenges briefly discussed above are only a small subset of the many open questions that remain to be addressed to demonstrate high-throughput plasma separation as a practical process (see Refs. [286],[300] for a detailed discussion). Yet, we believe that the promise plasma separation holds for many outstanding societal challenges is a compelling motivation to tackle these questions. In addition, progress towards many of these scientific and engineering goals will benefit applications beyond plasma separation. For instance, the basic question of perpendicular conductivity in a magnetized plasma is also central to recently proposed promising schemes for magnetic confinement fusion [309][310]. Understanding how perpendicular electric fields can be externally applied will more generally benefit the large number of applications making use of cross-field configurations.

# 9.  Validation and Verification procedures for discharge modeling

Anne Bourdon[1], Pascal Chabert[1], Alexander V. Khrabrov[2], Andrew Tasman  Powis[2], Ioannis G. Mikellides[3], Igor D. Kaganovich[2]

[1]*Laboratoire de Physique des Plasmas, CNRS, Ecole Polytechnique, Sorbonne Université, 91120 Palaiseau, France*
[2]*Princeton Plasma Physics Laboratory, Princeton NJ 08543 USA*
[3]*Jet Propulsion Laboratory, California Institute of Technology, Pasadena, CA, 91109, USA*

*State of the art and recent progress*

The ultimate objective for developing computer simulations of complex physical systems is to use these simulations as a predictive tool for science and engineering design. This would reduce the number of costly experiments required for the development of applications involving low-temperature plasmas. For example, currently, the design and development of electric thrusters are semi empirical with long (on the order of 10 000 hours) and expensive lifetime tests. Applications that employ low-temperature plasmas generally involve complex multiphysics and multiscale processes. Developing accurate and predictive simulation tools remains an active area of research. Critical to these goals is the need to verify and validate codes used for the predictive modeling of low-temperature plasmas.

In other fields, rigorous verification and validation (V&V) procedures have been implemented for decades. Computational fluid dynamics (CFD) codes that are used for the design of commercial airplanes, cooling systems and to simulate wind loads around buildings and bridges apply tests of code reliability by regulation and to adhere to professional engineering standards, [311],[312]. Prominent engineering journals require V&V testing as part of their editorial policy before acceptance of papers [313].

In order to develop predictive capabilities for low-temperature plasmas, several necessary steps have to be performed:

* Develop minimum complexity self-consistent mathematical model that describes the effects to be predicted;
* Develop numerical codes and check these codes for bugs by benchmarking with other codes, and with analytical or manufacturing solutions, and verify that convergence is achieved;
* Assemble a reliable set of atomistic physics and plasma-surface interaction data required for modeling;
* Perform validation tests of the assumptions and simplifications of the model by comparing simulation results with available or dedicated experiments capable of exploring a wide range of parameters. The comparisons must be sufficiently comprehensive [315],[316] in order to avoid the bias which may occur when "measurements data and modeling results agree if they are performed in the same Lab," and because a narrow set of measurements data can be always matched by variation of adjustable parameters available in the models.

Code verification [317] is used to assess the numerical accuracy of a mathematical model. The first test should include convergence tests with number of grid points, values of time steps for fluid, hybrid and particle-in-cell codes, and number of particles for particle-in-cell and hybrid codes. Note that some of the studies, for example the number of particles in particle-in-cell (PIC) codes, are challenging due to slow convergence, see e.g. Ref. [314]. Proving the convergence of a solver can be accomplished through comparison with simple limiting test-cases or analytical solutions. Another way to verify the code is to use manufacturing solutions, see e.g. Ref. [314].

Code validation proves the predictive capabilities of the developed model by comparing simulation results with experimental data [317]. This is usually more challenging and demanding because it requires a well-organized systematic campaign of combined experimental and numerical studies in order to convincingly demonstrate validation [315],[316].

An early example of such an effort for low-temperature plasmas is the work of Surendra [318] for a capacitively-coupled radiofrequency discharge, in which results from twelve different codes (including four particle-in-cell codes) were compared with each other (benchmarking) and with experimental data (validation).  At the end of the 90s, a standardized RF discharge experiment - so-called, GEC Reference Cell [320] was also specifically designed for validation and is still used as shown in the recent study performed in Ref. [321]. Recently, there has been different works on the benchmarking of codes. In particular, Turner *et al.* [319] performed a parametric study of several 1D particle-in-cell codes for verification of the Poisson solver, the particle pusher, the wall boundary conditions and the ionization and a Monte Carlo collision modules. The comparison of the codes was performed until an error of less than 1% was obtained. Therefore, the results of this benchmark can be considered as a reference solution for 1D PIC codes. An international effort is underway to preform 2D benchmarks for (non-magnetized) low-pressure low-temperature plasmas [322]. Recently, there has also been a renewed interest to carry out studies on plasma sheaths in low-pressure low-temperature plasmas and to use PIC simulations to guide the development of theoretical models [323].

The benchmarking of fluid codes could be more involved than that of PIC codes. Indeed, a large variety of fluid models are used, ranging from drift-diffusion models to those which maintain a larger number of moments for fluid closures (13 and up). Furthermore, there is a large variety of numerical schemes used to solve the set of fluid equations. For example, six fluid simulation codes were recently benchmarked for simulation of positive streamers at atmospheric pressure [324]. Three test-cases of increasing complexity have been studied. A reasonable agreement between the results of the different codes has been obtained, also showing the difficulty to obtain a reference solution under given conditions.

For low-pressure low-temperature magnetized plasmas, of particular interest for electric propulsion or magnetrons, a significant international effort is currently underway to benchmark PIC and fluid codes on test-cases [325] corresponding to typical conditions of Hall thrusters. The problem has, however, been sufficiently simplified for ease of benchmarking. The first results obtained by different international groups on the three test-cases have been presented at the E×B workshop organized at Princeton in November 2018 [326]. In particular, the most significant effort has been placed on the 2D test-case dedicated to the simulation of micro-instabilities induced by the large E×B electron drift in the acceleration zone of a Hall thruster. The test-case is two-dimensional in the axial-azimuthal direction, close to that proposed in Ref. [327]. Seven international groups have compared their independently developed PIC code results for this 2D test-case and obtained good agreement within 5% difference between all the codes; the results of this study are published recently in Ref. [209]. Therefore, these results can be considered as a reference solution for other PIC codes.

The next step is the validation of codes by comparison with experiments, which is usually an interesting but long iterative process between experiments and simulations. Two examples are given here.

The first one is related to the work of Carlsson *et al.* [328] who also benchmarked two different PIC codes and validated them against a well-diagnosed glow discharge experiment [329]. The DC glow discharge represents a challenging test for simulations, even in one dimension. The model must correctly handle the electrons within a wide range of energies, namely from several hundred eV in the cathode fall to a fraction of an eV for those trapped in the negative glow. For the former group, one has to correctly treat anisotropic scattering in electron-neutral collisions (in a way still suitable for the MC approach) [330], and the Coulomb collisions between the two groups need to be properly reproduced in order to maintain the energy balance in the negative glow. An accurate model for charge exchange cross sections is also needed [331],[332], to obtain the correct ion density in the cathode fall as well as ambipolar diffusion in the plasma. Furthermore, the Debye scale for cold electrons must be resolved, to simulate the plasma - cathode fall boundary and the anode sheath. The experiment by Den Hartog *et al.* [329] provides spatially resolved measurements of the electric field in the cathode fall, with diagnostics of plasma density and electron temperature in the negative glow, for several values of discharge current and corresponding potential. Carlsson *et al.* simulated two of those cases, reporting good agreement in terms of the spatial structure and physical parameters of the discharge.

The second example is related to the effort to validate fluid-based solvers for low-temperature plasma discharges at the Jet Propulsion Laboratory, an effort that started there almost two decades ago. The longstanding code development and validation effort began with OrCa2D, a 2-D axisymmetric fluid solver of the partially ionized gas in hollow cathode discharges [335]-[337], and benefited immensely from a very close collaboration between experimentalists and theorists. After several years of code validation with plasma measurements, OrCa2D then served as the framework for the development of a new code for Hall thruster simulations. Dubbed Hall2De, the code was the first to solve the fluid conservation laws for ions and electrons in these devices on a magnetic-field-aligned-mesh, in a 2-D axisymmetric (r-z) plane [338]. To reduce errors associated with the large collisional relaxation times of ions in the near-plume region, the code also employed a multi-fluid algorithm [339],[340] that allows for different ion populations in up to three distinct ion-energy bins. Also, up to triply charged ions can be included for each of the ion fluids yielding a total of nine ion momentum and nine ion continuity equations. Though much progress has been made in the last decade on the anomalous electron transport known to exist in these devices, a physical basis for the underlying mechanisms that drive it has not been established yet. Hence, in the absence of a validated fluid closure model, generalized Ohm's law in Hall2De employs a multi-variable model of the anomalous collision frequency that is combined with detailed plasma measurements to obtain a near-continuous, empirically-derived, piecewise spatial variation of the frequency everywhere in the device [229]. The combination of simulations and extensive measurements, in a variety of different thrusters and operating conditions, has yielded some of the closest agreements we have achieved with plasma and wear measurements so far. A few representative examples are provided in Fig. 25. Persistent efforts to validate codes like Hall2De has allowed some of them to support directly thruster development and life qualification programs for space flight, and identify new needs for laboratory diagnostics while providing critical insight into the physics of these devices (Fig. 25).

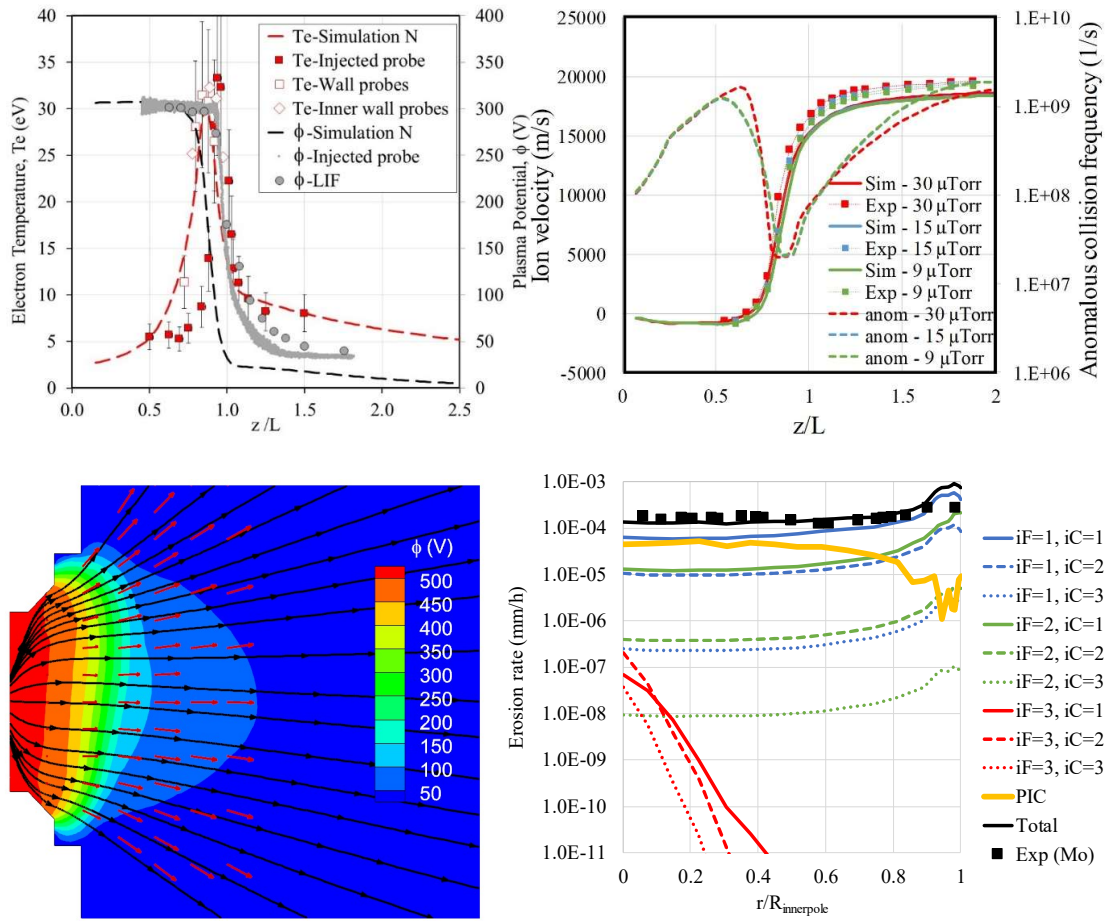

Fig. 25. Top left: Comparison of plasma measurements and simulations along the channel centreline (CL) of an unshielded 6-kW laboratory Hall thruster [229], with z denoting distance from the anode and L is the length of the acceleration channel. The differences between simulations and measurements revealed that plasma-perturbing diagnostics (like injected probes) must be avoided in the validation of Hall thruster codes [229], [341]. Reproduced with permission from Plasma Sources Science & Technology **28**, 075001 (2019). Copyright 2019 IOP Publishing.

Top right: Comparison of plasma measurements and simulations along the channel CL of a 4.5-kW SPT-140. These thrusters will enable NASA's Psyche mission the first to use this technology beyond lunar orbit. The code validation effort reduced margins for the mission and shed new insight into the effects of facility backpressure on Hall thrusters operating with external cathodes [342],[343]. Reproduced with permission from Plasma Sources Science & Technology **29**, 035011 (2020). Copyright 2020 IOP Publishing.

Bottom left: Comparison of the ion velocity fields from simulations (black traces) and Laser-Induced Fluorescence (LIF) measurements (red traces) in the Magnetically Shielded Miniature Hall Thruster (MaSMi) at 1 kW [344]. Reproduced with permission from Proceedings of International Conference on Electric Propulsion, Vienna, Austria 2019, Electric Rocket Propulsion Society, IEPC 2019-281. Copyright International Conference on Electric Propulsion 2019.

Bottom right: Comparison of computed and measured erosion rates along a molybdenum (Mo) cover of the inner magnetic pole in a magnetically shielded 6-kW laboratory Hall thruster [345]. The validation effort explained and quantified the source of the pole cover erosion observed during wear tests [346] of one of the first magnetically shielded Hall thrusters. The simulations used three ion-fluid (iF) populations and three ion charge (iC) states; r is the radial distance from the thruster CL and $R_{innerpole}$ is the radius of the pole cover. Reproduced from Journal of Applied Physics **125**, 033302 (2019), with the permission of AIP Publishing.

### *Current and future challenges*

Recently, there has been a renewed interest in the verification of codes through benchmarking. For low-temperature partially magnetized plasmas, most significant results have been obtained on 2D PIC codes but on a limited number of test-cases. Currently, there is a need to work on the verification of fluid codes for low-temperature magnetized plasmas. It is important to emphasize that the definition of the test-cases is a significant part of the problem. Indeed, it is important to develop a comprehensive set of test cases to provide benchmarking of codes for the regimes of interest that sufficiently characterize the relevant physics that are important for the studied plasmas. For low-temperature magnetized plasmas these are: anomalous transport, low-frequency oscillations such as breathing modes and plasma spokes, see sections 2-6. A detailed description of validation and benchmark test-cases will be the subject of future dedicated E×B workshops and the "Frontiers in Low-temperature Plasma Simulations" workshop. These workshops will facilitate discussions and idea exchanges between international groups to better define, discuss and compare results.

For validation of models and codes a comprehensive set of measurements is needed. This is challenging for a compact and energetic Hall thrusters where probes can strongly perturb plasma and it is difficult to measure plasma parameters in the ionization and acceleration zones. A possible approach is to validate codes on specially designed plasma systems that allow for easier access for diagnostics, similar to the GEC cell for RF discharge [320]. An example of such a study for E×B discharges is the Penning discharge [92]. For Hall thrusters, possibly a scaled-up Hall thruster with improved access is the desired for accurate validation of simulation results. Another approach is to use a wall-less Hall thruster where the acceleration zone is outside the thruster channel [166]. Comprehensive measurements by several diagnostics probes, LIF and CTS, see section 3 is also needed for complete characterization of anomalous transport and turbulent oscillation spectra.

*Advances in science and technology to meet challenges*

Currently most simulations are carried out in 2D and the number of 3D simulations remains limited. The development of computing resources and high-performance computing already enables us to carry out parametric studies in 2D which were out of reach 10 years ago. In the coming years, with the ongoing development of computing resources and the interest in small-size thrusters, the number of 3D simulations of thrusters without scaling of geometry to speedup simulations will increase, which will also require the need to carefully define 3D benchmarks. It will also be a very interesting opportunity to carry out parametric studies that are currently still out of reach in particular to study the effects of plasma-wall interaction and instabilities. New means of communication, systems of code version control, means of data exchanges and ease the sharing of results and discussions will greatly enable V&V activities of the future.

The development of more refined time and space-resolved measurement techniques is also expected. In particular as described in section 3, there are still too few measurements of high-frequency oscillations in low-temperature partially magnetized plasmas.

*Conclusions*

Currently, the design and development of electric thrusters remains semi-empirical with long and expensive lifetime tests. In the future, with the use of electric propulsion for constellations of small satellites in low-earth orbit or high-power electric propulsion systems for full orbit raising and orbit transfer, there is a need to reduce the time to design and to develop new generations of electric thrusters. This will be achieved only by combining more closely engineering experiments, measurement campaigns in academic experiments, and modeling and simulations. The recent developments of computing resources and high-performance computing already enables us to carry out parametric studies that were out of reach only a few years ago. The most significant difficulty is probably to develop a 3D hybrid and full PIC codes that can be used for engineering design, with a reasonable computing time and which would be validated on a comprehensive set of test cases for the regimes of interest for electric propulsion applications: anomalous transport, breathing oscillations and spokes. Recent efforts in the community on the definition of test-cases and benchmarking and validation of codes should be pursued.

## 10. Magnetic Nozzles for Electric Propulsion

Eduardo Ahedo[1], Mario Merino[1], Rod W. Boswell[2], Amnon Fruchtman[3], and Igor D. Kaganovich[4]

1 Equipo de Propulsión Espacial y Plasmas, Universidad Carlos III de Madrid, Leganés 28911, Spain
2 Space Plasma, Power and Propulsion Laboratory, Research School of Physics and Engineering, The Australian National University, Canberra, ACT, 2601, Australia.
3 Holon Institute of Technology, Holon 58102, Israel
4 Princeton Plasma Physics Laboratory, Princeton NJ 08543 USA

*State of the art and recent progress*

An axisymmetric magnetic nozzle (MN) is constituted by the magnetic streamtubes in form of a nozzle created by a set of coils or permanent magnets around a plasma source. Magnetic nozzles are being proposed as the accelerator stage of a family of innovative plasma thrusters [347]. The underlying principle is that the MN channels magnetically a hot plasma jet and accelerates it supersonically while controlling its radial divergence, in a similar way as a deLaval nozzle operates on a hot ideal gas [348]. Beyond that basic similarity, plasma physics in a MN are much more complex and varied than gas physics in a solid nozzle.

The main devices relying on plasma acceleration by MNs, are the Helicon Plasma Thruster (HPT) [349],[350], the Electron Cyclotron Resonance Thruster (ECRT) [351], the VAriable Specific Impulse Magnetoplasma Rocket (VASIMR) [352] and the Applied Field MagnetoPlasma Dynamic Thruster (AFMPDT) [353]. The first three ones are electrodeless devices, relying on electromagnetic waves for plasma production and heating, while the AFMPDT relies on DC annular electrodes for these functions. The plasma beam injected into the MN is globally current-free in the four devices, so no additional neutralizing device (e.g. a hollow cathode) is needed, thus simplifying the whole thruster system. MN operation is illustrated in Fig. 26.

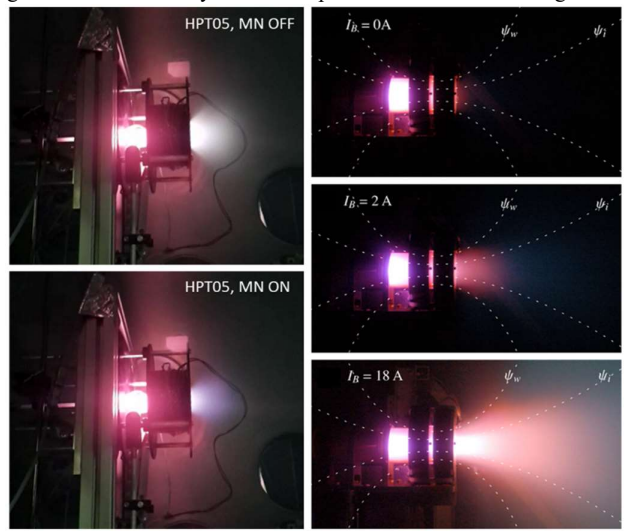

Fig. 26. Helicon plasma thruster HPT-05 operating with MN off (upper left) and on (bottom left), courtesy of SENER-EP2 consortium. In both cases, a base magnetic field exists in the thruster. (Right) A helicon plasma thruster operating at three magnetic strengths, given by the coil current (0 A, 2 A, and 18 A), from Ref. [433] Reproduced with permission from IEEE Trans on Plasma Scie. **43**, 277 (2014). Copyright 2014 IEEE. A stronger MN collimates more the plasma jet.

For these new plasma thrusters to be minimally efficient and competitive, the plasma produced by the source must be highly ionized and hot with the temperature of tens of eV. This implies that the expansion in a propulsive MN is expected to be near-collisionless, which is not the case in other applications (such as material processing [355] or plasma wind tunnels [354]). Theoretical efforts to understand these propulsive magnetic nozzles involve both fluid and kinetic models. Some of these models admit semi-analytical solutions, which have been essential to understand the MN physics explained below.

In particular, two-dimensional (2D) axisymmetric fluid models have shown to be a powerful tool to understand the main phenomena in the MN [358]. However, the closure of a fluid model for a collisionless plasma (at the heat flux level, for instance) is an open, elusive problem. One-dimensional (1D), stationary, kinetic models of a paraxial MN based on solving the Vlasov equation [359],[360] complement well fluid models. In particular, kinetic models have allowed analyzing the downstream ion and electron heat fluxes and the response to non-Maxwellian features of the ion and electron velocity distribution functions (VDFs).

From the energy point of view, plasma thrusters relying on a propulsive MN, are electrothermal, the MN converting the plasma internal energy into axially directed (ion) energy. However, these thrusters differ in the way the internal energy is stored in the plasma: as electron thermal energy on the HPT and ECRT, mostly as thermal ion energy on the VASIMR, and as a mix of energy types (electron thermal, ion swirling) on the AFMPDT [361]. Besides, internal energy (i.e. temperature) is anisotropic in the cases of the ECRT and the VASIMR. Hence, the energy conversion mechanism of the MN from internal to axially directed one is device-dependent. When energy must be transferred from electrons to ions, the conversion agent is the ambipolar electric field [362],[363], while the collective mechanism is mainly gas-dynamical when the conversion is from thermal to axially-directed ion energy [352].

Magnetic channeling of the plasma beam is based mainly on electron magnetization, which requires a large Hall parameter (i.e. the ratio between the electron gyrofrequency and the collision frequency) and a small electron gyroradius (compared to the nozzle cross section) [347]. In HPTs, ECRTs, ECRTs, and (low power) AFMPDTs, with magnetic fields in the 0.01-0.1 Tesla range, ions are weakly magnetized and are bound to highly magnetized electrons through the ambipolar electric field. In the VASIMR, with magnetic fields above 1 Tesla and using light propellants, ions are highly magnetized and possess the thermal energy, and some of the MN phenomena described below (focused on hot electrons) can differ somewhat.

Ion and electron dynamics along the divergent MN lines are very different. In the collisionless scenario, both species are coupled through the ambipolar electric field, which develops to keep charge quasineutrality. The resulting electrostatic potential decreases both radially and axially from the MN throat, in order to confine most of the electron population while it accelerates ions freely downstream [358]. The total fall of the electric potential along the MN is self-adjusted to maintain the plasma globally current-free and determines the ultimate velocity of ions [359]. The heavier the propellant is, the higher that total potential fall is, following the same principle as in a Debye sheath next to a non-conducting wall [360].

Ion dynamics are simpler. As it expands downstream of the MN, the ion becomes more energetic and less magnetized, while the ambipolar electric field becomes weaker. Thus, when the local ion gyroradius (based on its macroscopic velocity) is not smaller than the MN cross section radius, the electric field is unable to bind perfectly the ion streamsurfaces to the magnetic and electron ones and the ion streamsurfaces detach *progressively and inwards* from magnetic surfaces, becoming conic eventually [365]. The lower is the angle of this cone the lower is the plasma beam divergence. This MN detachment mechanism, illustrated in Fig. 27 and based on ion demagnetization is supported experimentally [352],[366].

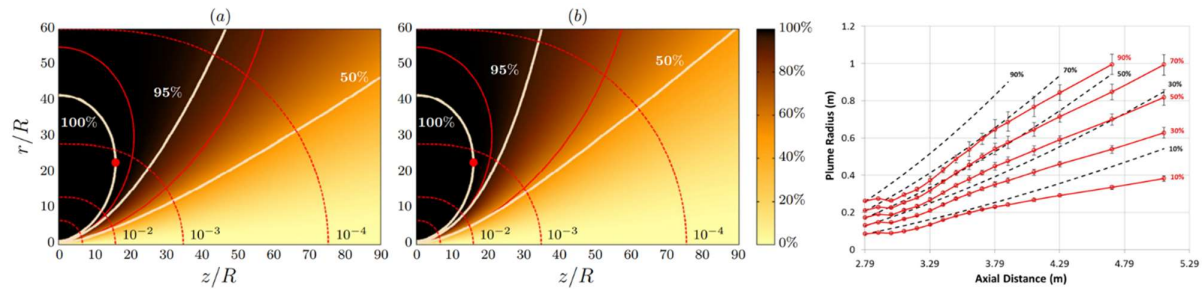

Fig. 27. Detachment of ion streamlines in a MN. The left and center plots show simulation results for moderate and high magnetic field strength in a helicon plasma thruster-like device. A higher magnetic field strength results in ions detaching further downstream and acquiring a larger divergence angle. From Ref. [365]. The figure to the right shows the experimentally-measured ion streamline detachment in a VASIMR prototype. From Ref.[352]. Reproduced with permission from Plasma Sources Sci. Technol. **23**, 032001 (2014). Copyright 2014 IOP Publishing.

Electrons, on the contrary, can remain magnetized for a very long distance. Collisions, electron inertia, and instabilities are diffusion processes favoring outwards electron detachment, which in fact is detrimental for beam collimation[367],[368]. Nonetheless, demagnetization and detachment of the confined cloud of electrons does not seem to affect much the far-downstream properties of the plasma beam.

Magnetic nozzles have no walls, thus avoiding thermal loads and erosion issues. But neither is there the mechanical thrust gain of a solid nozzle. Instead, the divergent MN is an electromagnetic device from the momentum point of view: magnetic thrust is achieved thanks to the diamagnetic drift developing on electrons [367],[369],[370],[371],[372]. These drifts form azimuthal current loops, of diamagnetic character, which induce a magnetic field acting on the thruster coils [373],[429],[430]. This coil/plasma configuration is equivalent to the basic setup of two wire loops with counterstreaming electric currents (of different magnitude), which repel each other, as shown in Fig. 28. In the moderate plasma-beta case (i.e. similar thermal and magnetic pressures), the induced magnetic field modifies the shape of the MN, increasing its divergence [374]. Ion azimuthal currents are created too, but (with possible VASIMR's exception) these are paramagnetic and decelerate the plasma beam [367],[375]. This paramagnetic effect is stronger the more magnetized the ions are and has been observed in propagation of an ion beam along a straight magnetic field [376]. Since high ion magnetization is negative in terms of paramagnetic effects and detachment, there must exist an intermediate magnetic field strength optimizing the propulsive features of a MN.

While a current-free beam assures that $\nabla \cdot \vec{j} = 0$, with $\vec{j}$ the electric current density, it does not assure local current ambipolarity, i.e. $\vec{j} = \vec{0}$ [390]. Indeed, the ion detachment implies that local current ambipolarity is not satisfied and small longitudinal (i.e. in the meridian plane) current loops develop, which nonetheless are several orders of magnitude smaller than the azimuthal electric currents [368].

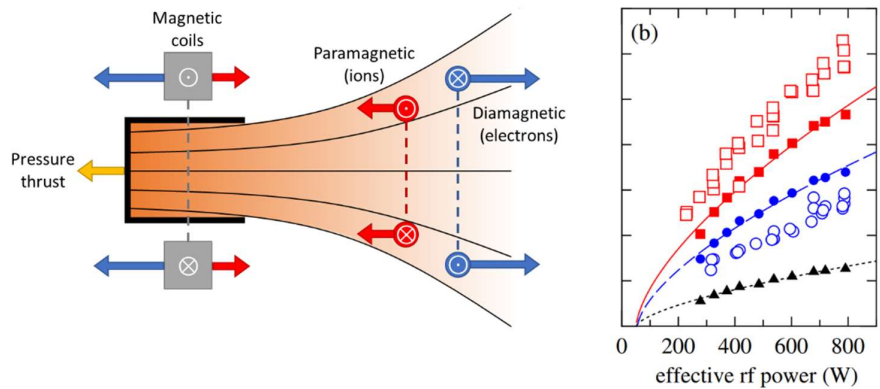

Fig. 28.(Left): Schematic of the azimuthal currents in the magnetic coils and in the plasma of a divergent MN. Paramagnetic azimuthal currents rotate in the same direction as coil currents and attract each other, producing magnetic drag (negative magnetic thrust). Ions develop a small paramagnetic current in e.g. HPTs and ECRTs. Diamagnetic azimuthal currents rotate in the opposite direction as coil currents, and repel each other, producing magnetic thrust. Electron azimuthal current is diamagnetic and dominates in the plasma. Additionally, the plasma creates pressure thrust on the chamber walls. (Right): Experimentally-measured magnetic thrust (open blue circles), and total thrust (open red squares) from [369] in a helicon plasma thruster. Reproduced with permission from Phys. Rev. Lett. **107**, 235001 (2011). Copyright 2011 American Physical Society. The solid blue circles and solid red squares correspond to the expected values from a model. Black triangles represent the modeled pressure thrust.

*Current and future challenges*

While the above MN physics can be considered reasonably established and understood, there are many other aspects requiring further investigation and definitely more experimental validation of MN physics.

The evolution of the electron velocity distribution function (VDF) in the divergent MN is a principal challenge, due to the relevance of the VDF and the thermal electron energy on the beam expansion in the MN. Paraxial kinetic models have identified regions in phase space where doubly trapped electrons, bouncing back and forth along magnetic lines in a limited region isolated from both upstream and downstream sources, can exist [359],[360]. These regions would become populated during the transient period of formation of the MN and by sporadic collisions [391],[392]. If the repopulation is large doubly trapped electrons can be the main electron subpopulation in certain MN regions and shape the plasma expansion there.

Doubly-trapped electrons are confined in a limited region by the combined and opposite effects of the magnetic mirror and the electrostatic barrier [360]. Interestingly, similar doubly-trapped electrons are also found in other collisionless plasmas. A well-known case is the Langmuir probe, where the role of the magnetic moment invariant is substituted by the angular momentum one [393]. A closely related case is a slowly divergent, unmagnetized, plasma plume where the magnetic moment is replaced by the radial-action integral invariant [394].

Temperature anisotropy is a relevant feature in most cases and a real challenge for diagnostics too. The problem has several faces. First, there is the case of the expansion of an isotropic (e.g. Maxwellian) VDF where recent analyses show that the electron population develops only a small temperature anisotropy along the nozzle and, as a consequence the collective magnetic mirror effect is weak on it [360]. On the contrary, ions, if hot and magnetized, develop a large temperature anisotropy. Both non-intuitive conclusions deserve experimental confirmation. Second, there is the case of the expansion of a Maxwellian plasma with a certain amount of suprathermal electrons, a case reported in some HPT experiments and leading to the possible formation of current-free double layers [395],[396],[397]. And third there is the evolution of the anisotropic VDFs of (a) energetic electrons in the ECRT case and (b) energetic ions in the VASIMR case, and both of them have not been well characterized experimentally yet.

The evolution of the electron VDF is closely related to the much-discussed problem of the nature the observed beam cooling along the MN. Experimental data usually assess the electron temperature decrease relative to the density one through an empirical polytropic coefficient, $\gamma$. Values in the range $\gamma$=1.15-1.25 seem to be common [351],[398],[399],[432], but larger $\gamma$, corresponding to adiabatic behaviors have been reported too [400]. Recent experiments capable of characterizing the VDFs of the different electron subpopulations suggest that free electrons (i.e. those flowing downstream with the ions) tend to be adiabatic (i.e. $\gamma$=5/3 in three dimensions), while confined ones (included double-trapped) are isothermal [401],[402]. Further evidences are needed. Interestingly, electron cooling in unmagnetized plumes fits also with low values of $\gamma$, around 1.1-1.3 [403],[404], supporting the postulated similarities with magnetized plumes. It is reminded that in any expanding plume plasma cooling is consubstantial to a finite electric potential fall along the nozzle [364],[405].

The low polytropic index is often associated with large heat conductivity [364], but there is an open debate on the essence of this heat conduction. On the one hand, some authors have proposed a Fourier-Spitzer-Harm law with the heat flux proportional to the electron temperature gradient [398]. This model indicates that the heat keeps being conducted by the electrons to a large distance and is not transferred to the ions. The lack of coupling between ions and electrons resulted in that calculation in a low thrust generation [409]. On the other hand, the nearly collisionless plasma is very far from local thermodynamic equilibrium and the electron response is nonlocal [356],[357],[359],[399],[406]. Furthermore, the collisionless kinetic models find electron energy fluxes can evolve with a Nusselt number lower than one, that is with heat conduction dominating over heat convection [360],[392]. Both polytropic fluid models [364], [405] and kinetic models [360] predict that most of the electron energy does transform into ion kinetic energy, so that all energy carried by the electrons is used for magnetic thrust generation. Thus, these different predictions result from describing different operation regimes. The understanding of which part of the electron energy is transformed into ion kinetic energy will also clarify whether a MN and a solid nozzle of the same shape and plasma/gas upstream pressure have similar efficiencies for propulsion.

In actual plasma thrusters, the plasma beam does not reach the divergent MN fully ionized due to either large plasma recombination at the walls of the source or low electron energy, so elastic and ionization collisions within the MN expansion can be relevant. This phenomenon is more acute inside a vacuum facility due to the additional effect of the background pressure. Other experiments have shown a significant decrease of ion energy (and thus thrust) as the background pressure is increased due to both new born ions and resistive deceleration [410]. The assessment of the effect of ion-neutral collisions in the beam expansion and the thrust would require a dedicated research, since other configurations have shown that for specified potential drop, ion-neutral collisions during the ion acceleration can increase the thrust [411],[412]. However, a high electron collisionality could reduce the potential drop across the magnetic nozzle, as suggested in in [398] and thus it also reduces the thrust.

In vacuum chamber testing, background pressure effects are coupled to the phenomenon of the electrical connection between the plasma beam and the (metallic) chamber walls, which can affect the electric potential profile, the VDF of confined electrons, and the amount of electron cooling. Indeed, kinetic models with semi-infinite, fully magnetized beams yield a much larger cooling than the one measured in (finite-size) vacuum chamber experiments. In summary, the quantitative assessment of facility effects in the MN expansion and thruster performances is very attractive line of research.

Another interesting experiment in a vacuum chamber has found that the radial edge of a MN-guided plasma presents a non-monotonic electrostatic potential profile, reaching subsequently a local minimum and a local maximum [413]. The work emphasizes the reversal in the direction of the electric field perpendicular to the magnetic field and its confining role on the ion beam. Theoretically, this potential profile has been found when ions are hot [364] and to have implications on edge azimuthal electron currents [358]. Nonetheless, more experiments are needed to understand the universality of these potential extrema and the possible influence of facility effects on their development.

Intertwined sets of magnetic coils at different angles can create three-dimensional (3D) MNs with adaptable shapes, capable of steering the plasma beam contactless, and thus have been proposed as a non-mechanical way of thrust vector control [414]. On the theoretical side, plasma beam steering has been proved under full ion and electron magnetization, but the more relevant case of weakly magnetized ions remains unexplored.

As in other magnetized plasmas with different ion and electron currents, a variety of plasma instabilities can develop in a MN expansion, with potentially large impact in electron diffusion, perpendicular transport, detachment, and thrust. This research field has been little explored so far, both theoretically and experimentally. Recently, high frequency oscillations have been observed in the VASIMR plume, which are thought to be related to the lower hybrid drift instability and are believed to expedite plasma detachment [415]. Also, kHz-range oscillations have been observed in the MN of an HPT with double layer, which seem to be driven by perpendicular plasma density gradients [433]. These instabilities vary with applied power, magnetic field, and operating pressure, and their amplitude has been successfully reduced using kHz modulation of the applied power. Finally, anisotropic VDFs are subject to the Weibel instabilities [423],[424],[425], [426].

The physics of MNs shows some similarities with so-called magnetic lenses or Robertson lenses [379],[380], used for quasineutral ion beam focusing and a comparative study could be illuminating on their respective physics. Both devices rely on magnetized electrons to follow the magnetic lines and to create a radial electric field that forces ions to conform with the shape of the resulting electron cloud [377],[378]. Also, both devices are globally current-free but not locally current free. However, while magnetic lenses rely mainly on the convergent part of a solenoidal magnetic field to concentrate the ions, a MN uses mainly the divergent part to expand and accelerate the plasma, creating magnetic thrust. Note that a diamagnetic plasma in a convergent magnetic field does not create magnetic thrust, but magnetic drag. Other types of magnetic lenses, such as Morozov lenses that include multiple electrodes [382], have been successfully used for focusing of ion beams [381], [383], and have been proposed for the narrowing of Hall thruster beams [380]. A comparative study could be illuminating on their respective physics.

The cylindrical Hall thruster (CHT), the Highly Efficient Multistage Plasma Thruster (HEMPT) [386] and the Divergent Cusped Field Hall Thruster (DCHT) [387] are propulsion technologies utilizing a divergent magnetic field in the plasma plume. These devices include a hollow cathode, which is biased relative to anode in the center of the thruster. The voltage drop across a diverging magnetic field is the main element that controls the thruster current, and the plume voltage fall, and consequently the ion acceleration. Plasma dynamics downstream the hollow cathode resemble the MN ones [388], where electrons are magnetized, and ions are not. Analyses of the CHT plume suggest that centrifugal forces on rotating electrons also have a role on narrowing plume divergence [388] [385],[389],[83]. Further study of the CHT far plume from the present MN perspective will surely provide a complementary insight on plasma detachment, electron thermodynamics, and other relevant physics.

*Advances in science and technology to meet challenges*

On the technology side, advances on MNs are linked to the advances needed in the plasma-generating sources. Indeed, the future importance of MNs depends on the HPT and the other MN-based thrusters becoming competitive, which, among other things, requires to improve the efficiency on plasma production, heating and confinement. These processes are still incompletely understood. Fortunately, the larger availability of prototypes means more testing opportunities and more data. In a first stage, this testing must confirm and assess more accurately central MN features such as the amount of magnetic thrust or the beam divergence (as a consequence of beam detachment).

The influence of magnetic coil strength on MN performances must be further assessed, since the prevalent MN theory for hot-electron plasmas indicates that minimum plume divergence (i.e. better detachment in a way) is achieved with intermediate magnetic strengths yielding highly-magnetized electrons and weakly-magnetized ions (i.e. easier to detach) [365], [434]. The actual relevance of electron demagnetization on the downstream plume divergence requires to be confirmed experimentally.

Another trade-off affecting beam divergence is related to the MN divergence rate and weight. A low divergence rate of the magnetic streamsurfaces is obviously beneficial but implies bigger coils and thus significant penalties on size, weight, and electric power. Hence, technological advances on MN designs can be critical for the practical implementation on MN-based thrusters. This includes materials to be used, possible cryogenic cooling, as well as the MN design itself. Light superconductors are already used in the VASIMR, but the needed cryogenic cooling adds more complexity to the thruster system, which can be a serious drawback. Permanent-magnets are being used as an alternative choice for MNs, mainly on low-power devices. Compared to coils, magnets present some differences in the nozzle's magnetic topology that must be further assessed.

The unique properties of the plasma expansion in the VASIMR's MN require to be further analyzed, in order to understand how the ion azimuthal currents, the ambipolar field, the detachment process, and the plasma cooling behave. Similar studies are needed for the AFMPDT, a device generally considered as a subsidiary of high-field MPDT. At powers of the order of a few tens of kilowatts and below, the self-field is negligible, while no enough attention has been given to the characterization of the MN.

Facility effects have been seen to be very important when testing MN expansions. Their quantitative assessment would also allow to infer the behavior of the plasma-MN-thruster system in the high-vacuum, space environment, where collisionality is much lower, cooling presumably larger, and the far plume will not interfere with any wall.

Many of these measurements to be carried out in the laboratory require high-quality, sophisticated, diagnostics equipment, such as accurate thrust balances, magnetic probes, and non-intrusive and intrusive devices able to characterize the non-Maxwellian VDFs of the ions and (mainly) the electrons. For instance, although there is some experimental evidence of the presence of doubly trapped electrons [401],[402], a good characterization of them under different conditions is primordial due to their probable large influence in the plasma expansion profiles. Also, measuring the beam temperature anisotropy has been proven to be very challenging and seems to require designing specific probes.

The previously mentioned analogies between plasma cooling in magnetically channeled and no-channeled plumes can be contrasted in parallel experiments, using HPT/ECRTs on the one side and conventional Hall thrusters of similar specific impulses on the other side. Incidentally, the adjective conventional is important, since some cylindrical Hall thrusters [416] and the cusped-field thruster [417] bear a MN-like magnetic topology at the thruster exit, but this one receives already a highly supersonic plasma beam.

Regarding instabilities in MN-guided plasma, the same techniques and the same theoretical framework used in turbulence on Hall thrusters and other electromagnetic thrusters and discussed in other sections of this paper can be applied. On top of this, a novel and stimulating line of research in RF based thrusters is the possible coupling of instabilities to the harmonics of the RF emission, or the excitation of new modes by the applied high-frequency fields.

On the theoretical side, paraxial MN kinetic models must complete the research on the path to a consistent model of non-local heat conduction, looking for support on the third moment collisionless fluid equation [407], [408], [418]. After that, results need to be extended to 2D kinetic or fluid/kinetic expansions [419], where the plasma azimuthal currents arise and the contribution of downstream electron energy to thrust can be assessed.

Beyond amenable semi-analytical kinetic and fluid models, 2D and 3D heavy-computing simulations are needed if most of the MN physics discussed in this section want to be treated consistently at the same time. Fluid [421], hybrid (kinetic/fluid) or fully kinetic formulations, as the ones discussed in other sections of this paper, are needed tackle with these advanced problems. These simulations operate by necessity a finite domain. Since the response of confined electrons is nonlocal [356],[357], [406], the definition of consistent downstream (and lateral) boundary conditions is a challenging problem, not fully solved and shared again by magnetized and unmagnetized plumes [391],[422].

On the innovation field, first, practical realizations of 3D MNs in order to test and quantify their beam steering capabilities are pending development. Indeed, these are very interesting devices to study magnetic confinement in open plasmas and raise novel questions such as the possible distortion of the previous azimuthal loops of electron current into non-closed streamlines and its effect in confinement [420]. Second, an HPT with two open source exits and two MNs has demonstrated recently bi-directional plasma ejection [426]. The control capabilities on both strength and direction of MN-guided flows make this device very suitable for the Ion Beam Shepherd concept for space debris removal [428]. And third, another advanced idea, with elements of the two previous ones, is a U-shaped plasma source with near-zero magnetic dipole and full magnetic shielding of the internal walls, featuring dual MNs firing plasma in the same direction [435]. This concept can minimize the losses to the thruster walls, while, thanks to the two opposed signs, interacting MNs, could enable simple differential thrust vector control and a low beam divergence.

*Conclusions*

Magnetic nozzles constitute a exciting field of research requiring up-to-date techniques on both the theoretical and experimental sides. Much understanding has been acquired in the recent years, and it has opened the way to explore more complex questions for the coming years. Some of them and the ways to tackle these questions and issues are common to other subjects treated in other sections of this paper. MN studies and developments are strongly dependent on the advances and technological challenges on the plasma thruster they are implemented in. Since MNs guide near collisionless plasmas, there are also interesting commonalities with magnetically confined plasmas in fusion devices, especially in terms of diffusion processes and the scrape-off layer regions.

## Acknowledgments

Y. Raitses and I. D. Kaganovich gratefully acknowledge partial financial support by the AFOSR grant FA9550-17-1-0010 and assistance in preparation and fruitful discussions with I. Romadanov, Eduardo Rodriquez, and J. B. Simmonds. The work of A. Smolyakov was supported in part by Natural Sciences and Engineering Research Council of Canada (NSERC) Canada, by the AFOSR grant FA9550-18-1-0132, and by Compute Canada, and he acknowledges fruitful discussions with O. Chapurin, S. Janhunen, M. Jimenez, O. Koshkarov, I Romadanov, and D. Sydorenko. The contribution of E. Ahedo and M. Merino was supported by the Government of Spain, National Development and Research Program, Grant No. ESP2016-75887P, and they thank J. Navarro and P. Fajardo for their contribution. M. Keidar and I. Schweigert gratefully acknowledge AFSOR grant FA9550-19-1-0166. I. Schweigert was partly supported by Russian Science Foundation, grant 17-19-01375. S. Tsikata acknowledges support from CNES (Centre national d'études spatiales, France)". F. Taccogna gratefully acknowledges financial support of Italian minister of university and research (MIUR) under the project PON "CLOSE to the Earth" ARS ARS01-00141. R. Gueroult and N.J. Fisch were supported by DOE DE-SC0016072 and NSF PHY-1805316. A. Bourdon and P. Chabert gratefully acknowledge financial support of the French National Research Agency (L'Agence nationale de la recherche) ANR grant ANR-16-CHIN-003-01 and Safran Aircraft Engines within the project POSEIDON.

## Data Availability Statement
The data that support the findings of this study are available from the corresponding author upon reasonable request.

### References:

[1] Presentations of the Princeton 2018 E×B workshop are available at https://htx.pppl.gov/E×B 2018papers.html

[2] Workshop on Plasma Cross Field Diffusion at 65th Annual Gaseous Electronics Conference 2012, organized by Rod Boswell and Igor D Kaganovich, http://meetings.aps.org/Meeting/GEC12/Session/AM2 ; followed by special issue: Rod Boswell and Igor D Kaganovich, "Special issue on transport in B-fields in low-temperature plasmas" Plasma Sources Sci. Technol. **23** 040201 (2014).

[3] Workshop "E×B Plasmas for Space and Industrial Applications" Toulouse 2017, organized by Jean-Pierre Boeuf and Andrei Smoylakov, https://E×B -2017.sciencesconf.org/ followed by special issue: Jean-Pierre Boeuf and Andrei Smoylakov, "Preface to Special Topic: Modern issues and applications of E × B plasmas", Phys. Plasmas **25**, 061001 (2018).

[4] A.I. Morozov, V.V. Savelyev (2000) Fundamentals of Stationary Plasma Thruster Theory. In: B.B. Kadomtsev, V.D. Shafranov (eds.) Reviews of Plasma Physics vol. 21, pp 203-391, Kluwer Academic, New York (2000).

[5] A. I. Bugrova, A. I. Morozov, and V. K. Kharchevnikov, "Electron distribution function in a Hall accelerator", Fiz. Plazmy **18**, 963-975 (1992).

[6] Y. Raitses, J. Ashkenazy, G. Appelbaum, and M. Gualman, "Experimental investigation of Effect of Channel material on Hall Thruster Characteristics", in 25th International Conference on Electric Propulsion, Cleveland, OH, 1997 Electric Rocket Propulsion Society, Cleveland, OH, 1997, IEPC 97-056.

[7] N. Gascon, M. Dudeck, and S. Barral, "Wall material effects in stationary plasma thrusters. I. Parametric studies of an SPT-100", Phys. Plasmas **10**, 4123 (2003).

[8] S. Barral, K. Makowski, and Z. Peradzyn´ N. Gascon and M. Dudeck, "Wall material effects in stationary plasma thrusters. II. Near-wall and in-wall conductivity" Phys. Plasmas **10**, 4137 (2003).

[9] Y. Raitses, I. D. Kaganovich, A. Khrabrov, D. Sydorenko, N. J. Fisch, and A. Smolyakov, "Effect of Secondary Electron Emission on Electron Cross-Field Current in E × B Discharges", IEEE Trans on Plasma Scie. **39**, 995 (2011).

[10] Y. Raitses, D. Staack and N. J. Fisch, "Plasma Characterization of Hall Thruster with Active and Passive Segmented Electrodes", AIAA Paper 2002-3954, 38th Joint Propulsion Conference and Exhibit, Indianapolis, IN, July 7-10, 2002.

[11] V. I. Demidov, S. F. Adams, I. D. Kaganovich, M. E. Koepke, and I. P. Kurlyandskaya, "Measurements of low-energy electron reflection at a plasma boundary", Physics of Plasmas **22**, 104501 (2015);

[12] A. Fruchtman and N. J. Fisch, "Variational principle for optimal accelerated neutralized flow", Phys. Plasmas **8**, 56 (2001).

[13] A. Fruchtman, N. J. Fisch, and Y. Raitses, "Control of the electric-field profile in the Hall thruster", Phys. Plasmas **8**, 1048 (2001).

[14] A. Hecimovic, C. Corbella, C. Maszl, W. Breilmann, and A. von Keudell, "Investigation of plasma spokes in reactive high power impulse magnetron sputtering discharge", J. Appl. Phys. **121**, 171915 (2017).

[15] A. Dunaevsky, Y. Raitses, and N. J. Fisch, Secondary electron emission from dielectric materials of a Hall thruster with segmented electrodes, Physics of Plasmas **10**, 2574 (2003).

[16] M Belhaj, T Tondu, V Inguimbert, and J P Chardon, "A Kelvin probe-based method for measuring the electron emission yield of insulators and insulated conductors subjected to electron irradiation", J. Phys. D: Appl. Phys. **42** 105309, (2009).

[17] M. Belhaj, N. Guibert, K. Guerch, P. Sarrailh, N. Arcis, "Temperature effect on the electron emission yield of BN-SiO2 under electron irradiation" In Proceedings of Spacecraft Charging Technology Conference 2014 - 202 Paper, 2014.

[18] M. Villemant, M. Belhaj, P. Sarrailh, S. Dadouch, L. Garrigues, C. Boniface, "Measurements of electron emission under electron impact on BN sample for incident electron energy between 10eV and 1000eV", Europhysics Letters **127**, 23001 (2019).

[19] B. Vincent, S. Tsikata, S. Mazouffre, C. Boniface. Thomson scattering investigations of a low-power Hall thruster in standard and magnetically-shielded configurations. 36th International Electric Propulsion Conference, Sep 2019, Vienne, Austria. https://hal.archives-ouvertes.fr/hal-02346188/document

[20] E. Ahedo, J.M. Gallardo, M. Martínez-Sánchez ,"Effects of the radial plasma-wall interaction on the Hall thruster discharge", Physics of Plasmas **10**, 3397 (2003).

[21] E. Ahedo, V. de Pablo, "Combined effects of electron partial thermalization and secondary emission in Hall thruster discharges", Phys. Plasmas **14**, 083501 (2007).

[22] D. Sydorenko, A. Smolyakov, I. Kaganovich, and Y. Raitses, "Kinetic simulation of secondary electron emission effects in Hall thrusters", Phys. Plasmas **13**, 014501 (2006).

[23] I. D. Kaganovich, and Y. Raitses, D. Sydorenko, A. Smolyakov "Kinetic Effects in Hall Thruster Discharge", Phys. Plasmas **14**, 057104 (2007).

[24] A. Tavant, V. Croes, R. Lucken, T. Lafleur, A. Bourdon, P. Chabert, "The effects of secondary electron emission on plasma sheath characteristics and electron transport in an E×B discharge via kinetic simulations", Plasma Sources Science and Technology **27**, 124001 (2018).

[25] A Tavant, R Lucken, A Bourdon, P Chabert, "Non-isothermal sheath model for low-pressure plasmas", Plasma Sources Science and Technology **28**, 075007 (2019).

[26] A. Domínguez-Vázquez, F. Taccogna, E. Ahedo, 'Particle modeling of radial electron dynamics in a controlled discharge of a Hall Thruster', Plasma Sources Sci. Technol. **27**, 064006 (2018).

[27] D. Sydorenko, A. Smolyakov, I. Kaganovich, and Y. Raitses, "Effects of non-Maxwellian electron velocity distribution function on two-stream instability in low-pressure discharges", Phys. Plasmas **14**, 013508 (2007).

[28] E. Ahedo, F.I. Parra 'Partial trapping of secondary-electron emission in a Hall thruster plasma', Phys. Plasmas **12**, 073503 (2005).

[29] H. Wang, M. D. Campanell, I. D. Kaganovich, G.B. Cai, "Effect of asymmetric secondary emission in bounded low-collisional E x B plasma on sheath and plasma properties", J. Phys. D- Appl. Phys. **47**, 405204 (2014).

[30] F. Taccogna, R. Schneider, S. Longo, and M. Capitelli, 'Kinetic simulations of a plasma thruster, Plasma Sources Sci. Technol. **17**, 024003 (2008).

[31] D. Sydorenko, A. Smolyakov, I. Kaganovich, and Y. Raitses, "Plasma-sheath instability in Hall thrusters due to periodic modulation of the energy of secondary electrons in cyclotron motion", Phys. Plasmas **15**, 053506 (2008).

[32] D. Sydorenko, I. Kaganovich, Y. Raitses, A. Smolyakov, "Breakdown of a space charge limited regime of a sheath in a weakly collisional plasma bounded by walls with secondary electron emission.", Phys. Rev. Lett. **103**, 145004 (2009).

[33] M. D. Campanell, A. V. Khrabrov, and I. D. Kaganovich, "General Cause of Sheath Instability Identified for Low Collisionality Plasmas in Devices with Secondary Electron Emission", Phys. Rev. Lett. **108**, 235001 (2012).

[34] M. D. Campanell, A. Khrabrov, and I. D. Kaganovich, "Instability, Collapse and Oscillation of Sheaths Caused by Secondary Electron Emission", Phys. Plasmas **19**, 123513 (2012).

[35] F Taccogna, S Longo, M Capitelli and R Schneider, 'Anomalous transport induced by sheath instability in Hall effect thrusters', Applied Physics Letters **94**, 251502 (2009).

[36] M. D. Campanell, A. V. Khrabrov, and I. D. Kaganovich, "Absence of Debye Sheaths due to Secondary Electron Emission", Phys. Rev. Lett. **108**, 255001 (2012).

[37] J. P. Sheehan, N. Hershkowitz, I. D. Kaganovich, H. Wang, Y. Raitses, E. V. Barnat, B. R. Weatherford, and D. Sydorenko, "Kinetic Theory of Plasma Sheaths Surrounding Electron-Emitting Surfaces", Phys. Rev. Lett. **111**, 075002 (2013).

[38] B. F. Kraus and Y. Raitses, "Floating potential of emitting surfaces in plasmas with respect to the space potential", Physics of Plasmas **25**, 030701 (2018).

[39] A. Khrabry, I. D. Kaganovich, V. Nemchinsky, and A. Khodak, "Investigation of the short argon arc with a hot anode. I. Numerical simulations of non-equilibrium effects in the near-electrode regions", Phys. Plasmas **25**, 013521 (2018).

[40] , E. Ahedo, "Non-Maxwellian EVDF features in a Hall thruster chamber, presented at Princeton E×B Workshop 2018 (available at https://htx.pppl.gov/E×B 2018presentations/Friday/1%20Ahedo-evdf-het.pdf ).

[41] R L Stenzel, "Instability of plasma-sheath resonance", Phys. Rev. Lett. **60**, 704 1988.

[42] I V Schweigert, S J Langendorf, M L R Walker and M Keidar, "Sheath structure transition controlled by secondary electron emission", Plasma Sources Sci. Technol. **24**, 025012 (2015).

[43] A. V. Khrabrov, I. D. Kaganovich, Y. Raitses, D. Sydorenko, A. Smolyakov, "Excitation of Ion Acoustic Waves in Plasmas with Electron Emission from Walls" Joint Conference of 30th International Symposium on Space Technology and Science and 34th International Electric Propulsion Conference and 6th Nano-satellite Symposium, Hyogo-Kobe, Japan, July 4 – 10, 2015 https://erps.spacegrant.org/uploads/images/2015Presentations/IEPC-2015-340_ISTS-2015-b-340.pdf

[44] D. M. Goebel, I. Katz, Fundamentals of Electric Propulsion: ion and Hall thrusters, Willey, 2008.

[45] M. Keidar and I.I. Beilis, Plasma Engineering, 2nd Edition, Elsevier-Academic Press, Oxford UK, August 2018.

[46] J. Yim, M. Keidar, and I.D. Boyd, , "A hydrodynamic-based erosion model for Hall thrusters", 29[th] International Electric Propulsion Conference, Princeton NJ, IEPC Paper-05-013, October 2005.

[47] E. Ahedo, A. Antón, I. Garmendia, I. Caro, J. González del Amo, "Simulation of wall erosion in Hall thrusters", paper IEPC-2007-067, Electric Rocket Propulsion Society, available at https://erps.spacegrant.org/ (2007).

[48] D. Pérez-Grande, P. Fajardo, E. Ahedo, "Evaluation of Erosion Reduction Mechanisms in Hall Effect Thrusters", paper IEPC-2015-280, Electric Rocket Propulsion Society, available at https://erps.spacegrant.org/ (2015).

[49] I.G. Mikellides, I. Katz, R. Hofer, D.M. Goebel, K de Grys and A. Mathers, "Magnetic shielding of the channel walls in a Hall plasma accelerator", Phys. Plasmas **18**, 033501, (2011).

[50] D. M. Goebel, R. R. Hofer, I. G. Mikellides, I. Katz, J. E. Polk, and B. N. Dotson, "Conducting Wall Hall Thrusters", IEEE Trans. on Plasma Scie. **43**, 118 (2015).

[51] K. de Grys, A. Mathers, B. Welander, and V. Khayms, "Demonstration of 10,400 hours of operation on a 4.5 kW qualification model Hall thruster," in Proc. 46th Joint Propuls. Conf., Nashville, TN, USA, Jul. 2010.

[52] S. K. Absalamov, V. B. Andreev, T. Colberi, M. Day, V. V. Egorov, R. U. Gnizdor, H. Kaufman, V. Kim, A. I. Korakin, K. N. Kozubsky, S. S. Kudravzev, U. V. Lebedev, G. A. Popov, V. V. Zhurin, "Measurement of plasma parameters in the stationary plasma thruster (spt-100) plume and its effect on spacecraft components" AIAA/SAE/ASME/ASEE 28th Joint Propulsion Conference and Exhibit July 6-8, 1992 / Nashville, TN AIAA-92-31 56.

[53] Arhipov, B. A., Bober, A. S., Gnizdor, R. Y., Kozubsky, K. N., Korakin, A. I., Maslennikov, and S.Y. Pridannikov, "The results of 7000-hour life testing" Paper IEPC-95-39 in the Proceedings of the 24-th International Electric Propulsion Conference, Moscow 1995, pages 315-324.

[54] N. Koch , M. Schirra , S. Weis , A. Lazurenko , B. van Reijen , J. Haderspeck , A. Genovese , and P. Holtmann, R. Schneider , K. Matyash , and O. Kalentyev, The HEMPT Concept - A Survey on Theoretical Considerations and Experimental Evidences, IEPC-2011-236.

[55] Y. Raitses, N. J. Fisch, 'Parametric investigations of a nonconventional Hall thruster', Phys. Plasmas **8**, 2579 (2001).

[56] L.E. Zakharenkov, V. Kim, A.S. Lovtsov, A.V. Semenkin, and A.E. Solodukhin, "Modern trends and development prospects of thrusters with closed electron drift", Conference: Space Propulsion Seville Spain, 2018.

[57] M. Keidar and I. Beilis, "Sheath and boundary conditions for plasma simulations of a Hall thruster discharge with magnetic lenses", Appl. Phys. Lett. **94**, 191501 (2009).

[58] L. Brieda and M. Keidar, "Plasma-wall interaction in Hall thrusters with magnetic lens configuration", J. Appl. Phys. **111**, 123302 (2012).

[59] I. Schweigert and M. Keidar, Periodical plasma structures controlled by external magnetic field, Plasma Source Science & Technology **26**, 064001 (2017).

[60] I. V. Schweigert, M. L. R. Walker, M. Keidar, "Genesis of Non-Uniformity of Plasma Fluxes Over Emissive Wall in Low-Temperature Plasmas", Plasma Research Express **1**, 045007 (2019).

[61] E. P. Velikhov and A. M. Dykhne, VI Conference Internationale sure les Phenomenes d'Ionisation sur les Gaz, Paris **4**, 511 (1963).

[62] J. L. Kerrebrock, "Nonequilibrium ionization due to electron heating - ii – experiments", AIAA Journal 2, I072 0964.

[63] J N Lukas 2016 Ph.D. Dissertation, The George Washington University.

[64] A. Domínguez-Vázquez, F. Taccogna, P. Fajardo, E. Ahedo, 'Parametric study of the radial plasma-wall interaction in a hall thruster", J. Phys. D: Appl. Phys. **52**, 474003 (2019).

[65] A.I. Morozov, Y.V. Esipchuk, A.M. Kapulkin, V.A. Nevrovskii, V.A. Smirnov, Effect of the magnetic field on a closed-electron-drift accelerator, Sov. Phys.-Tech. Phys. **17**, 482 (1972).

[66] A.I. Morozov, "The Conceptual Development of Stationary Plasma Thrusters", Plasma Physics Reports **29**, 235 (2003).

[67] K. Thomassen, "Turbulent diffusion in a penning-type discharge," Phys. Fluids **9**, 1836 (1966)

[68] Y. Sakawa, C. Joshi, P.K. Kaw, F.F. Chen, and V.K. Jain, "Excitation of the modified Simon–Hoh instability in an electron beam produced plasma", Phys. Fluids B **5**, 1681 (1993).

[69] A. Anders and Y. Yang, Direct observation of spoke evolution in magnetron sputtering," Appl. Phys. Lett. **111**, 064103 (2017).

[70] J. -P. Boeuf and L. J. Garrigues, "Low frequency oscillations in a stationary plasma thruster", J. Appl. Phys. **84**, 3541 (1998).

[71] N. Yamamoto, K. Komurasaki and Y. Arakawa, "Discharge Current Oscillation in Hall Thrusters", J. Propul. Power **21**, 870 (2005).

[72] S. Barral, E. Ahedo, "Low-frequency model of breathing oscillations in Hall discharges", Phys. Rev. E **79**, 046401 (2009).

[73] J. -P. Boeuf and B. Chaudhury, "Rotating instability in low-temperature magnetized plasmas," Phys. Rev. Lett. **111**, 155005 (2013).

[74] A.I. Smolyakov, O. Chapurin, W. Frias, O. Koshkarov, I. Romadanov, T. Tang, M. Umansky, Y. Raitses, I.D. Kaganovich, V.P. Lakhin, "Fluid theory and simulations of instabilities, turbulent transport and coherent structures in partially-magnetized plasmas of E x B discharges", Plasma Phys. Control. Fusion **59**, 014041 (2017).

[75] A. Powis, J. Carlsson, I. Kaganovich, Y. Raitses, and A. Smolyakov, "Scaling of spoke rotation frequency within a Penning discharge", Phys. Plasmas **25**, 072110 (2018).

[76] R. Kawashima, K. Hara, and K. Komurasaki, "Numerical analysis of azimuthal rotating spokes in a crossed-field discharge plasma," Plasma Sources Sci. Technol. **27**, 035010 (2018).

[77] K. Hara, "An overview of discharge plasma modeling for Hall effect thrusters", Plasma Sources Sci. Technol. **28**, 044001 (2019).

[78] C. L. Ellison, Y. Raitses and N. J. Fisch, "Cross-field electron transport induced by a rotating spoke in a cylindrical Hall thruster", Phys. Plasmas **19**, 013503 (2012).

[79] T. Ito, C. V. Young, and M. A. Cappelli, "Self-organization in planar magnetron microdischarge plasmas," Appl. Phys. Lett. **106**, 254104 (2015).

[80] A. P. Ehiasarian, A. Hecimovic, T. de los Arcos, R. New, V. S. von der Gathen, M. Boke, and J. Winter, "High power impulse magnetron sputtering discharges: Instabilities and plasma self-organization," Appl. Phys. Lett. **100**, 114101 (2012).

[81] K. Hara, M. J. Sekerak, I. D. Boyd, and A. D. Gallimore, "Mode transition of a Hall thruster discharge plasma", J Appl. Phys. **115**, 203304 (2014).

[82] S. Mazouffre and G. Bourgeois, "Spatio-temporal characteristics of ion velocity in a Hall thruster discharge", Plasma Sources Sci. Technol. **19**, 065018 (2010).

[83] M. E. Griswold, C. L. Ellison, Y. Raitses and N. J. Fisch, "Feedback control of an azimuthal oscillation in the E×B discharge of Hall thrusters", Phys. Plasmas **19**, 053506 (2012).

[84] B. Jorns, C. Dodson, D. Goebel, and R. Wirz, "Propagation of Ion Acoustic Wave Energy in the Plume of a High-Current LaB6 Hollow Cathode", Physical Review E **96**, 023208 (2017).

[85] B. Jorns, I. Mikellides, A. Lopez-Ortega, and D. Goebel. Mitigation of Energetic Ions and Keeper Erosion in a High-Current Hollow Cathode. Paper presented at the 34th International Electric Propulsion Conference, Kobe, Japan, July 6-15, 2015. IEPC-2015-134.

[86] Y. Raitses, A. Smirnov, and N. J. Fisch, "Effects of enhanced cathode electron emission on Hall thruster operation", Phys. Plasmas **16**, 057106 (2009).

[87] Y. Shi, Y. Raitses, and A. Diallo, "Controlling azimuthal spoke modes in a cylindrical Hall thruster using a segmented anode", Plasma Sour. Sci. Technol. **27**, 104006 (2018).

[88] I. Romadanov, A. Smolyakov, Y. Raitses, I. Kaganovich, T. Tian and S. Ryzhkov, "Structure of nonlocal gradient-drift instabilities in Hall E × B discharges", Phys. Plasmas **23**, 122111(2016).

[89] G. S. Janes and R. S. Lowder, "Anomalous electron diffusion and ion acceleration in a low-density plasma", Phys. Fluids **9**, 1115 (1966).

[90] J. B. Parker, Y. Raitses, and N. J. Fisch, "Transition in electron transport in a cylindrical hall thruster", Appl. Phys. Lett. **97**, 091501 (2010).

[91] M. S. McDonald and A. D. Gallimore, "Rotating spoke instabilities in Hall thrusters", IEEE Trans. Plasma Sci. **39**, 2952 (2011).

[92] E. Rodriguez, V. Skoutnev, Y. Raitses, A. Powis, I. D. Kaganovich, A. Smolyakov, "Boundary-induced effect on the spoke-like activity in E×B plasma" Phys. of Plasmas **26**, 053503 (2019).

[93] M. J. Sekerak, B. W. Longmier, A. D. Gallimore, D. L. Brown, R. R. Hofer, and J. E. Polk, "Azimuthal spoke propagation in hall effect thrusters," IEEE Trans. Plasma Sci. **43**, 72 (2015).

[94] S. Jaeger, Th. Pierre, and C. Rebont, "Direct observation of a cross-field current carrying plasma rotating around an unstable magnetized plasma column", Phys. Plasmas **16**, 022304 (2009).

[95] A. Simon, "Instability of a partially ionized plasma in crossed electric and magnetic fields", Phys. Fluids **6** 382 (1963).

[96] F. C. Hoh, "Instability of Penning-type discharges", Phys. Fluids **6** 1184 (1963).

[97] Y. Sakawa, C. Joshi, P. K. Kaw, V. K. Jain, T. W. Johnston, F. F. Chen, and J. M. Dawson, "Nonlinear evolution of the modified Simon-Hoh instability via a cascade of sideband instabilities in a weak beam plasma system", Phys. Rev. Lett. **69**, 85 (1992).

[98] K. Hara, M. J. Sekerak, I. D. Boyd, and A. D. Gallimore, "Perturbation analysis of ionization oscillations in Hall effect thrusters", Phys. Plasmas **21**, 122103 (2014).

[99] A. Anders, "Tutorial: Reactive high-power impulse magnetron sputtering (R-HiPIMS)", J. Appl. Phys. **121**, 171101 (2017).

[100]     A. Hecimovic and A. von Keudell, "Spokes in high power impulse magnetron sputtering plasmas", J. Phys. D: Appl. Phys. **51**, 453001 (2018).

[101]     J Held, P A Maaß, V Schulz-von der Gathen and A von Keudell, "Electron density, temperature and the potential structure of spokes in HiPIMS", Plasma Sources Sci. Technol. **29**, 025006 (2020).

[102]     J Held abd  A von Keudell Held, "Pattern Formation in High Power Impulse Magnetron Sputtering (HiPIMS) Plasmas",, Plasma Chemistry and Plasma Processing **40**, 643–660 (2020).

[103]     M. Panjan and A. Anders, "Plasma potential of a moving ionization zone in DC magnetron sputtering", J. Appl. Phys. **121**, 063302 (2017).

[104]     J.-P. Boeuf and M. Takahashi, "Rotating Spokes, Ionization Instability, and Electron Vortices in Partially Magnetized E × B Plasmas", Phys. Rev. Lett. **124**, 185005 (2020).

[105]     A. Piels, E. Möbius, and G. Himmel, "The influence of the plasma inhomogeneity on the Critical Ionization Velocity phenomenon", Astrophysics and Space Science **72**, 211 (1980).

[106]     J.P. Boeuf, "Rotating structures in low temperature magnetized plasmas-insight from particle simulations", Front. Phys. **2**, 74 (2014).

[107]     O. Koshkarov, A. Smolyakov, Y. Raitses, and I. Kaganovich, "Self-Organization, Structures, and Anomalous Transport in Turbulent Partially Magnetized Plasmas with Crossed Electric and Magnetic Fields", Phys. Rev. Lett. **122**, 185001 (2019).

[108]    J. Carlsson, I. Kaganovich, A. Powis, Y. Raitses, I. Romadanov, and A. Smolyakov, "Particle-in-cell simulations of anomalous transport in a Penning discharge", Phys. Plasmas **25**, 061201 (2018).

[109]    W. Frias, A.I. Smolyakov, I.D. Kaganovich, Y. Raitses, "Long wavelength gradient drift instability in Hall plasma devices. Part I. Fluid theory, Phys. Plasmas **19**, 072112 (2012).

[110]    W. Frias, A. Smolyakov, I. Kaganovich, Y. Raitses, "Long wavelength gradient drift instability in Hall plasma devices. Part II, Applications", Phys. Plasmas **20**, 052108 (2013).

[111]    K. Hara, "Non-oscillatory quasineutral fluid model of cross-field discharge plasmas", Phys. Plasmas **25**, 123508 (2018).

[112]    J. M. Fife, "Hybrid-PIC modeling and electrostatic probe survey of Hall thrusters", Ph.D. dissertation (MIT, 1998).

[113]    A. I. Morozov, V.V. Savel'ev," One-dimensional hydrodynamic model of the atom and ion dynamics in a stationary plasma thruster", Plasma Phys. Rep. **26**, 219 (2000).

[114]    A. A. Litvak, N. J. Fisch, "Resistive instabilities in Hall current plasma discharge", Phys. Plasmas **8**, 648 (2001).

[115]    E. Fernandez, M.K. Scharfe, C.A. Thomas, N. Gascon, M.A. Cappelli, "Growth of resistive instabilities in E×B  plasma discharge simulations", Phys. Plasmas **15**, 012102 (2008).

[116]    S. Chable, F. Rogier, "Numerical investigation and modeling of stationary plasma thruster low frequency oscillations", Phys. Plasmas **12**, 033504 (2005).

[117]    O. Koshkarov, A. I. Smolyakov, A. Kapulkin, Y. Raitses, I. Kaganovich, "Nonlinear structures of lower-hybrid waves driven by the ion beam", Phys. Plasmas **25**, 061209 (2018).

[118]    O. Koshkarov, A. I. Smolyakov, I. V. Romadanov, O. Chapurin, M. V. Umansky, Y. Raitses, I. D. Kaganovich, "Current flow instability and nonlinear structures in dissipative two-fluid plasmas", Phys. Plasmas **25**, 011604 (2018).

[119]    F. F. Chen, "Microinstability and Shear Stabilization of a Low-β, Rotating, Resistive Plasma", Phys. Fluids **9**, 965 (1966).

[120]    R. Gueroult, J.-M. Rax and N. J. Fisch, "Centrifugal instability in the regime of fast rotation" Phys. Plasmas **24**, 082102 (2017).

[121]    G. J. M. Hagelaar "Modeling electron transport in magnetized low-temperature discharge plasmas", Plasma Sources Sci. Technol. **16**, S57 (2007).

[122]    V. Morin, A.I. Smolyakov, "Modification of the Simon-Hoh Instability by the sheath effects in partially magnetized E x B plasmas", Phys. Plasmas **25**, 084505 (2018).

[123]    D. Escobar and E. Ahedo, "Low frequency azimuthal stability of the ionization region of the Hall thruster discharge. I. Local analysis", Phys. Plasmas **21**, 043505 (2014).

[124]    D. Escobar and E. Ahedo, "Global Stability Analysis of Azimuthal Oscillations in Hall Thrusters", IEEE Trans Plasma Sci. **43**, 149 (2015).

[125]    S. Barral, Z. Peradzynski, "Ionization oscillations in Hall accelerators", Phys. Plasmas **17**, 014505 (2010).

[126]    E. Chesta, C. M. Lam, N. B. Meezan, D. P. Schmidt, and M. A. Cappelli, "A characterization of plasma fluctuations within a hall discharge," IEEE Trans. Plasma Sci. **29**, 582 (2001).

[127]    I. Romadanov, Y. Raitses, and A. Smolyakov, "Control of Coherent Structures via External Drive of the Breathing Mode", Plasma Phys. Rep. **45**, 134 (2019).

[128]    K. Hara, "Non-oscillatory quasineutral fluid model of cross-field discharge plasmas", Phys. Plasmas **25**, 123508 (2018).

[129]    K. Hara and K. M. Hanquist, "Test Cases for Grid-Based Direct Kinetic Modeling of Plasma Flows", Plasma Sour. Sci. Technol. **27**, 065004 (2018).

[130]    K. Hara, I. D. Boyd, and V. I. Kolobov, "One-dimensional hybrid-direct kinetic simulation of the discharge plasma in a Hall thruster", Physics of Plasmas **19**, 113508 (2012).

[131]    C. Rebont, N. Claire, Th. Pierre, and F. Doveil, "Ion Velocity Distribution Function Investigated Inside an Unstable Magnetized Plasma Exhibiting a Rotating Nonlinear Structure", Phys. Rev. Lett. **106**, 225006 (2011).

[132]    S. Mazouffre, "Laser-induced fluorescence diagnostics of the cross-field discharge of Hall thrusters", Plasma Sources Sci. Technol. **22**, 013001 (2013).

[133]    W. A. Hargus, Jr., and M. A. Cappelli, "Laser-induced fluorescence measurements of velocity within a Hall discharge," Appl. Phys. B, Lasers Opt. **72**, 961 (2001).

[134]    C. V. Young, A. Lucca Fabris, and M. A. Cappelli, "Ion dynamics in an E × B Hall plasma accelerator", Appl. Phys. Lett. **106**, 044102 (2015).

[135]    C. J. Durot, A. D. Gallimore, and T. B. Smith, "Validation and evaluation of a novel time-resolved laser-induced fluorescence technique", Rev. Sci. Instrum. **85**, 013508 (2014).

[136]    A. Diallo, S. Keller, Y. Shi, Y. Raitses, and S. Mazouffre, "Time-resolved ion velocity distribution in a cylindrical Hall thruster: Heterodyne-based experiment and modeling", Rev. Sci. Instrum. **86**, 033506 (2015).

[137]    B. Vincent, S. Tsikata, S. Mazouffre, T. Minea and J. Fils, "A compact new incoherent Thomson scattering diagnostic for low-temperature plasma studies", Plasma Sources Sci. Technol. **27**, 055002 (2018).

[138]    F. Chu, R. Hood, and F. Skiff, "Measurement of wave-particle interaction and metastable lifetime using laser-induced fluorescence", Phys. Plasmas **26**, 042111 (2019).

[139]    I. Romadanov, Y. Raitses, A. Diallo, K. Hara, I. D. Kaganovich, and A. Smolyakov, "On limitations of laser-induced fluorescence diagnostics for xenon ion velocity distribution function measurements in Hall thrusters", Phys. Plasmas **25**, 033501 (2018).

[140]    S. Tsikata, J. Cavalier, A. Héron, C. Honoré, N. Lemoine, D. Grésillon, and D. Coulette, "An axially propagating two-stream instability in the Hall thruster plasma", Phys. Plasmas **21**, 072116 (2014).

[141]    Y. B. Esipchuk, A. I. Morozov, G. N. Tilinin, and A. V. Trofimov, "Plasma oscillations in closed-drift accelerators with an extended acceleration zone", Sov. Phys. Tech. Phys. **18**, 928 (1974).

[142]    J-P. Boeuf, "Tutorial: Physics and modeling of hall thrusters", Journal of Applied Physics **121,** 011101 (2017).

[143]    J.C. Adam, A. Héron, and G. Laval, "Study of stationary plasma thrusters using two-dimensional fully kinetic simulations" Phys. Plasmas **11**, 295-305 (2004).

[144]    I. Mikellides, I. Katz, D. Goebel, and K. Jameson, "Evidence of nonclassical plasma transport in hollow cathodes for electric propulsion", J. Appl. Phys. **101**, 063301 (2007).

[145]    S. Tsikata, N. Lemoine, V. Pisarev, and D. M. Grésillon, "Dispersion relations of electron density fluctuations in a Hall thruster, observed by collective light scattering", Phys. of Plasmas **16**, 033506 (2009).

[146]    Z. Brown and B. Jorns, "Spatial evolution of small wavelength fluctuations in a Hall thruster," Phys. Plasmas **26**, 113504 (2019).

[147]    S. Tsikata, A. Héron, C. Honoré, "Oscillatory discharge behavior in Hall thrusters: relationships between the discharge current, electric field and microturbulence", $35^{th}$ International Electric Propulsion Conference, Atlanta, *GA,* IEPC-2017-443 (2017).

[148]    A. Héron and J. C. Adam, "Anomalous conductivity in Hall thrusters: effects of the non-linear coupling of the electron-cyclotron drift instability with secondary electron emission of the walls", Phys. of Plasmas **20**, 082313 (2013).

[149]    S. Tsikata and T. Minea, "Modulated electron cyclotron drift instability in a high-power pulsed magnetron discharge", Phys. Rev. Lett. **144**, 185001 (2015).

[150]    Z. Brown, E. Dale, and B. Jorns, "Experimental Correlation between AnomalousElectron Collision Frequency and Plasma Turbulence in a Hall Effect Thruster," 36th International Electric Propulsion ConferenceUniversity of Vienna, Austria, IEPC-2019-843.

[151]    B. Jorns, I. Mikellides, and D. Goebel, "Ion acoustic turbulence in a 100-A LaB₆ hollow cathode", Phys. Rev. E **90**, 063106 (2014).

[152]    B. Jorns, S. Cusson, Z. Brown, and E. Dale, "Non-classical electron transport in the cathode plume of a Hall effect thruster," Phys. Plasmas **27**, 022311 (2020).

[153]    S. Tsikata, A. Héron, and C. Honoré, "Hall thruster microturbulence under conditions of modified electron wall emission", Phys. Plasmas **24**, 053519 (2017).

[154]    I. M. DesJardin, K. Hara and S. Tsikata, "Self-Organized Standing Waves Generated By AC-Driven Electron Cyclotron Drift Instabilities" Appl. Phys. Lett. **115**, 234103 (2019).

[155]    Ioannis Mikellides, Alejandro Lopez Ortega, Dan M.Goebel, and Giulia Becatti, "Dynamics of a hollowcathode discharge in the frequency range of 1-500 kHz", Plasma Sources Sci. and Technol. **29** 035003 (2020).

[156]    A. Ducrocq, "Rôle des instabilités électroniques de dérive dans le transport électronique du propulseur à effet Hall", Ph. D. Thesis, Ecole Polytechnique (2006).

[157]    V. Kouznetsov, K. Macak, J. M. Schneider, U. Helmersson and I. Petrov, "A novel pulsed magnetron sputter technique utilizing very high target power densities", Surf. Coat. Technol. **122**, 290 (1999).

[158]    J. Cavalier, N. Lemoine, G. Bonhomme, S. Tsikata, C. Honoré and D. Grésillon, "Hall thruster plasma fluctuations identified as the E x B electron drift instability: Modeling and fitting on experimental data." Phys. Plasmas **20**, 082107 (2013).

[159]    S. Tsikata, C. Honoré, N. Lemoine and D. M. Grésillon, "Three-dimensional structure of electron density fluctuations in the Hall thruster plasma: E x B mode", Phys. Plasmas **17**, 112110 (2010).

[160]    S. Tsikata, "Small-scale electron density fluctuations in the Hall thruster, investigated by collective light scattering", Ph. D. Thesis, Ecole Polytechnique (2009).

[161]    B. Vincent, S. Tsikata, G.-C. Potrivitu, L. Garrigues, G. Sary and S. Mazouffre, "Electron properties of an emissive cathode: analysis with incoherent Thomson scattering, fluid simulations and Langmuir probe measurements", accepted J. Phys. D (2020), https://iopscience.iop.org/article/10.1088/1361-6463/ab9974

[162]    B. Vincent, S. Tsikata and S. Mazouffre, "Incoherent Thomson scattering measurements of electron properties in a conventional and magnetically-shielded Hall thruster", Plasma Sources Sci. Technol. **29**, 035015 (2020).

[163]    I. G. Mikellides, I. Katz, R. R. Hofer and D. M. Goebel, "Magnetic shielding of walls from the unmagnetized ion beam in a Hall thruster", Appl. Phys. Lett. **102**, 023509 (2013).

[164]    S. Tsikata, B. Vincent, T. Minea, A. Revel and C. Ballage, "Time-resolved electron properties of a HiPIMS argon discharge via incoherent Thomson scattering", Plasma Sources Sci. Technol. **28**, 03LT02 (2019).

[165]    A. M. Kapulkin, A. D. Grishkevic and V. F. Prisnyakov, "Outside Electric Field Thruster", Space Technol. **15**, 391 (1995).

[166]    S. Mazouffre, S. Tsikata and J. Vaudolon, "Development and experimental characterization of a wall-less Hall thruster", J. Appl. Phys. **116**, 243302 (2014).

[167]    S. Janhunen, A. Smolyakov, O. Chapurin, D. Sydorenko, I. Kaganovich and Y. Raitses, "Nonlinear structures and anomalous transport in partially magnetized E×B plasmas", Phys. Plasmas **25**, 011608 (2018).

[168]    I. G. Mikellides, I. Katz, D. M. Goebel, K. K. Jameson and J. E. Polk, "Wear mechanisms for electron sources in ion propulsion, 2: discharge hollow cathode", J. Prop. Power **24**, 866 (2008).

[169]    G. Sary, L. Garrigues and J.-P. Boeuf, "Hollow cathode plasma modeling: I. A coupled plasma thermal two-dimensional model", Plasma Sources Sci. Technol. **26**, 055007 (2017).

[170]    E. Dale and B. Jorns, "Non-invasive time-resolved measurements of anomalous collision frequency in a Hall thruster", Phys. Plasmas **26**, 013516 (2019).

[171]    A. Revel, T. Minea and S. Tsikata, "Pseudo-3D modeling of drift-induced spatial inhomogeneities in planar magnetron plasmas", Phys. Plasmas **23**,100701 (2016).

[172]    B. Jorns, "Predictive data-driven model for the anomalous electron collision frequency in a Hall effect thruster", Plasma Sources Sci. Technol. **27**, 104007 (2018).

[173]    K. Matyash, R. Schneider, S. Mazouffre, S. Tsikata and L. Grimaud, Plasma Sources Sci. Technol.  **28**, 044002 (2019).

[174]    F. Taccogna and P. Minelli, "Three-dimensional particle-in-cell model of a Hall thruster: The discharge channel", Phys. Plasmas **25**, 061208 (2018).

[175]    A. A. Litvak, Y. Raitses and N. Fisch, "Experimental studies of high-frequency azimuthal waves in Hall thrusters", Phys. Plasmas **11**, 1701 (2004).

[176]    A. Lazurenko, V. Vial, M. Prioul and A. Bouchoule, "Experimental investigation of high-frequency drifting perturbations in Hall thrusters", Phys. Plasmas **12**, 013501 (2005).

[177]    A. Ducrocq, J. C. Adam, A. Heron, and G. Laval, "High-frequency electron drift instability in the cross-field configuration of Hall thrusters", Phys. of Plasmas **13**, 102111 (2006).

[178]    P. Coche and L. Garrigues, "A two-dimensional (azimuthal-axial) particle-in-cell model of a Hall thruster", Phys. of Plasmas **21**, 023503, (2014).

[179]    T. Lafleur, S. D. Baalrud, and P. Chabert. Theory for the anomalous electron transport in Hall effect thrusters. i. insights from particle-in-cell simulations", Phys. of Plasmas **23**, 053502 (2016).

[180]    T. Lafleur, S. D. Baalrud, and P. Chabert, "Theory for the anomalous electron transport in hall effect thrusters. ii. kinetic model, Phys. of Plasmas **23**, 053503 (2016).

[181]    T. Lafleur and P. Chabert, "The role of instability-enhanced friction on 'anomalous' electron and ion transport in Hall-effect thrusters", Plasma Sources Sci. Technol. **27**, 015003 (2018).

[182]    S. P. Gary and D. Biskamp, "Instabilities in perpendicular collisionless shock waves", Journal of Physics: Part A General **4**, L27 (1971).

[183]    D. Forslund, C. Nielson, R. Morse, and J. Fu, "Electron-cyclotron drift instability and turbulence", Physics of Fluids **15**, 1303 (1972).

[184]    M. Lampe, J. B. McBride, W.M. Manheimer, R. N. Sudan, R. Shanny, J. H. Orens, and K. Papadopolous, "Theory and simulation of beam cyclotron instability", Physics of Fluids **15**, 662 (1972).

[185]    D. Biskamp and R. Chodura, "Collisionless dissipation of a cross-field electric-current", Physics of Fluids **16**, 893–901, (1973).

[186]    O. Buneman, "Instability of electrons streaming through ions across a magnetic field", Journal of Nuclear Energy. Part C, Plasma Physics, Accelerators, Thermonuclear Research **4**, 111, (1962).

[187]    S. P. Gary and J. J. Sanderson, "Longitudinal waves in a perpendicular collisionless plasma shock", 1. cold ions, Journal of Plasma Physics **4**, 739 (1970).

[188]    D. W. Forslund, R. L. Morse, and C. W. Nielson, "Anomalous resistance due to cross-field electron-ion streaming instabilities", Physics of Fluids **15**, 2363–2366, (1972).

[189]    L. Muschietti, B. Lembege, "Microturbulence in the electron cyclotron frequency range at perpendicular supercritical shocks", Journal of Geophysical Research-Space Physics **118**, 2267-2285 (2013).

[190]    D. Biskamp, R. Chodura, "Nonlinear electron-cyclotron drift instability", Nuclear Fusion **12**, 485 (1972).

[191]    C.K. Birdsall, A.B Langdon, "Plasma Physics via Computer Simulation", CRC Press (2004).

[192]    J. P. Boeuf and L. Garrigues, "E×B electron drift instability in hall thrusters: Particle-in-cell simulations vs. theory", Phys. of Plasmas **25**, 061204 (2018).

[193]    V. Croes, T. Lafleur, Z. Bonaventura, A. Bourdon, and P. Chabert, "2d particle-in-cell simulations of the electron drift instability and associated anomalous electron transport in hall-effect thrusters", Plasma Science & Technology **26**, 034001 (2017).

[194]    V. Croes, A. Tavant, R. Lucken, R. Martorelli, T. Lafleur, A. Bourdon, and P. Chabert, "The effect of alternative propellants on the electron drift instability in hall-effect thrusters: Insight from 2d particle-in-cell simulations", Physics of Plasmas **25**, 063522 (2018).

[195]    S. Janhunen, A. Smolyakov, D. Sydorenko, M. Jimenez, I. Kaganovich, and Y. Raitses, "Evolution of the electron cyclotron drift instability in two-dimensions", Phys. Plasmas **25**, 082308 (2018).

[196]    Z. Asadi, F. Taccogna, and M. Sharifian, "Numerical study of electron cyclotron drift instability: Application to hall thruster", Front. Phys. **7**, 140 (2019).

[197]    A. Smolyakov, T. Zintel, L. Couedel, D. Sydorenko, A.M. Umnov, E. Sorokina, N. Marusov, "Anomalous electron transport in one-dimensional Electron Cyclotron Drift Turbulence", Plasma Physics Reports to be published (2020).

[198]    T. Lafleur, R. Martorelli, P. Chabert, and A. Bourdon, "Anomalous electron transport in hall-effect thrusters: Comparison between quasi-linear kinetic theory and particle-in-cell simulations", Physics of Plasmas **25**, 061202 (2018).

[199]    I. Katz, I. G. Mikellides, R. R. Hofer, and Alejandro Lopez Ortega, "Hall2De simulations with an anomalous transport model based on the electron cyclotron drift instability", *34th International Electric Propulsion Conference*, pages IEPC–2015–402, 2015.

[200]    Ira Katz, Alejandro Lopez Ortega, Benjamin Jorns, and Ioannis G. Mikellides, "Growth and Saturation of Ion Acoustic Waves in Hall Thrusters", AIAA Propulsion and Energy Forum. American Institute of Aeronautics and Astronautics, -4534 2016. https://doi.org/10.2514/6.2016-4534

[201]    Z. Brown, B.A. Jorns, "Spatial Evolution of Plasma Waves in the Near-field of a Magnetically Shielded Hall Thruster", 2018 Joint Propulsion Conference, American Institute of Aeronautics and Astronautics 2018.

[202]    A. B. Langdon, "Kinetic-theory for fluctuations and noise in computer-simulation of plasma", Physics of Fluids **22**, 163–171 (1979).

[203]    E. L. Lindman, "Numerical-simulation of ion-acoustic turbulence", Journal of Statistical Physics **39**, 769–782 (1985).

[204]    D. Biskamp and R. Chodura, "*Computer Simulation of Anomalous Resistance"*, volume 2 of *Plasma Physics and Controlled Fusion Research*. IAEA, Vienna, (1971).

[205]    C. T. Dum and R. N. Sudan, "Saturation of nonlinear explosive instabilities", Physical Review Letters **23**, 1149 (1969).

[206]    T. Lafleur, S. D. Baalrud, and P. Chabert, "Characteristics and transport effects of the electron drift instability in hall-effect thrusters", Plasma Sources Science and Technology **26**, 024008 (2017).

[207]    C. F. F. Karney, "Stochastic ion heating by a lower hybrid wave", Physics of Fluids **21**, 1584–1599 (1978).

[208]    LANDMARK: Low temperAture magNetizeD plasMA benchmaRKs, https://www.landmark-plasma.com/

[209]    T Charoy, J P Boeuf, A Bourdon, J A Carlsson, P Chabert, B Cuenot, D Eremin, L Garrigues, K Hara, I D Kaganovich, A T Powis, A Smolyakov, D Sydorenko, A Tavant, O Vermorel, and W Villafana, "2D axial-azimuthal particle-in-cell benchmark for low-temperature partially magnetized plasmas", Plasma Sources Science and Technology **28**, 105010, (2019).

[210]    D. W. Forslund, R. L. Morse, and C. W. Nielson, "Electron Cyclotron Drift Instability," Physical Review Letters, **25**, 1266 (1970).

[211]    P. Coche, and L. Garrigues, "A two-dimensional (azimuthal-axial) particle-in-cell model of a Hall thruster," Physics of Plasmas **21**, 023503 (2014).

[212]    F. Taccogna, R. Schneider, S. Longo, and M. Capitelli, "Kinetic simulations of a plasma thruster," Plasma Sources Science and Technology **17**, 024003 (2008).

[213]    C. Vivien, L. Trevor, B. Zdeněk, B. Anne, and C. Pascal, "2D particle-in-cell simulations of the electron drift instability and associated anomalous electron transport in Hall-effect thrusters," Plasma Sources Science and Technology **26**, 034001(2017).

[214]    F. Taccogna, and P. Minelli, "Three-dimensional particle-in-cell model of Hall thruster: The discharge channel," Physics of Plasmas **25**, 061208 (2018).

[215]    J. M. Fife, "Hybrid-PIC modeling and electrostatic probe survey of Hall thrusters," Ph.D., Massachusetts Institute of Technology, 1998.

[216]    F. I. Parra, E. Ahedo, J. M. Fife, and M. Martinez-Sanchez, "A two-dimensional hybrid model of the Hall thruster discharge," Journal of Applied Physics **100**, 023304 (2006).

[217]    E. Sommier, M. K. Scharfe, N. Gascon, M. A. Cappelli, and E. Fernandez, "Simulating Plasma-Induced Hall Thruster Wall Erosion with a Two-Dimensional Hybrid Model," Transactions on Plasma Science, IEEE **35**, 1379 (2007).

[218]    G. J. M. Hagelaar, J. Bareilles, L. Garrigues, and J. P. Boeuf, "Two-dimensional model of a stationary plasma thruster," Journal of Applied Physics **91**, 5592 (2002).

[219]    D. J. Bohm, E. Burhop, and H. Massey, "The Characteristics of Electrical Discharges in Magnetic Fields," National Nuclear Energy Series, Manhattan Project Technical Section, Division I, 5, A. Guthrie and R. K. Wakerling, eds., pp. 13-76, New York: McGraw-Hill, 1949.

[220]    I. G. Mikellides, "The Effectiveness of Magnetic Shielding in High-Isp Hall Thrusters," in 49th AIAA/ASME/SAE/ASEE Joint Propulsion Conference, San Jose, CA, AIAA-2013-3885, July 2013.

[221]    A. Lopez Ortega, and I. G. Mikellides, "The importance of the cathode plume and its interactions with the ion beam in numerical simulations of Hall thrusters," Physics of Plasmas **23**, 043515 (2016).

[222]    I. G. Mikellides, and I. Katz, "Numerical simulations of Hall-effect plasma accelerators on a magnetic-field-aligned mesh," Physical Review E **86**, 046703 (2012).

[223]    L. Garrigues, G. J. M. Hagelaar, C. Boniface, and J. P. Boeuf, "Anomalous conductivity and secondary electron emission in Hall effect thrusters," Journal of Applied Physics **100**, 123301 (2006).

[224]    M. K. Scharfe, N. Gascon, M. A. Cappelli, and E. Fernandez, "Comparison of hybrid Hall thruster model to experimental measurements," Physics of Plasmas **13**, 083505 (2006).

[225]    B. Jorns, "Predictive, data-driven model for the anomalous electron collision frequency in a Hall effect thruster," Plasma Sources Science & Technology **27**, 104007 (2018).

[226]    A. Lopez Ortega, I. Katz, and V. H. Chaplin, "Application of a first-principles anomalous transport model for electrons to multiple Hall thrusters and operating conditions," 2018 Joint Propulsion Conference, AIAA Propulsion and Energy Forum: American Institute of Aeronautics and Astronautics, 2018.

[227]     I. G. Mikellides, B. Jorns, I. Katz, and A. Lopez Ortega, "Hall2De Simulations with a First-principles Electron Transport Model Based on the Electron Cyclotron Drift Instability," 52nd AIAA/SAE/ASEE Joint Propulsion Conference, AIAA Propulsion and Energy Forum: American Institute of Aeronautics and Astronautics, 2016.

[228]     V. Joncquieres, F. Pechereau, A. Alvarez Laguna, A. Bourdon, O. Vermorel, and B. Cuenot, "A 10-moment fluid numerical solver of plasma with sheaths in a Hall Effect Thruster," 2018 Joint Propulsion Conference, AIAA Propulsion and Energy Forum: American Institute of Aeronautics and Astronautics, 2018.

[229]     I. G. Mikellides, and A. L. Ortega, "Challenges in the development and verification of first-principles models in Hall-effect thruster simulations that are based on anomalous resistivity and generalized Ohm's law," Plasma Sources Science & Technology **28**, 075001 (2019).

[230]     S.N. Abolmasov, "Physics and engineering of crossed-field discharge devices", Plasma Sources Science & Technology **21**, 035006 (2012).

[231]     A.V. Brantov, V.Y. Bychenkov, V.T. Tikhonchuk, W. Rozmus, "Nonlocal electron transport in laser heated plasmas", Physics of Plasmas **5**, 2742 (1998).

[232]     S. Chable, F. Rogier, "Numerical investigation and modeling of stationary plasma thruster low frequency oscillations", Physics of Plasmas **12**, 033504 (2005)

[233]     A.M. Dimits, G. Bateman, M.A. Beer, B.I. Cohen, W. Dorland, G.W. Hammett, C. Kim, J.E. Kinsey, M. Kotschenreuther, A.H. Kritz, L.L. Lao, J. Mandrekas, W.M. Nevins, S.E. Parker, A.J. Redd, D.E. Shumaker, R. Sydora, J. Weiland, "Comparisons and physics basis of tokamak transport models and turbulence simulations", Physics of Plasmas **7**, 969 (2000).

[234]     A.M. Dimits, I. Joseph, M.V. Umansky, A fast non-Fourier method for Landau-fluid operators, Physics of Plasmas **21**, 10 (2014).

[235]     Y.V. Esipchuk, G.N. Tilinin, "Drift Instability in a Hall-current plasma accelerator", Sov. Phys. Tech. Phys. **21**, 417 (1976).

[236]     E. Fernandez, M.K. Scharfe, C.A. Thomas, N. Gascon, M.A. Cappelli, "Growth of resistive instabilities in E×B  plasma discharge simulations", Physics of Plasmas **15**, 012102 (2008)

[237]     W. Frias, A. Smolyakov, I.D. Kaganovich, Y. Raitses, "Wall current closure effects on plasma and sheath fluctuations in Hall thrusters", Physics of Plasmas **21**, 062113 (2014).

[238]     G.J.M. Hagelaar, "Modeling electron transport in magnetized low-temperature discharge plasmas", Plasma Sources Science & Technology **16**, S57 (2007).

[239]     G.W. Hammett, F.W. Perkins, "Fluid moment models for Landau damping with applications to the on-temperature gradient instability", Physical Review Letters **64**, 3019 (1990).

[240]     B.A. Jorns, R.R. Hofer, "Plasma oscillations in a 6-kW magnetically shielded Hall thruster", Physics of Plasmas **21**, 053512 (2014).

[241]     M. Keidar, I.I. Beilis, "Electron transport phenomena in plasma devices with E x B drift", IEEE T Plasma Sci. **34**, 804 (2006).

[242]     S. Kolev, G.J.M. Hagelaar, G. Fubiani, J.P. Boeuf, "Physics of a magnetic barrier in low-temperature bounded plasmas: insight from particle-in-cell simulations", Plasma Sources Science & Technology **21**, 025002 (2012).

[243]     A.I. Morozov, Y. Esipchuk, V. A.M. Kapulkin, V.A. Nevrovskii, V.A. Smirnov, "Azimuthally Asymmetric Modes and Anomalous Conductivity in Closed Electron Drift Accelerators", Soviet Phys -Tech Phys. **18**, 615 (1973).

[244]     A.I. Morozov, Y.V. Esipchuk, A.M. Kapulkin, V.A. Nevrovskii, V.A. Smirnov, "Effect of the magnetic field on a closed-electron-drift accelerator", Sov. Phys. Tech. Phys. **17**, 482 (1972).

[245]     A.I. Morozov, V.V. Savelyev, Reviews of plasma physics, in: B.B. Kadomtsev, V.D. Shafranov (Eds.) Kluver, New York, 2000, pp. 203.

[246]     V. Nikitin, D. Tomilin, A. Lovtsov, A. Tarasov, "Gradient-drift and resistive mechanisms of the anomalous electron transport in Hall effect thrusters", EPL **117**, 45001 (2017).

[247]     B.N. Rogers, P. Ricci, "Low-Frequency Turbulence in a Linear Magnetized Plasma", Physical Review Letters **104**, 225002 (2010).

[248]     Y. Sakawa, C. Joshi, P.K. Kaw, F.F. Chen, V.K. Jain, "Excitation of the modified Simon-Hoh instability in an electron-beam produced plasma", Physics of Fluids B-Plasma Physics **5**, 1681 (1993).

[249]     Y. Sakawa, C. Joshi, P.K. Kaw, V.K. Jain, T.W. Johnston, F.F. Chen, J.M. Dawson, "Nonlinear evolution of the modified Simon-Hoh instability via a cascade of side-band instabilities in a weak beam-plasma system", Physical Review Letters **69**, 85 (1992).

[250]     M. K. Scharfe, N. Gascon, M.A. Cappelli, E. Fernandez, "Comparison of hybrid Hall thruster model to experimental measurements", Physics of Plasmas **13**, 083505 (2006).

[251]     T.A. van der Straaten, N.F. Cramer, "Transverse electric field and density gradient induced instabilities in a cylindrical magnetron discharge", Physics of Plasmas 7, 391 (2000).

[252]     T. Passot, P.L. Sulem, E. Tassi, "Electron-scale reduced fluid models with gyroviscous effects", Journal of Plasma Physics **83**, 30402 (2017).

[253]     E.Y. Choueiri, "Plasma oscillations in Hall thrusters", Physics of Plasmas **8**, 1411 (2001).

[254]     O. Koshkarov, PhD Thesis, University of Saskatchean, 2018.

[255]    J.P. J.P. Boeuf, "Micro instabilities and rotating spokes in the near-anode region of partially magnetized plasmas", "Micro instabilities and rotating spokes in the near-anode region of partially magnetized plasmas", Physics of Plasmas **26**, 072113 (2019).

[256]    M. Mavridis, H. Isliker, L. Vlahos, T. Gorler, F. Jenko, D. Told, "A study of self-organized criticality in ion temperature gradient mode driven gyrokinetic turbulence", Physics of Plasmas **21**, 102312 (2014).

[257]    A. Lopez Ortega, I. G. Mikellides, and I. Katz, "Hall2De Numerical Simulations for the Assessment of Pole Erosion in a Magnetically Shielded Hall Thruster," in 34th International Electric Propulsion Conference, Hyogo-Kobe, Japan, IEPC-2015-249, July 2015.

[258]    V.H. Chaplin, B.A. Jorns, A. Lopez Ortega, I. G. Mikellides, R.W. Conversano, R.B. Lobbia, R. R. Hofer, "Laser-induced Fluorescence Measurements of Acceleration Zone Scaling in the 12.5 kW HERMeS Hall Thruster," Journal of Applied Physics **124**, 183302 (2018).

[259]    I. Katz, V. H. Chaplin and A. López Ortega, "Particle-in-cell Simulations of Hall Thruster Acceleration and Near Plume Regions", Physics of Plasmas **25**, 123504, (2018).

[260]    F. Taccogna and L. Garrigues, "Latest progress in Hall thrusters plasma modeling," Rev. Mod. Plasma Phys. **3**, 12 (2019).

[261]    V. V. Zhurin, H. R. Kaufman, and R. S. Robinson, "Physics of closed drift thrusters," Plasma Sources Sci. Technol. **8**, R1 (1999).

[262]    F. Taccogna, P. Minelli, Z. Asadi, and G. Bogopolsky, "Numerical studies of the E×B electron drift instability in Hall thrusters," Plasma Sources Sci. Technol. **28**, 064002 (2019).

[263]    M. M. Turner, "Kinetic properties of particle-in-cell simulations compromised by Monte Carlo collisions," Phys. Plasmas **13**, 033506 (2006).

[264]    M. M. Turner, "Numerical effects on energy distribution fonctions in particle-in-cell simulations with Monte Carlo collisions: Choosing numerical parameters," Plasma Sources Sci. Technol. **22**, 055001 (2013).

[265]    P. Y. Lai, T. Y. Lin, Y. R. Lin-Liu, and S. H. Chen, "Study of discrete-particle effects in a one-dimensional plasma simulation with the Krook type collision model," Phys. Plasmas **21**, 122111 (2014).

[266]    R. A. Fonseca, J. Vieira, F. Fiuza, A. Davidson, F. S. Tsung, W.B. Mori, and L. O. Silva, "Exploiting multi-scale parallelism for large scale numerical modeling of laser wakefield accelerators," Plasma Physics and Controlled Fusion **55**, 124011 (2013).

[267]    I. A. Surmin, S. I. Bastrakov, E. S. Efimenko, A. A. Gonoskova, A. V. Korzhimanova, and I. B. Meyerov, "Particle-in-Cell laser-plasma simulation on Xeon Phi coprocessors," Computer Physics Communications **202**, 204 (2016).

[268]    J. Derouillat, A. Beck, F. Pérez, T. Vinci, M. Chiaramello, A. Grassi, M. Flé, G. Bouchard, I. Plotnikov, N. Aunai, J. Dargent, C. Riconda, and M. Grech, "Smilei: A collaborative, open-source, multi-purpose particle-in-cell code for plasma simulation," Computer Physics Communications **222**, 351 (2018).

[269]    H. Vincenti, M. Lobet, R. Lehe, J.-L. Vay, and J. Deslippe, in Exascale Scientific Applications, Scalability and Performance Portability, Edited by T. Straatsma, K. B. Antypas, T. J. Williams, CRC Press, p. 375 (2018).

[270]    see http://computation.llnl.gov/project/linear_solvers/software.php (2015).

[271]    see https://www2.cisl.ucar.edu/resources/legacy/fishpack (2011).

[272]    see http://crd-legacy.lbl.gov/~xiaoye/SuperLU/ (2015).

[273]    see http://www.mcs.anl.gov/petsc (2015).

[274]    see http://www.research.ibm.com/projects/wsmp (2015).

[275]    K. J. Bowers, "Accelerating a Particle-in-Cell simulation using a hybrid counting sort" Journal of Computational Physics **173**, 393 (2001).

[276]    V. K. Decyk and T. V. Singh, "Particle-in-Cell algorithms for emerging computer architectures," Computer Physics Communications **185**, 708 (2014).

[277]    P. Minelli and F. Taccogna, "How to Build PIC-MCC Models for Hall Microthrusters," IEEE Transactions on plasma science **46**, 219 (2017).

[278]    I. Levchenko, K. Bazaka, Y. Ding, Y. Raitses, S. Mazouffre, T. Henning, P. J. Klar, S. Shinohara, J. Schein, L. Garrigues, M. Kim, D. Lev, F. Taccogna, R. W. Boswell, C. Charles, H. Koizumi, Y. Shen, C. Scharlemann, M. Keidar, and S. Xu, "Space micropropulsion systems for Cubesats and small satellites: from proximate targets to furthermost frontiers," Applied Physics Reviews **5**, 011104 (2018).

[279]    W. Bleakney and J. A. Hipple, "A New Mass Spectrometer with Improved Focusing Properties", Phys. Rev. **53**, 521 (1938).

[280]    E. O. Lawrence (1958), "Calutron system", US patent #2847576.

[281]    M. W. Grossman and T. A. Shepp, "Plasma isotope separation methods", IEEE Trans. Plasma Sci. **19**, 1114 (1991).

[282]    B. Lehnert, "Rotating Plasmas", Nucl. Fusion **11**, 485 (1971).

[283]    B. Lehnert, "The Partially Ionized Plasma Centrifuge", Phys. Scr. **7**, 102 (1973).

[284]    M. Krishnan, M. Geva and J. L. Hirshfield, "Plasma Centrifuge", Phys. Rev. Lett. **46**, 36(1981).

[285]    R. Gueroult, D. T. Hobbs and N. J. Fisch, "Plasma filtering techniques for nuclear waste remediation", J. Hazard. Mater. **297**, 153 (2015).

[286]    D. A. Dolgolenko and Y. A. Muromkin, "Separation of mixtures of chemical elements in plasma", Phys. Usp. **60**, 994. (2017).

[287]    A. V. Timofeev, "On the theory of plasma processing of spent nuclear fuel", Sov. Phys. Usp. **57**, 990 (2014).

[288]    R. Gueroult and N. J. Fisch, "Plasma mass filtering for separation of actinides from lanthanides", Plasma Sources Sci. Technol. **23**, 035002 (2014).

[289]    N. A. Vorona, A. V. Gavrikov, A. A. Samokhin, V. P. Smirnov and Y. S. Khomyakov, "On the possibility of reprocessing

spent nuclear fuel and radioactive waste by plasma methods", Phys. At. Nucl. **78**, 1624 (2015).

[290]   V. B. Yuferov, S. V. Shariy, T. I. Tkachova, V. V. Katrechko, A. S. Svichkar, V. O. Ilichova,  M. O. Shvets and E. V. Mufel, "The magnetoplasma separation method of spent nuclear fuel", Problems Atomic Sci. Tech. Ser.: Plasma physics **107**, 223 (2017).

[291]   R. Gueroult, J. M. Rax and N. J. Fisch, "Opportunities for plasma separation techniques in rare earth elements recycling", J. Clean. Prod. **182**, 1060 (2018).

[292]    R. Gueroult, J.-M. Rax, S. Zweben and N. J. Fisch, "Harnessing mass differential confinement effects in magnetized rotating plasmas to address new separation needs", Plasma Phys. Control. Fusion **60**,  014018 (2018).

[293]   R. Gueroult, S. J. Zweben, N. J. Fisch and J.-M. Rax, "E x B configurations for high-throughput plasma mass separation: An outlook on possibilities and challenges", Phys. Plasmas **26**, 043511 (2019).

[294]   T. Ohkawa and R. L. Miller, "Band gap ion mass filter", Phys. Plasmas **9**, 5116 (2002).

[295]   A. J. Fetterman and N. J. Fisch, "The magnetic centrifugal mass filter", Phys. Plasmas **18**, 094503 (2011).

[296]   R. Gueroult, J.-M. Rax and N. J. Fisch, "The double well mass filter", Phys. Plasmas **21**, 020701 (2014).

[297]   A. I. Morozov and V. V. Savel'ev, "Axisymmetric plasma-optic mass separators", Plasma Phys. Rep. **31**, 417 (2005).

[298]   V. M. Bardakov, G. N. Kichigin and N. A. Strokin, "Mass separation of ions in a circular plasma flow", Tech. Phys. Lett. **36**, 185 (2010).

[299]   V. P. Smirnov, A. A. Samokhin, N. A. Vorona and A. V. Gavrikov, "Study of charged particle motion in fields of different configurations for developing the concept of plasma separation of spent nuclear fuel", Plasma Phys. Rep.  **39**, 456 (2013).

[300]   S. J. Zweben, R. Gueroult and N. J. Fisch, "Plasma mass separation", Phys. Plasmas **25**, 090901 (2018).

[301]   R. Gueroult, J.-M. Rax and N. J. Fisch, "A necessary condition for perpendicular electric field control in magnetized plasmas", Phys. Plasmas **26**, 122106 (2019).

[302]   W. Horton, "Drift waves and transport", Rev. Mod. Phys. **71**, 735 (1999).

[303]   M. E. Fenstermacher, "Effects of random fluctuations in external magnetic field on plasma conductivity", Phys. Fluids **26**, 475 (1983).

[304]   V. Rozhansky, "Mechanisma of Transverse Conductivity and Generation of Self-Consistent Electric Fields in Strongly Ionized Magnetized Plasma", In V. D. Shafranov, editor, *Reviews of Plasma Physics*, volume 24.  Springer-Verlag Berlin Heidelberg (2008).

[305]   J. M. Rax, E. J. Kolmes, I. E. Ochs, N. J. Fisch and R. Gueroult, "Nonlinear ohmic dissipation in axisymmetric DC and RF driven rotating plasmas", Phys. Plasmas **26**,  012503 (2019).

[306]   E. J. Kolmes, I. E. Ochs, M. E. Mlodik, J.-M. Rax, R. Gueroult and N. J. Fisch, "Radial current and rotation profile tailoring in highly ionized linear plasma devices", Phys. Plasmas **26**, 082309 (2019).

[307]   G. D. Liziakin, A. V. Gavrikov, Y. A. Murzaev, R. A. Usmanov and V. P. Smirnov, "Parameters influencing plasma column potential in a reflex discharge",  Phys. Plasmas **23**, 123502 (2016).

[308]   H. Alvén, "Collision between a Nonionized Gas and a Magnetized Plasma", Rev. Mod. Phys **32**, 710 (1960).

[309]   J. M. Rax, R. Gueroult and N. J. Fisch, "Efficiency of wave-driven rigid body rotation toroidal confinement", Phys. Plasmas **24**, 032504 (2017).

[310]   E. J. Kolmes, I. E. Ochs and N. J. Fisch, "Strategies for advantageous differential transport of ions in magnetic fusion devices", Phys Plasmas **25**, 032508 (2018).

[311]   P. J. Roache, "Verification and Validation in Computational Science and Engineering" (Hermosa, Albuquerque, 1998).

[312]   W. L. Oberkampf and T. G. Trucano, "Verification and validation in computational fluid dynamics", Prog. Aerosp. Sci. **38**, 209 (2002).

[313]   W. L. Oberkampf, T. G Trucano, C. Hirsch, "Verification, validation, and predictive capability in computational engineering and physics", Appl. Mech. Rev. **57**,  345-384 (2004) https://doi.org/10.1115/1.1767847

[314]   F. Riva, C. F. Beadle, and P. Ricci, "A methodology for the rigorous verification of Particle-in-Cell simulations", Phys. Plasmas **24**, 055703 (2017); https://doi.org/10.1063/1.4977917

[315]   M. Greenwald, "Verification and validation for magnetic fusion", Physics of Plasmas **17**, 058101 (2010); https://doi.org/10.1063/1.3298884

[316]   P. Ricci , F. Riva , C. Theiler, A. Fasoli, I. Furno, F. D. Halpern, and J. Loizu, "Approaching the investigation of plasma turbulence through a rigorous verification and validation procedure: A practical example", Phys. Plasmas **22**, 055704 (2015); https://doi.org/10.1063/1.4919276

[317]   W. L. Oberkampf and T. G. Trucano, "Verification and validation benchmarks", Nucl. Eng. Des. 238, 716 (2008).

[318]   M. Surendra, "Radiofrequency discharge benchmark model comparison", Plasma Sources Science and Technology **4**, 56 (1995).

[319]   M. M. Turner, A. Derzsi, Z. Donko, S. J. Kelly,  T. Lafleur, and T. Mussenbrock, "Simulation benchmarks for low-pressure plasmas: Capacitive discharges", Physics of Plasmas **20**, 013507 (2013).

[320]   M. L. Brake, J. Pender, and J. Fournier, "The Gaseous Electronic Conference (GEC) reference cell as a benchmark for understanding microelectronics processing plasmas", Phys. Plasmas **6**, 2307 (1999).

[321]   Z Donkó, A Derzsi, I Korolov, P Hartmann, S Brandt, J Schulze, B Berger, M Koepke, B Bruneau, E Johnson, T Lafleur, J-P Booth, A R Gibson, D O'Connell, T Gans, "Experimental benchmark of kinetic simulations of capacitively coupled plasmas in molecular gases", Plasma Physics and Controlled Fusion **60**, 014010 (2018).

[322]   M Turner, D Eremin, P Hartmann, A Derszi, Z Donko, R Lucken, P Chabert,  T Mussenbrock, P Stolz, "Benchmarks for two-dimensional electrostatic particle-in-cell simulations", 71st Annual Gaseous Electronics Conference, Portland November 5-9, 2018.

[323]   R Lucken, V Croes, T Lafleur JL Raimbault, A. Bourdon and P. Chabert, "Edge-to-center plasma density ratios in two-dimensional plasma discharges", Plasma Sources Sci. Technol. **27**, 035004, (2018).

[324]   B Bagheri, J Teunissen, U. Ebert, M M Becker, S Chen, O Eichwald, D. Loffhagen, A. Luque, M. Yousfi, "Comparison of six simulation codes for positive streamers in air", Plasma Sources Sci. Technol. **27**, 095002, (2018).

[325]     https://www.landmark-plasma.com/services

[326]     T. Charoy, J-P Bœuf, W. Villafana, A. Tavant, A. Bourdon, P. Chabert, "In LANDMARK : The 2D axialazimuthal Particle-In-Cell benchmark on E×B discharges" https://htx.pppl.gov/E×B 2018presentations/Thursday/6%20Charoy_E×B _workshop.pdf

[327]     J.P. Bœuf, L. Garrigues,  "E x B electron drift instability in Hall thrusters: Particle-in-cell simulations vs. theory", Phys. Plasmas **25**, 061204, (2018).

[328]     J. Carlsson, A. Khrabrov, T. J. Sommerer, and D. Keating, "Validation and benchmarking of two particle-in-cell codes for a glow discharge", Plasma Sources Sci. Technol. **26**, 14003 (2017).

[329]      E. A. Den Hartog, D. A. Doughty, and J. E. Lawler, "Laser optogalvanic and fluorescence studies of the cathode region of a glow discharge", Phys. Rev. A, 38, 2471 (1988).

[330]     A. V. Khrabrov and I. D. Kaganovich," Electron scattering in helium for Monte Carlo simulations", Physics of Plasmas **19**, 093511 (2012).

[331]     H. Wang, V.S. Sukhomlinov, I.D. Kaganovich, and A.S. Mustafaev, "Simulations of Ion Velocity Distribution Functions Taken into Account Both Elastic and Charge Exchange Collisions", Plasma Sources and Technology **26**, 024001 (2017).

[332]     H. Wang, V. S. Sukhomlinov, I. D. Kaganovich, and A. S. Mustafaev, "Ion velocity distribution functions in argon and helium discharges: detailed comparison of numerical simulation results and experimental data." Plasma Sources and Technology **26**, 024002 (2017).

[333]     H. Wang, V.S. Sukhomlinov, I.D. Kaganovich, and A.S. Mustafaev, "Simulations of Ion Velocity Distribution Functions Taken into Account Both Elastic and Charge Exchange Collisions", Plasma Sources and Technology **26**, 024001 (2017).

[334]     H. Wang, V.S. Sukhomlinov, I. D. Kaganovich, and A. S. Mustafaev, "Ion velocity distribution functions in argon and helium discharges: detailed comparison of numerical simulation results and experimental data." Plasma Sources and Technology **26**, 024002 (2017).

[335]     I. G. Mikellides, I. Katz, D. M. Goebel, and J. E. Polk, "Hollow cathode theory and experiment. II. A two-dimensional theoretical model of the emitter region," Journal of Applied Physics **98**, no. 11, Dec 1, 2005.

[336]     I. G. Mikellides, D. M. Goebel, B. A. Jorns, J. E. Polk, and P. Guerrero, "Numerical Simulations of the Partially Ionized Gas in a 100-A LaB6 Hollow Cathode," IEEE Transactions on Plasma Science **43**, 173 (2015).

[337]     I. G. Mikellides, A. Lopez Ortega, D. M. Goebel, and G. Becatti, "Dynamics of a hollow cathode discharge in the frequency range of 1-500 kHz," Plasma Sources Sci. Technol. **29**, 035003 (2020).

[338]     I. G. Mikellides, and I. Katz, "Numerical simulations of Hall-effect plasma accelerators on a magnetic-field-aligned mesh," Physical Review E **86**, 046703 (2012).

[339]     I. G. Mikellides, A. L. Ortega, and B. A. Jorns, "Assessment of Pole Erosion in a Magnetically Shielded Hall Thruster," in 50th AIAA/ASME/SAE/ASEE Joint Propulsion Conference, Cleveland, OH, AIAA-2014-3897, July 2014.

[340]     A. Lopez Ortega, and I. G. Mikellides, "A New Cell-Centered Implicit Numerical Scheme for Ions in the 2-D Axisymmetric Code Hall2De," in 50th AIAA/ASME/SAE/ASEE Joint Propulsion Conference, Cleveland, OH, AIAA-2014-3621, July 2014.

[341]     B. A. Jorns, D. M. Goebel, and R. R. Hofer, "Plasma Perturbations in High-Speed Probing of Hall Thruster Discharge Chambers: Quantification and Mitigation," in 51st AIAA/SAE/ASEE Joint Propulsion Conference Orlando, FL, AIAA-2015-4006, July 2015, pp. 52.

[342]     A. Lopez Ortega, I. G. Mikellides, V. H. Chaplin, J. S. Snyder, and G. Lenguito, "Facility pressure effects on a Hall thruster with an external cathode, I: numerical simulations," Plasma Sources Sci. Technol. **29**, 035011 (2020).

[343]     I. G. Mikellides, A. Lopez Ortega, V. H. Chaplin, J. S. Snyder, and G. Lenguito, "Facility pressure effects on a Hall thruster with an external cathode, II: a theoretical model of the thrust and the significance of azimuthal asymmetries in the cathode plasma," Plasma Sources Sci. Technol. **29**, 035010 (2020).

[344]     A. Lopez Ortega, I. G. Mikellides, R. W. Conversano, R. Lobbia, and V. H. Chaplin, "Plasma Simulations for the Assessment of Pole Erosion in the Magnetically Shielded Miniature Hall Thruster (MaSMi)," in 36th International Electric Propulsion Conference, Vienna, Austria, IEPC-2019-281, September, 2019.

[345]     A. Lopez Ortega, I. G. Mikellides, M. J. Sekerak, and B. A. Jorns, "Plasma simulations in 2-D (r-z) Geometry for the assessment of pole erosion in a magnetically shielded Hall thruster," Journal of Applied Physics **125**, 033302 (2019).

[346]     M. J. Sekerak, R. R. Hofer, J. E. Polk, B. A. Jorns, and I. G. Mikellides, "Wear testing of a magnetically shielded Hall thruster at 2000-s specific impulse," in 34th International Electric Propulsion Conference, Hyogo-Kobe, Japan, IEPC 2015-155, July 2015.

[347]     M. Merino, E. Ahedo, 'Space Plasma Thrusters: Magnetic Nozzles for', in 'Encyclopedia of Plasma Technology', 1st Edition, Shohet, J., Ed.; Taylor & Francis: New York, Vol. 2, 1329-1351 (2016).

[348]     G. P. Sutton, O. Biblarz, 'Rocket Propulsion Elements', Wiley, 2010

[349]     K. Takahashi, 'Helicon-type radiofrequency plasma thrusters and magnetic plasma nozzles', Reviews of Modern Plasma Physics **3**, 3 (2019).

[350]     J. Navarro-Cavallé, M. Wijnen, P. Fajardo, E. Ahedo, 'Experimental characterization of a 1 kW Helicon Plasma Thruster', Vacuum **149**, 69 (2018).

[351]     S. Correyero, J. Jarrige, D. Packan, E. Ahedo, 'Plasma beam characterization along the magnetic nozzle of an ECR thruster', Plasma Sources & Science Technology **28**, 095004 (2019)

[352]     C. S. Olsen et al., 'Investigation of Plasma Detachment from a Magnetic Nozzle in the Plume of the VX-200 Magnetoplasma Thruster', IEEE Trans. Plasma Sci. **43**, 252 (2015).

[353]     G. Krülle, M. Auweter-Kurtz, and A. Sasoh, 'Technology and application aspects of applied field magnetoplasmadynamic propulsion', J. Propul. Power **14**, 754 (1998).

[354]     S. Andersen, V. Jensen, P. Nielsen, N. D'Angelo, 'Continuous supersonic plasma wind tunnel', Phys. Fluids **12**, 557 (1969).

[355]     K. Schoenberg, R. Gerwin, R. Moses, J. Scheuer, H. Wagner, 'Magnetohydrodynamic flow physics of magnetically nozzled plasma accelerators with applications to advanced manufacturing', Phys. Plasmas **5**, 2090 (1998).

[356]    I.D. Kaganovich, M. Misina, R. Gijbels and S.V. Berezhnoi, "Electron Boltzmann kinetic equation averaged over fast electron bouncing and pitch-angle scattering for fast modeling of electron cyclotron resonance discharge", Phys. Rev. E. **61**, 1875 (2000).

[357]    I. D. Kaganovich, V. I. Demidov, S. F. Adams, Y. Raitses,"Non-local collisionless and collisional electron transport in low-temperature plasma", Plasma Physics and Controlled Fusion **51**, 124003 (2009).

[358]    E. Ahedo, M. Merino, 'Two-dimensional supersonic plasma expansion in a magnetic nozzle', Phys. Plasmas **17**, 073501 (2010).

[359]    M. Martínez Sánchez, J. Navarro-Cavallé, E. Ahedo, 'Electron cooling and finite potential drop in a magnetized plasma expansion', Phys. Plasmas **22**, 053501 (2015).

[360]    E. Ahedo, S. Correyero, J. Navarro, M. Merino, 'Macroscopic and parametric study of a kinetic plasma expansion in a paraxial magnetic nozzle', Plasma Sources Science Technology **29**, 045017 (2020).

[361]    A. Sasoh, 'Simple formulation of magnetoplasmadynamic acceleration', Phys. Plasmas **1**, 464 (1994).

[362]    B. W. Longmier et al., 'Ambipolar ion acceleration in an expanding magnetic nozzle', Plasma Sources Science and Technology **20**, 015007 (2011).

[363]    C.S. Corr, R.W. Boswell, C. Charles, J. Zanger, 'Spatial evolution of an ion beam created by a geometrically expanding low-pressure argon plasma', Appl. Phys. Lett. **92**, 221508 (2008).

[364]    M. Merino, E. Ahedo, 'Influence of Electron and Ion Thermodynamics on the Magnetic Nozzle Plasma Expansion', IEEE Transactions on Plasma Science **43**, 244-251 (2015).

[365]    M. Merino, E. Ahedo, 'Plasma detachment in a propulsive magnetic nozzle via ion demagnetization', Plasma Sources Sci. Technol.  **23**, 032001 (2014).

[366]    C. A. Deline et al. 'Plume detachment from a magnetic nozzle', Phys. Plasmas **16**, 033502 (2009).

[367]    E. Ahedo, M. Merino, 'On plasma detachment in propulsive magnetic nozzles', Phys. Plasmas **18**, 053504  (2011).

[368]    E. Ahedo, M. Merino, '2D plasma expansion in a magnetic nozzle: separation due to electron inertia', Physics of Plasmas **19**, 083501 (2012).

[369]    K. Takahashi, T. Lafleur, C. Charles, P. Alexander, R. W. Boswell, 'Axial force imparted by a current-free magnetically expanding plasma', Phys. Rev. Lett. **107**, 235001 (2011).

[370]    T. Vialis, J. Jarrige, D. Packan. 'Separate measurements of magnetic and pressure thrust contributions in a magnetic nozzle electron cyclotron resonance plasma thruster', Paper SP2018-00499, 6th Space Propulsion Conference, Sevilla, May (2019).

[371]    S. Pottinger, V. Lappas, C. Charles, R. Boswell, 'Performance characterization of a helicon double layer thruster using direct thrust measurements', J. Phys. D: Appl. Phys. **44**, 235201 (2011).

[372]    A. Fruchtman, K. Takahashi, C. Charles, R. W. Boswell, 'A magnetic nozzle calculation of the force on a plasma', Phys. Plasmas **19**, 033507 (2012).

[373]    B. R. Roberson, R. Winglee, J. Prager, 'Enhanced diamagnetic perturbations and electric currents observed downstream of the high-power helicon', Phys. Plasmas **18**, 053505-1–11 (2011).

[374]    M. Merino, E. Ahedo, 'Effect of the Plasma-induced Magnetic Field on a Magnetic Nozzle', Plasma Sources & Science Technology **25**, 045012 (2016).

[375]    B. N. Breizman, M. R. Tushentsov, and A. V. Arefiev, 'Magnetic nozzle and plasma detachment model for a steady-state flow', Phys. Plasmas **15**, 057103 (2008).

[376]    I. D. Kaganovich, E. A. Startsev, A. B. Sefkow, R. C. Davidson, "Controlling Charge and Current Neutralization of an Ion Beam Pulse in a Background Plasma by Application of a Solenoidal Magnetic Field", Phys. Plasmas **15**, 103108 (2008).

[377]    M. Dorf, I.D. Kaganovich, E.A. Startsev, and R.C. Davidson, "Enhanced Self-Focusing of an Ion Beam Pulse Propagating Through a Background Plasma Along a Solenoidal Magnetic Field", Phys. Rev. Lett. **103**, 075003 (2009).

[378]    M. Dorf, I. D. Kaganovich, E. A. Startsev, and R. C. Davidson, "Collective focusing of intense ion beam pulses for high-energy density physics applications", Phys. Plasmas **18**, 033106 (2011).

[379]    S. Robertson, "Collective focusing of a charge-neutral ion beam with warm electrons", J. Appl. Phys. **59**, 1765 (1986).

[380]    R. Kraft, B. Kusse, and J. Moschella, "Collective focusing of an intense pulsed ion beam", Phys. Fluids **30**, 245 (1987).

[381]    A. Gincharov, "Invited Review Article: The electrostatic plasma lens", Rev. Sci. Instrum. **84**, 021101 (2013).

[382]    A. I. Morozov,  "Focusing of cold quasineutral beams in electromagnetic fields," Dokl. Acad. Nauk. USSR **163**, 1363 (1965).

[383]    A. A. Goncharov and I. G. Brown, "Plasma devices based on the plasma lens—A review of results and applications", IEEE Trans. Plasma Sci. **35**, 986 (2007).

[384]    M. E Griswold, Y. Raitses, and N J Fisch, "Cross-field plasma lens for focusing of the Hall thruster plume", Plasma Sources Sci. Technol. **23** 044005 (2014)

[385]    A. Smirnov, Y. Raitses, and N. J. Fisch, "Experimental and theoretical studies of cylindrical Hall thrusters", Phys. Plasmas **14**, 057106 (2007).

[386]    G. Kornfeld, N. Koch, H.P. Harmann, "Physics and Evolution of HEMP-Thrusters", Proc. 30th Int. Electric Propulsion Conf. (Florence, Italy, 2007), IEPC 2007-108

[387]    D. G. Courtney, M Martinez-Sanchez, "Diverging cusped-field Hall thruster", Proc. 30th Int. Electric Propulsion Conf. (Florence, Italy, 2007), IEPC 2007-039

[388]    N. J. Fisch, A. Fruchtman, and Y. Raitses, "Ion acceleration in supersonically rotating magnetized-electron plasma", Plasma Physics and Controlled Fusion **53**, 124038 (2011).

[389]    Y. Raitses, A. Smirnov, and N. J. Fisch, "Enhanced performance of cylindrical Hall thrusters", Applied Physics Letters **90**, 221502 (2007).

[390]    I.D. Kaganovich, V.A. Rozhansky, L.D. Tsendin and I.Yu. Veselova, "Fast expansion of a plasma beam controlled by short-circuiting effects in a longitudinal magnetic field", Plasma Sources Sci. Technol. **5**, 743 (1996).

[391]    G. Sánchez-Arriaga, J. Zhou, E. Ahedo, M. Martínez-Sánchez, J. J. Ramos, 'Kinetic features and non-stationary electron trapping in paraxial magnetic nozzles', Plasma Sources Sci. Technol. **27**, 035002 (2018).

[392]    J. Zhou, G. Sánchez-Arriaga, E. Ahedo, M. Martínez-Sánchez, J. Ramos, 'Collisional Effects in Non-stationary Plasma Expansions along Convergent-Divergent Magnetic Nozzles', paper SP2018-00332, 6th Space Propulsion Conference, Sevilla, May (2019)

[393]    J. G. Laframboise, 'Theory of spherical and cylindrical Langmuir probes in a collisionless, Maxwellian plasma at rest', No. UTIAS-100. Toronto Univ. Downsview (Ontario) Inst. Aerospace Studies (1966).

[394]    M. Merino, J. Mauriño, E. Ahedo, 'Kinetic electron model for plasma thruster plumes', Plasma Sources Sci. Technol. **27**, 035013 (2018).

[395]    C. Charles, R.W. Boswell, 'Laboratory evidence of a supersonic ion beam generated by a current-free helicon double-layer', Phys. Plasmas **11**, 1706 (2004).

[396]    M. Merino, E. Ahedo, '2D quasi-double-layer in a two-electron-temperature, magnetically-expanded, current-free plasma', Physics of Plasmas **20**, 023502 (2013).

[397]    R. T. S. Chen, N. Hershkowitz, 'Multiple Electron Beams Generated by a Helicon Plasma Discharge', Phys. Rev. Lett. **80**, 4677 (1998).

[398]    J. M. Little and E. Y. Choueiri, 'Electron Cooling in a Magnetically Expanding Plasma', Phys. Rev. Lett. **117**, 225003 (2016).

[399]    Y. Zhang, C. Charles, and R. Boswell, 'Thermodynamic study on plasma expansion along a divergent magnetic field', Phys. Rev. Lett. **116**, 025001 (2016).

[400]    J. P. Sheehan et al, 'Temperature gradients due to adiabatic plasma expansion in a magnetic nozzle', Plasma Sources Sci. Technol. **23**, 045014 (2014).

[401]    J. Y. Kim, K. S. Chung, S. Kim, J. H. Ryu, K.-J. Chung, Y. S. Wang, 'Thermodynamics of a magnetically expanding plasma with isothermally behaving confined electrons', New J. Phys. **20**, 063033 (2018).

[402]    J. Y. Kim, J. Y. Jang, K. S. Chung, K. J. Chung, Y. S. Hwang, 'Time-dependent kinetic analysis of trapped electrons in a magnetically expanding plasma'. Plasma Sources Science and Technology **28**, 07LT01 (2019).

[403]    M. Merino et al., 'Collisionless electron cooling in a plasma thruster plume: experimental validation of a kinetic model', Plasma Sources Science and Technology **29**, 035029 (2020).

[404]    K. Dannenmayer, S. Mazouffre, 'Electron flow properties in the far-field plume of a Hall thruster', Plasma Sources Sci. Technol. **22**, 035004 (2013).

[405]    E. Ahedo, J. Navarro, 'Helicon thruster plasma modeling: two-dimensional fluid-dynamics and propulsive performances', Phys. Plasmas **20**, 043512 (2013).

[406]    R. W. Boswell, K. Takahashi, C. Charles, I. D. Kaganovich, 'Non-local electron energy probability function in a plasma expanding along a magnetic nozzle', Frontiers in Physics **3**, 14 (2015).

[407]    S. Robertson, A reduced set of gyrofluid equations for plasma flow in a diverging magnetic field", Phys. Plasmas **23**, 043513 (2016).

[408]    Zehua Guo, Xian-Zhu Tang, Chris McDevitt, "Parallel heat flux and flow acceleration in open field line plasmas with magnetic trapping", Physics of Plasmas **21**, 102512 (2014).

[409]    A. Fruchtman, 'Limits on the Efficiency of a Helicon Plasma Thruster', paper IEPC-2015-158, the 34th IEPC, Kobe, Japan, July 6-10 (2015).

[410]    B. Wachs, B. A. Jorns, 'Background pressure effects on ion dynamics in a low-power magnetic nozzle thruster', Plasma Sources Science and Technology, in press (2020).

[411]    G. Makrinich and A. Fruchtman, "Experimental study of a radial plasma source", Phys. Plasmas **16**, 043507 (2009).

[412]    G. Makrinich and A. Fruchtman, "Enhanced momentum delivery by electric force to ions due to collisions of ions with neutrals", Phys. Plasmas **20**, 043509 (2013).

[413]    J. M. Little, E. Y. Choueiri, 'Electron Demagnetization in a Magnetically Expanding Plasma', Physical Review Letters **123**, 145001 (2019).

[414]    M. Merino, E. Ahedo, 'Contactless steering of a plasma jet with a 3D magnetic nozzle', Plasma Sources & Science Technology **26**, 095001 (2017).

[415]    M. Giambusso et al., 'Investigation of Electric Field Oscillations related to VASIMR® VX-200 Plasma Plume Detachment', paper AIAA 2018-4505 (2018).

[416]    Y. Raitses, E. Merino, N. J. Fisch. 'Cylindrical Hall thrusters with permanent magnets', Journal of Applied Physics **108**, 093307 (2010).

[417]    D. G. Courtney and M. Martínez-Sánchez, 'Diverging Cusped-Field Hall Thruster', paper IEPC 2007-39, 30th International Electric Propulsion Conference, Florence, Italy, 17-20 September (2007).

[418]    J. J. Ramos, 'Dynamic evolution of the heat fluxes in a collisionless magnetized plasma', Phys. Plasmas **10**, 3601 (2003).

[419]    J. Nuez, M. Merino, E. Ahedo, 'Fluid-kinetic propulsive magnetic nozzle model in the fully magnetized limit', paper IEPC-2019-254, 36th International Electric Propulsion Conference, Vienna (2019).

[420]    J.J. Ramos, M. Merino, E. Ahedo, 'Three-dimensional fluid-kinetic model of a magnetically guided plasma jet', Physics of Plasmas **25**, 061206 (2018).

[421]    H. Lorzel, P. G. Mikellides. 'Three-dimensional modeling of magnetic nozzle processes', AIAA Journal **48**, 1494 (2010).

[422]    M. Li, M. Merino, E. Ahedo, H. Tang, 'On electron boundary conditions in PIC plasma thruster plume simulations', Plasma Sources Science Technology **28**, 034004 (2019).

[423]    Ronald C. Davidson, Igor D. Kaganovich, Edward A. Startsev, H. Qin, M. Dorf, Adam B. Sefkow, Dale R. Welch, D.V. Rose, and S.M. Lund, "Multispecies Weibel instability for intense charged particle beam propagation through neutralizing background plasma", Nuclear Instruments and Methods in Physics Research A **577**, 70 (2007).

[424]    M. Dorf, I. D. Kaganovich, E. A. Startsev, and R. C. Davidson, "Whistler Wave Excitation and Effects of Self-Focusing on Ion Beam Propagation through a Background Plasma along a Solenoidal Magnetic Field", Phys. Plasmas **17**, 023103 (2010).

[425]    X Z Tang "Kinetic magnetic dynamo in a sheath-limited high-temperature and low-density plasma", Plasma Phys. Control. Fusion **53**, 082002 (2011).

[426]    Z. Guo, X Z tang, "Ambipolar Transport via Trapped-Electron Whistler Instability Along Open Magnetic Field Lines", Physical Review Letters, **109**, 135005 (2012).

[427]    K. Takahashi et al., 'Demonstrating a new technology for space debris removal using a bi-directional plasma thruster." Scientific reports **8**, 14417 (2018).

[428]    C. Bombardelli, H. Urrutxua, M. Merino, J. Peláez, E. Ahedo, 'The ion beam shepherd: A new concept for asteroid deflection', Acta Astronautica **90**, 98 (2013).

[429]    K. Takahashi, A. Chiba, A. Komuro, A. Ando, 'Experimental identification of an azimuthal current in a magnetic nozzle of a radiofrequency plasma thruster', Plasma Sources Sci. Technol. **25**, 055011 (2016).

[430]    S. Correyero, M. Merino, P.-Q. Elias, J. Jarrige, D. Packan, E. Ahedo, 'Characterization of diamagnetism inside an ECR thruster with a diamagnetic loop', Phys. Plasmas **26**, 053511 (2019).

[431]    M. Merino, E. Ahedo, 'Fully magnetized plasma flow in a magnetic nozzle', Phys. Plasmas **23**, 023506 (2016).

[432]    K. Takahashi, C. Charles, R.W. Boswell, A. Ando, 'Adiabatic expansion of electron gas in a magnetic nozzle', Phys. Rev. Lett. **120**, 045001 (2018).

[433]    J. M. Little and E. Y. Choueiri, 'Critical Condition for Plasma Confinement in the Source of a Magnetic Nozzle Flow', IEEE Trans. on Plasma Scie. **43**, 277 (2014).

[434]    S.J. Doyle, A. Bennet, D. Tsifakis, J. P. Dedrick, R. W. Boswell, C.  Charles, 'Characterization and Control of an Ion-Acoustic Plasma Instability Downstream of a Diverging Magnetic Nozzle', Frontiers in Physics **8**, 24 (2020).

[435]    M. Merino, 'Motor espacial de plasma sin electrodos con geometría en U' ('Electrodeless plasma space thruster with U-shape'), Spanish Patent Office, PCT patent ES2733773 (2019).